\documentclass{aa}

\usepackage{graphicx}
\usepackage[varg]{txfonts}
\usepackage{natbib}
\bibpunct{(}{)}{;}{a}{}{,}
\usepackage{tikz}
\usepackage{caption,subcaption}

\usepackage{multirow}
\usepackage{hyperref}
\usepackage{gensymb}
\usepackage{lscape}
\usepackage{longtable}
\usepackage{diagbox}
\usepackage{supertabular,booktabs}
\usepackage{cleveref}
\usepackage{multicol}
\usepackage{amsmath}
\hypersetup{colorlinks=true, citecolor=blue, linkcolor=blue}
\definecolor{lightblue}{RGB}{51,131,255}

\begin{document} 

\renewcommand{\thefootnote}{\arabic{footnote}}
\let\ACMmaketitle=\maketitle
\renewcommand{\maketitle}{\begingroup\let\footnote=\thanks \ACMmaketitle\endgroup}

\title{Radial velocity homogeneous analysis of M dwarfs observed with HARPS\\ I. Exoplanet detection and candidates\footnote{All radial velocity time series and the full Table B.5 are only available at the CDS via anonymous ftp to cdsarc.u-strasbg.fr (130.79.128.5) or via http://cdsarc.u-strasbg.fr/viz-bin/cat/J/A+A/XXX/AX}}

\titlerunning{RV analysis of an M-dwarf H.A.R.P.S. sample. I.}

  \author{L. Mignon \inst{1,2}
  \and X. Delfosse \inst{1}
  \and X. Bonfils \inst{1}
  \and N. Meunier \inst{1}
  \and N. Astudillo-Defru \inst{3}
  \and G. Gaisne \inst{1}
  \and T. Forveille \inst{1}
  \and F. Bouchy \inst{2}
  \and G. Lo Curto \inst{4}
  \and S. Udry \inst{3}
  \and D. Segransan \inst{2}
  \and N. Unger \inst{2}
  \and C. Lovis \inst{2}
  \and N. C. Santos \inst{5,6} 
  \and M. Mayor \inst{2}
  }
  \authorrunning{L. Mignon et al.}

  \institute{Univ. Grenoble Alpes, CNRS, IPAG, F-38000 Grenoble, France 
\\ correspondence: \texttt{Lucile.Mignon@unige.ch}
      \and
      Observatoire astronomique de l'Université de Genève, 51 chemin Pegasi, 1290 Versoix, Switzerland
      \and
      Departamento de Matemática y Física Aplicadas, Universidad Católica de la Santísima Concepción, Alonso de Rivera 2850, Concepción, Chile
      \and European Southern Observatory, Casilla 19001, Santiago 19, Chile
      \and Instituto de Astrofísica e Ciências do Espaço, Universidade do Porto, CAUP, Rua das Estrelas, 4150-762 Porto, Portugal
      \and Departamento de Física e Astronomia, Faculdade de Ciências, Universidade do Porto, Rua Campo Alegre, 4169-007 Porto, Portugal
  }

  \date{Received ; accepted }

 
 \abstract
  {
The census of planets around M dwarfs in the solar neighbourhood meets two challenges: detecting the best targets for the future characterisation of planets with ELTs,
and studying the statistics of planet occurrence that are crucial to formation scenarios. The radial velocity (RV) method remains the most appropriate for such a census as it is sensitive to the widest ranges of masses and periods. HARPS, mounted on the 3.6 m telescope at La Silla Observatory (ESO, Chile), has been obtaining velocity measurements since 2003, and can therefore be used to analyse a very large and homogeneous dataset.
  }
  {We  performed a homogeneous analysis of the RV time series of 200 M dwarfs observed with HARPS from 2003 to 2019 (gathering more than 15000 spectra), with the aim of understanding detectable signals such as stellar and planetary companions and activity signals.
  }
  {The RVs were computed with a template matching method before carrying out the time series analysis. First, we focused on the systematic analysis of the presence of a dominant long-term pattern in the RV time series (linear or quadratic trend and sine function). Then, we analysed higher-frequency perdiodic signals using periodograms of the residual time series and Keplerian function fitting.
  }
  {We found long-term variability in 57 RV time series (28.5\%). This led to the revision of the parameters of the massive planet (GJ~9482~b), as well as the detection of four substellar and stellar companions (around GJ~3307, GJ~4001, GJ~4254, and GJ~9588), for which we characterised inclinations and masses by combining RV and astrometry. The periodic analysis allowed us to recover 97\% of the planetary systems already published in this sample, but also to propose three new planetary candidates orbiting GJ~300 (7.3M$_{\oplus}$), GJ~654(5M$_{\oplus}$), and GJ~739 (39M$_{\oplus}$), which require additional measurements before they can be confirmed.
  }
  {}


  \maketitle


\section{Introduction}\label{intro}

M dwarfs represent 80\% of the stellar neighbourhood \cite[e.g.][]{henry1994solar}.
Due to their favourable star--planet contrast and separation, planetary systems orbiting the closest M dwarfs are the most promising targets for future atmospheric characterisation by the new generation of facilities.
They provide the only opportunity in the coming decade to obtain reflected spectra of Earth-like or Neptune-type planets by combining high-contrast imaging and high-dispersion spectroscopy on extremely large telescopes (ELTs) \citep{Snellen2015, marconi2020}.
Although many planets have been identified in the very close solar neighbourhood \citep[e.g.][]{bonfils2011harps,bonfils2018temperate,delfosse2013,anglada2016terrestrial,astudillo2017harps,Diaz2019}, the census of planets around these stars is far from complete.

Such a census is also necessary for statistical studies of planets orbiting around very low-mass stars and their unbiased mass--period distribution \citep{bonfils2013harps,sabotta2021,Pinamonti2022}.
Increasing the size of the probed sample provides more precise occurrence rates, and thus provides even stronger constraints on the formation scenarios, and in particular on the dependence on the central mass \citep[e.g.][]{Burn2021}.

The radial velocity (RV) method remains the most complete way to identify planetary systems in the solar neighbourhood as it is not limited to particular configurations, unlike the transit method. 
It is particularly well suited to the search for systems around very low-mass stars, especially for planets located in their habitable zone (HZ).
The orbital distances of the HZs strongly depend on the luminosity (hence the mass) of the host stars: fainter stars have closer HZs.
Exo-Earths in the HZs of M dwarfs thus exhibit much larger RV oscillations on much shorter timescales than those orbiting Sun-like stars.

Since 2003 the HARPS spectrograph \citep{Mayor2003}, installed on the 3.6-meter telescope at La Silla Observatory (ESO, Chile), has been producing precise RV measurements for thousands of stars, including many M dwarfs in the solar neighbourhood.
Our study is a systematic and homogeneous analysis of the RV time series of 200~M~dwarfs observed on at least ten nights with HARPS (gathering more than 15000 velocities).
After computing the RVs using a template matching method, we applied an analysis based on the method of \cite[][]{cumming1999lick}, \cite[][]{cumming2008keck}, \cite[][]{zechmeister2009}, \cite[][]{zechmeister2009generalised}, and \cite[][]{bonfils2013harps}.
To summarise, we first searched for a long-term signal (modelled by different functions) in the time series, then for higher-frequency periodic signals; finally, we analysed the signals found to determine whether they are due to stellar activity or a planetary candidate.

The outline of this paper is as follows. 
In Sect.~2 we present the sample of M dwarfs observed by HARPS.
In Sect.~3 we describe the observations, our procedure to extract accurate RVs, and the automatic corrections made to build the final sample.
In Sect.~4 we present the long-term models that tested the different statistical validations used.
In Sect.~5 we present the frequency analysis of the sample, the tools used, and the identification of the Keplerian signals found and subtracted.
In Sect.~6 we present a first analysis of the stellar and planetary candidates detected.
Finally, we conclude in Sect.~7, and discuss the limitations of this study in Sect.~8.


\section{Sample}\label{secSample}
Since our study is based on a dataset entirely derived from HARPS \citep{Mayor2003}, it is 
subject to similar systematic effects, and homogeneous analysis can be performed using a consistent approach. 
The first RV follow-up programme for M dwarfs was conducted as soon as HARPS came into operation, allocating 10\% of the 500 nights of the Guaranteed Times to characterising the planetary population orbiting stars with masses below 0.6 M$_{\odot}$ \citep{Mayor2003,bonfils2013harps}.
Since then, HARPS has played a major role in the search for planets around M dwarfs, contributing to the detection of a large fraction of the M-dwarf planets thanks to numerous programmes dedicated entirely or in part to M dwarfs.
This possibly includes the first super-Earths in HZ \citep{bonfils2013GJ163, delfosse2013, anglada2013dynamically, astudillo2017harps0, bonfils2018temperate}, and the closest planet to us \citep{anglada2016terrestrial}. 
HARPS RVs were also used to confirm the planetary nature and to measure the masses of transiting planet candidates around M dwarfs originally identified by transit surveys such as K2, MEarth, and now TESS \citep[e.g.][]{Charbonneau2009, Berta-Thompson2015, Almenara2015, dittmann2017temperate, ment2019second, Winters2019, Astudillo2020,lillo2020A,lillo2020B,vaneylen2021}.

Our first objective in this study is to collect the spectra obtained by HARPS for all the programmes targeting M dwarfs. We restrict our sample to stars located within 26~pc. 

\subsection{Input catalogues}\label{subsec1}

\subsubsection{Initial selection}

Our reference catalogues of M dwarfs in the solar neighbourhood are from a compilation of \citet{gaidos2014}, \citet{winters2014solar}, \citet[the release of the Gliese and Jahreiss catalogue]{stauffer2010accurate}, \citet{henry2018}, and \citet{winters2021}.
We cross-referenced each object in the combined catalogue with the second release of the Gaia catalogue \cite[Gaia DR2][]{gaia2018vizier}\footnote{Gaia DR3\citep{GAIA2022} was not yet available when we initiated the project. After verification, the Gaia parameters used for our study did not change significantly between Gaia DR2 and DR3.} to extract very precise coordinates, parallaxes, proper motions, and G-band photometry.
Finally, we retained 1362~stars, the southern ($\delta < +20^{\circ}$) M dwarfs within 26~pc.
We cut the catalogue slightly beyond the traditional \citep{Gliese1991} 25~pc limit of the solar neighbourhood to provide a margin that includes some true members of the 25~pc volume with very slightly overestimated distances.

\begin{figure}[!h]
\includegraphics[width=0.99\linewidth]{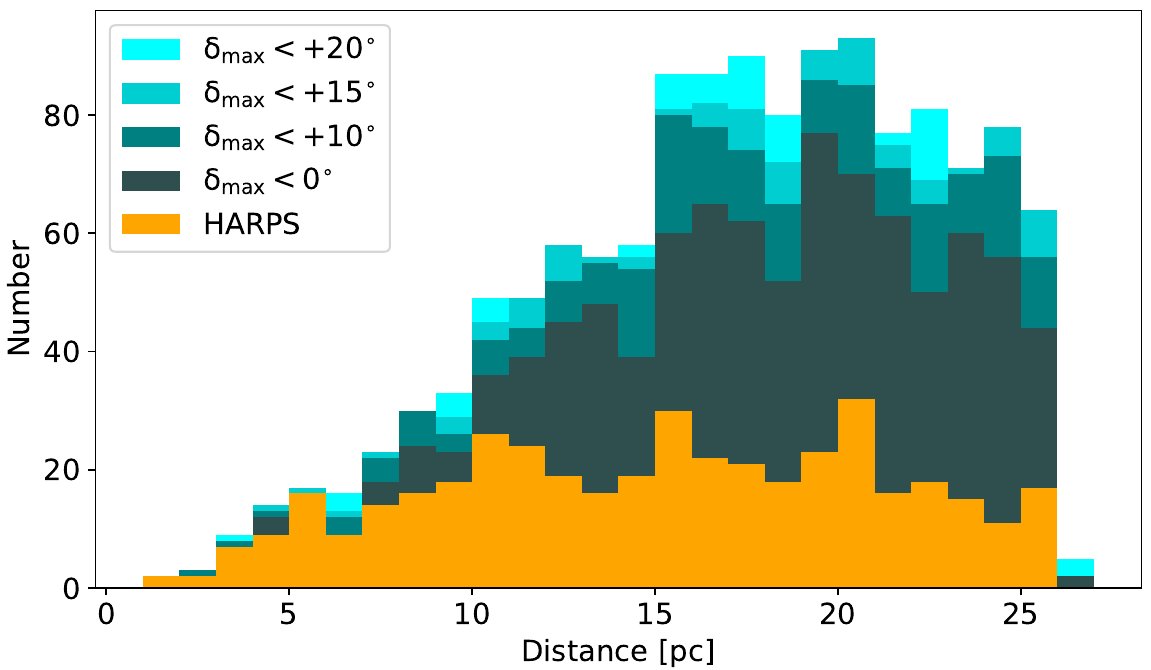}
\caption{Distance distribution of the sample. In green is the distribution of our reference catalogue for different declination limits. In orange is the distance distribution of the M dwarfs with HARPS measurements.}
\label{dist}
\end{figure}

We cross-matched the positions of these 1362~stars with those of the public HARPS spectra observed before 2019\footnote{Corresponding to data for which the proprietary time had expired when we interrogated the ESO archive in 2021; we note that HARPS has acquired few measurements of M dwarfs since 2019.} available in the ESO archives.
Of these 1362~stars, 425 have at least one HARPS spectrum, for a total of 12984~spectra.
Figure~\ref{dist} superposes the distance distributions of the 1362~M dwarfs of our combined catalogue with the 425 observed with HARPS; all the programmes used are listed in Table~\ref{progidtable}.

As detailed in the caption, the figure proposes several alternative declination cuts for the combined catalogue.
Most of the southern M dwarfs within 10~pc were observed with HARPS, as this corresponds to the distance limit of the sample studied during the GTO programme \citep{bonfils2013harps}.
Beyond this distance, only a minority of the catalogued M dwarfs were observed with HARPS, and overall 31.2\% of the stars in our combined catalogue have HARPS spectra.
Completeness rates are detailed in Table~\ref{tabcompletude} as a function of distance and declination . 

\begin{table}
\caption{Fraction of stars observed with HARPS.}
\label{tabcompletude}
\begin{center}
\begin{tabular}{l|c|c|c|c}\hline
\diagbox[width=7em]{Distance}{$\delta_{max}$}&
 $+0^{\circ}$ & $+10^{\circ}$ & $+15^{\circ}$ & $+20^{\circ}$\\ \hline 
5 pc & 71,43\% & 73,08\% & 74,07\% & 72,41\% \\ 
7 pc & 80,0\% & 81,13\% & 80,70\% & 75,0 \% \\ 
10 pc & 63,64\% & 63,91\% & 64,28\% & 62,34 \% \\ 
15 pc & 45,03\% & 48,74\% & 49,73\%&48,64\% \\ 
26 pc & 31,33\% & 34,39\% & 35,13\% &31,25\% \\ 
\hline
\end{tabular}
\end{center}
\tablefoot{Fraction of observed stars from the combined catalogue as a function of the declination and the distance.}
\end{table}

Pre-Gaia solar neighbourhood catalogues are known to be incomplete beyond 15~pc, and miss over half of the stars between 20 and 25~pc \citep[see][and in particular their figure 10]{gaia2021}.
A much more complete list of stars can now be obtained directly from the Gaia DR3 catalogue, as was done by e.g. \cite{Gillon2020} and \cite{Reyle2021}.
This, however, is not directly relevant to the scope of this paper since all observing programmes that obtained M dwarf spectra with HARPS before 2019 built upon pre-Gaia input catalogues.
The more complete Gaia catalogues can therefore only include solar neighbourhood M dwarfs that have not yet been observed by HARPS.

\subsubsection{Complementary data}

As explained above, coordinates, proper motions, and parallaxes are mainly from the Gaia catalogue \cite[Gaia DR2][]{gaia2018vizier} and completed by various sources for the 11 missing stars\footnote{\cite{van2007validation}: GJ~257, GJ~2033, GJ~4206, GJ~9163, HD196982; \cite{weinberger2016trigonometric}: GJ~406; \cite{riedel2014solar}: GJ~3332; \cite{finch2018urat}: GJ~3813, GJ~4038; \cite{winters2016solar}: L43-72; \cite{riaz2006identification}: LP~993-116}

sV-band magnitudes are from \cite{koen2010ubv} for 57\% of the sample, supplemented by \cite{zacharias2013fourth} for 15\%.
A further 15\% come from a dozen other catalogues, leaving 13\% of the targets without a V-band magnitude value.
The G-band magnitudes come from the Gaia DR2 catalogue \citet{gaia2018vizier} for 97.5\% of the sample.
The H-, J-, and K-band magnitudes come from the 2MASS database \citep{cutri2003vizier}, supplemented by H- and J-band magnitudes of GJ~205 from \citet{ducati2002vizier}, and the H-band magnitude of GJ~1 and K-band magnitude of GJ~803 from \citet{koen2010ubv}.

For our study we also needed to determine the masses of the 425 M dwarfs in our sample. 
We computed them using \citet{delfosse2000accurate} empirical mass-luminosity relation with parallaxes and photometry collected.
Since the magnitude V is not available for the entire sample and the mass--luminosity relations in the optical bands depend strongly on the metallicity, we used only the three near-infrared relationships (mass versus $M_K$, $M_H$, and $M_J$).
The mass value for each star chosen and used for this study is the average of the three values.
Table~\ref{parameters_sample} lists all the stellar parameters collected and computed for the 425 stars in the sample.

\subsection{Magnitude and mass distributions}

\begin{figure}
\includegraphics[width=\linewidth]{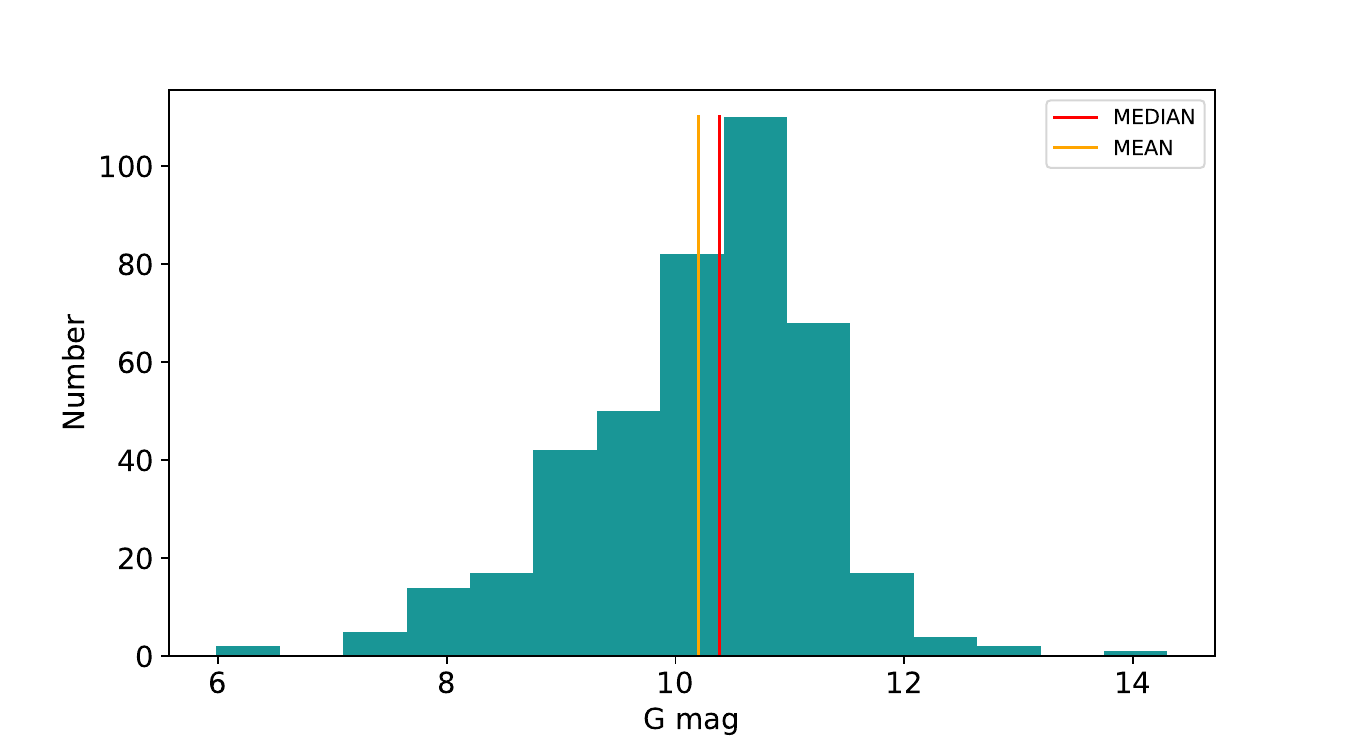}
\caption{G magnitude distribution of the sample.}
\label{fig:VG}
\end{figure}

The G-band distribution in Fig.~\ref{fig:VG} has a mean (resp. median) value of 10.20 mag (resp. 10.38 mag), and ranges from 5.98 to 14.30 mag.
Regarding the magnitude V, with a median value of 11.28 mag, our sample is almost 1 magnitude fainter than the 11~pc volume-limited sample studied by \citet{bonfils2013harps}, which will naturally impact the median photon noise of our RV measurements.

\begin{figure}
\includegraphics[width=0.99\linewidth]{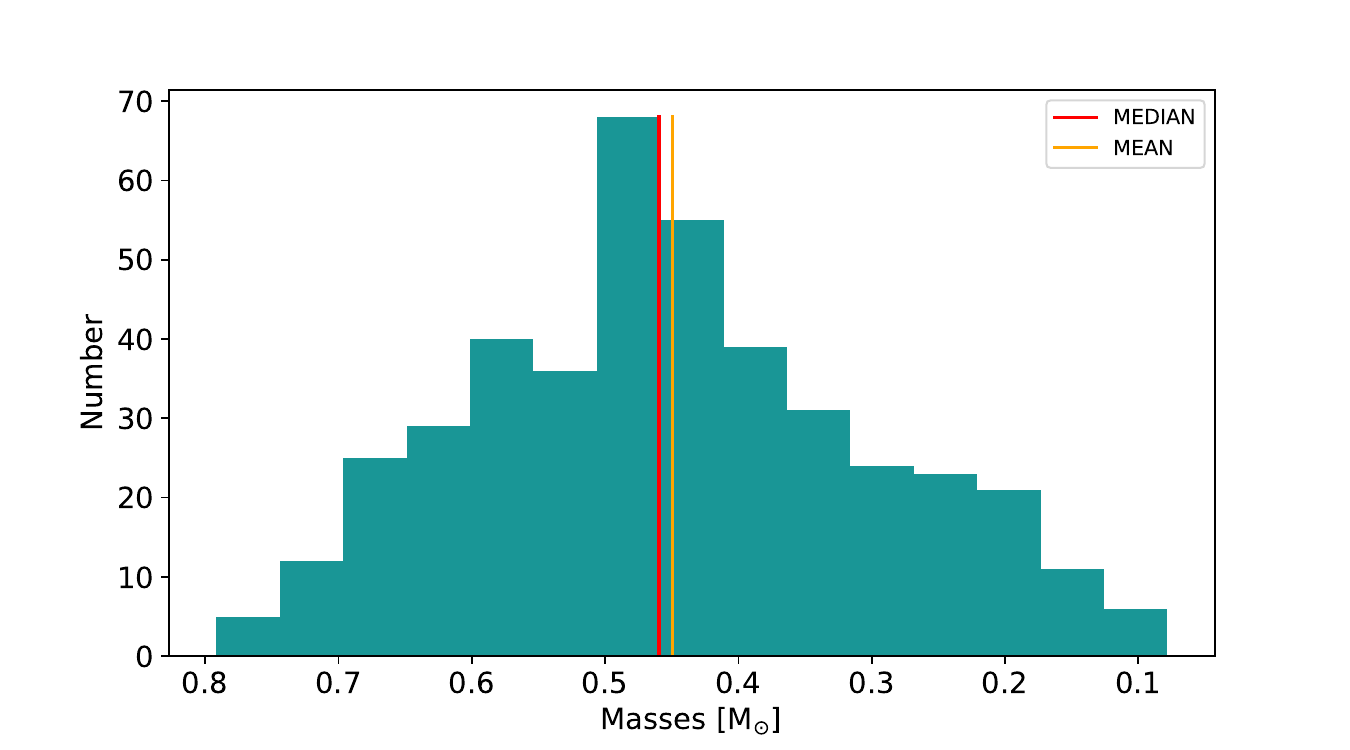}
\caption{Mass distribution of the sample.}
\label{massemoy}
\end{figure}

The mass distribution in our sample (Fig.~\ref{massemoy}) ranges from 0.79 to 0.08~M$_{\odot}$ with an average (resp. median) value of 0.45 M$_{\odot}$ (resp. 0.46 M$_{\odot}$). 
The distribution peaks between 0.5 and 0.45~M$_{\odot}$, whereas the stellar mass function of the solar neighbourhood is known to increase monotonically (on a linear scale) up to the hydrogen burning limit \citep[e.g. Figure 5 in][]{chabrier2001}. 
This simply illustrates a selection bias in our sample: the underlying observing programmes are magnitude-limited, and therefore under-represent the faint late M dwarfs.

\subsection{The 7~pc subsample}\label{7pc}

Planets orbiting the nearest M dwarfs represent the only practical short-term opportunities for characterising Earth-like planets.
In addition to transmission spectroscopy of transiting planets (e.g. with JWST \citealp[][]{doyon2014, beichman2014}, with NIRPS \citealp[][]{bouchy2017}), a new strategy using ELTs is being developed for the more general case of non-transiting planets.
\citet{Snellen2015} demonstrates that the combination of high-contrast imaging and high-dispersion spectroscopy with ELTs can be used to characterise the rocky planets around our nearest neighbours.
This ambitious objective can only be achieved from the ground in the most favourable case: lowest planet-to-star contrast and widest angular separation. 
Adopting $\rm 3\lambda / D$ in the J-band as the practical limit for resolving a planetary system by this method, and for a planet located at the outer limit of the HZ of a mid-M dwarf \citep[$\sim 0.3$ au of the central star; see e.g.][]{kopparapu2013}, the distance limit for the method on an ELT is 7~pc. 

Monitoring this sample of 61 M dwarfs at less than 7~pc (for $\rm \delta<+20^{\circ}$) is therefore a particularly important issue at the moment.
Furthermore, it is better observed than the complete sample, with 46 out of the 61 targets already observed with HARPS (75\%).
The 16 unobserved stellar systems illustrate the selection biases of RV monitoring by falling into three categories: very late-M dwarfs that are too faint (V $>$ 15), binary stars that cannot be resolved with HARPS (separation smaller than the diameter of the fibre entrance), and fast rotators already identified before the beginning of HARPS observations high RV jitter expected from magnetic activity.
Table~\ref{tabnonobs} lists the 16 closest southern systems without HARPS observations and provides the reason identified for each.

\begin{table*}
\caption{Nearest stars not observed by HARPS programmes (distance < 7 pc )}
\label{tabnonobs}
\begin{center}
\renewcommand{\footnoterule}{} 
\begin{tabular}{ l || c| c| c| c| c |c }
\hline
Star & Distance (pc) & Spectral Type & Explanation & V flux & Vsini & Separation \\
\hline
GJ~1005 & 4.99$^1$ & M4V & Close Binary & 11.483$^4$ & 4.30$^{11}$ & - \\
GJ~65 A/B & 2.7195 - 2.6749 & M5.5V - M6V & Fast Rotator \& Binary & 12.08$^4$ & 29.5$^{12}$ - 37.9$^{12}$ & 2.31 \\
GAT1370 & 3.8315 & M7 & faint late-M dwarf & 15.13 & - & - \\
LP944-20 & 6.4268 & M9 & faint late-M dwarf & 18.69$^6$ & - & - \\
GJ~1116 A/B & 5.1508 & M7V - M7V & Faint close binary & 13.93$^7$ & - & 2.48 \\
GJ~3522 & 4.39$^{3}$ & M4 & Close Binary & 10.98$^5$ & - & - \\
GJ~1128 & 6.5036 & M4 & \textit{no explanation} & 12.44 & - & - \\
GJ~3622 & 4.5593 & M6.5 & faint late-M dwarf & 15.784$^5$ & 3.3$^{12}$ & - \\
2MUCD 20385 & 4.0450 & M9V & faint late-M dwarf & 17.532$^8$ & - & - \\
GJ~1156 & 6.4641 & M4.5V & Fast Rotator & 13.90$^5$ & 15.6$^{12}$ & - \\
GJ~473 A/B & 4.3267 - 4.4747 & M5.5V - & Close Binary & 12.467$^9$ & - & - \\
UCAC4 195-119117 & 5.3268 & M6.5 & faint late-M dwarf & - & - & - \\
SCRJ1845-6357 & 4.0053 & M8.5 & faint late-M dwarf & 17.40$^{10}$ & - & - \\
GJ~829 A & 6.7798 - 3.4$^2$ & M4 & SB2 Binary & 10.303$^4$ & - & - \\
GJ~866 B & & M5 & SB2 Binary & 12.38$^5$ & - & - \\
GJ~896 A/B & 6.2631 - 6.2535 & M3.5 - M4 & Fast Rotator & 10.173$^4$ & 14.5$^{12}$ & 4.34 \\
\hline
\end{tabular}
\end{center}
\tablefoot{Distances are from \cite{gaia2020} except marked: $^1$ \cite{koen2010ubv}, $^{2}$ \cite{torres2010accurate}, and $^{3}$ \cite{riedel2014solar}. Spectral types are from \cite{gaidos2014}, V flux : $^{4}$ \cite{koen2010ubv}, $^5$ \cite{zacharias2013fourth}, $^6$ \cite{lurie2014solar}, $^7$ \cite{lepine2005}, $^8$ \cite{costa2005solar}, $^9$ \cite{landolt1992ubvri}, and $^{10}$ \cite{winters2010solar}. Vsini values are from: $^{11}$ \cite{hojjatpanah2019} and $^{12}$ \cite{fouque2018}. Separations were obtained via ExoFOP tools: \url{https://exofop.ipac.caltech.edu/tess/}.}
\end{table*}


\section{HARPS observations}\label{sec3}

\subsection{Radial velocity extraction}\label{subsec31}

12984 HARPS spectra\footnote{E2DS from \url{http://archive.eso.org/wdb/wdb/adp/phase3_spectral/form}} of the 425 M dwarfs in our sample were retrieved from the ESO archive.
Each spectrum covers the 380--690 nm spectral range, spread over 71 orders with a median resolving power of 115,000 and
is wavelength calibrated with very high accuracy.
The data were analysed by the HARPS pipeline \citep{Lovis2007}, determining the RV by cross-correlation with a binary mask \citep{Pepe2002} giving a typical precision of around 1~m/s.
The cross-correlation function (CCF) obtained by the pipeline can be considered an average spectral line.

To fully exploit the Doppler information of the spectra, we used a template-matching code that extracts the RV through a maximum likelihood analysis between a template and each individual spectrum \citep[see][for more details]{astudillo2017harps}.
A template is built for each star by computing the median of all its spectra; according to \cite{astudillo2015harps}, at least ten spectra are needed for this operation.
We therefore focused on the 218 M dwarfs observed at least ten times with HARPS.
RV uncertainties (hereafter $svrad$) were computed as described in \cite{bouchy2001}, and mostly depend on the read-out noise and photon noise.
To account for the long-term instabilities of HARPS, we quadratically added 0.8m/s to this value to obtain the total svrad \citep{Lovis2008, coffinet2019}.

\subsection{Fibres change on HARPS}\label{fibre}

\begin{figure}
\includegraphics[width=\linewidth]{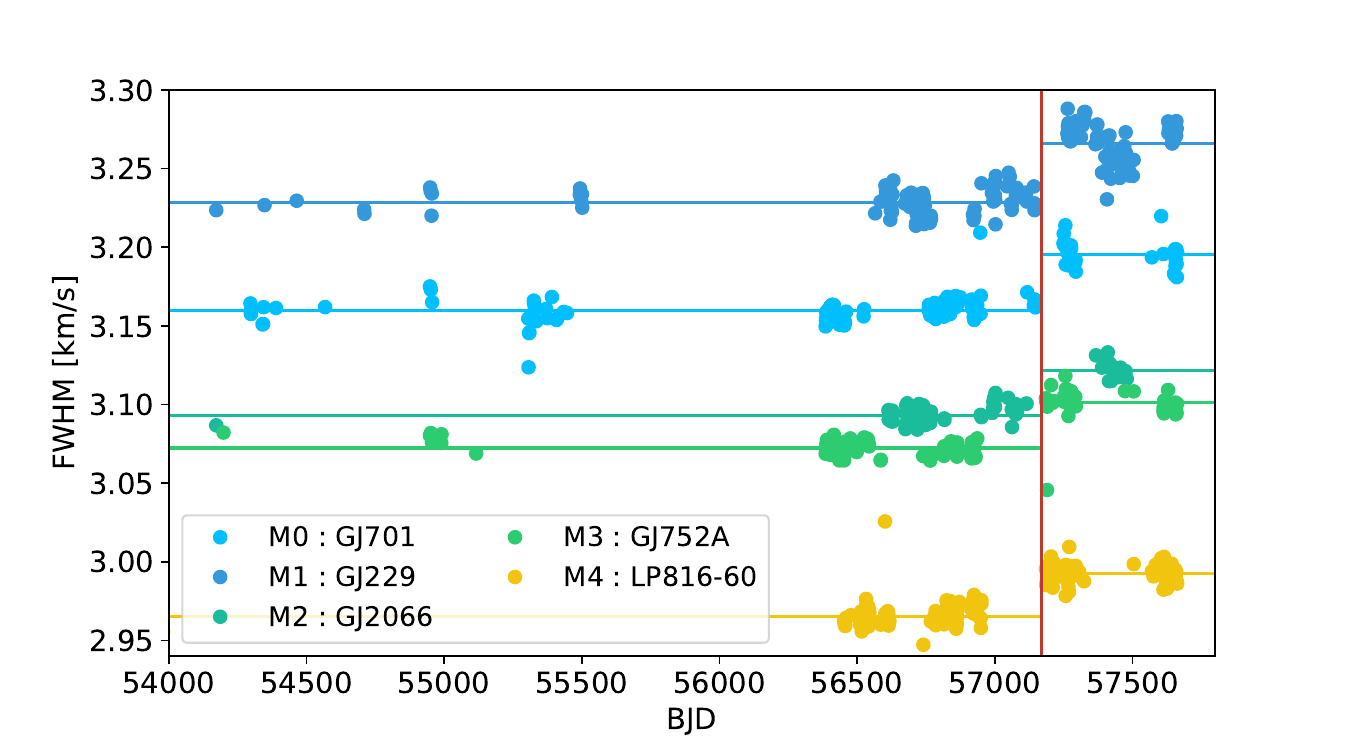}
\caption{Evolution of the CCF-FWHM for five different spectral subtypes. The red vertical line gives the date (BJD: Barycentric Julian Day) of the fibre change. The horizontal solid lines show the median of the FWHM of the CCF with the M2 mask before and after the fibre change.}
\label{evolutionFWHM}
\end{figure}

HARPS was upgraded on 28 May 2015, when the vacuum vessel was opened to change the fibres and improve the RV stability \citep{curto2015}.
It introduced an RV offset that depends on the spectral type of the observed target, and varies from several hundred m/s for F dwarfs to just a few m/s for M dwarfs.
This upgrade modifies the shape of the spectral lines, and hence the full width at half maximum (FWHM) of the CCF.
Figure~\ref{evolutionFWHM} highlights this temporal evolution of FWHM values for the most observed M dwarfs (before and after the change) in five different subspectral types: GJ~701 (M0), GJ~229 (M1), GJ~2066 (M2), GJ~752 (M3), and LP~816-60 (M4).

The width of the CCF of the M dwarfs changes by a few tens of m/s (1.003\% relative), slightly above its dispersion before or after the upgrade. 
Among the 218 stars with at least ten HARPS spectra, 71 were observed before and after the fibre change.
Since the shape of the lines changed, we conservatively analysed the spectra before and after the upgrade separately for each star affected by the fibre change. 

For the 37 stars with at least ten measurements both before and after the fibre change, we extracted the RV using two independent templates.
The RV series for these stars corresponds to two successive series HARPS03 and HARPS15, and the RV offset between them is modelled together with other possible variations (Keplerian, activity effects). A total of 
32 stars have at least ten~measurements either before or after the upgrade (but not both), and for these we chose to analyse the longest series and ignore the shortest.
Finally, GJ~79 and GJ~3192 were observed less than ten times both before and after the upgrade, and consequently were rejected from this study, leaving a remaining sample of 216 stars.

\subsection{Selection process}

Before carrying out our systematic analysis, we discarded some lower-quality spectra to guarantee the robustness of the RV time series.

\subsubsection{Selection based on the signal-to-noise ratio}

Firstly, we rejected all exposures that did not reach an RV precision of 5 m/s (corresponding to a signal-to-noise ratio, S/N$<10$ in spectral order 50 = 550 nm), a typical limit for detecting low-mass planets.
Secondly, we quantified the chromospheric activity by measuring the ionised calcium doublet lines (Ca H\&K at 393,3 nm and 396,3 nm) to identify whether any velocity variation is due to stellar activity.
We therefore imposed a S/N greater than 1 in the order continuum corresponding to this spectral region of these wavelengths (order 7).
After this selection, two stars ended up with fewer than ten spectra and were therefore excluded from the sample.

We also automatically excluded spectra penalised by poor weather conditions by imposing a relative criterion on the S/N (in the spectral order 50)
of each star.
We computed the dispersion of the S/N value ($\rm \sigma_{S/N_{50}}$) of all observations on a target, and rejected all spectra whose S/N was more than 5~$\sigma$ below the median value.
Once the outliers were rejected, we carried out a new iteration after calculating a new dispersion and median.

\subsubsection{Selection based on the FWHM}

As shown in Fig.~\ref{evolutionFWHM}, for a given star and instrumental set-up (before or after the fibre change), the FWHM of the CCF varies slightly, with a typical dispersion $\rm \sigma_{FWHM}$ of a few m/s (0.3\% relative).
Thus, a large variation in the FWHM indicates a problem in the RV measurement, an incorrect star identification, for example.
For each star we rejected spectra whose FWHM value was more than $\rm 3\sigma_{FWHM}$ from the median value.

\subsection{Average data per night} \label{bin}
The targets of this study were monitored via various observation programmes (listed in Table~\ref{progidtable}).
Time sampling can therefore vary considerably from one star to another. 
In order to standardise the datasets, we binned the RVs per night, using an average weighted by the uncertainties.
After this final step, we rejected the 14 stars that had not been observed on at least ten different nights.

\subsection{Final sample} \label{subsecsample}

\begin{figure}
   \includegraphics[width=\linewidth]{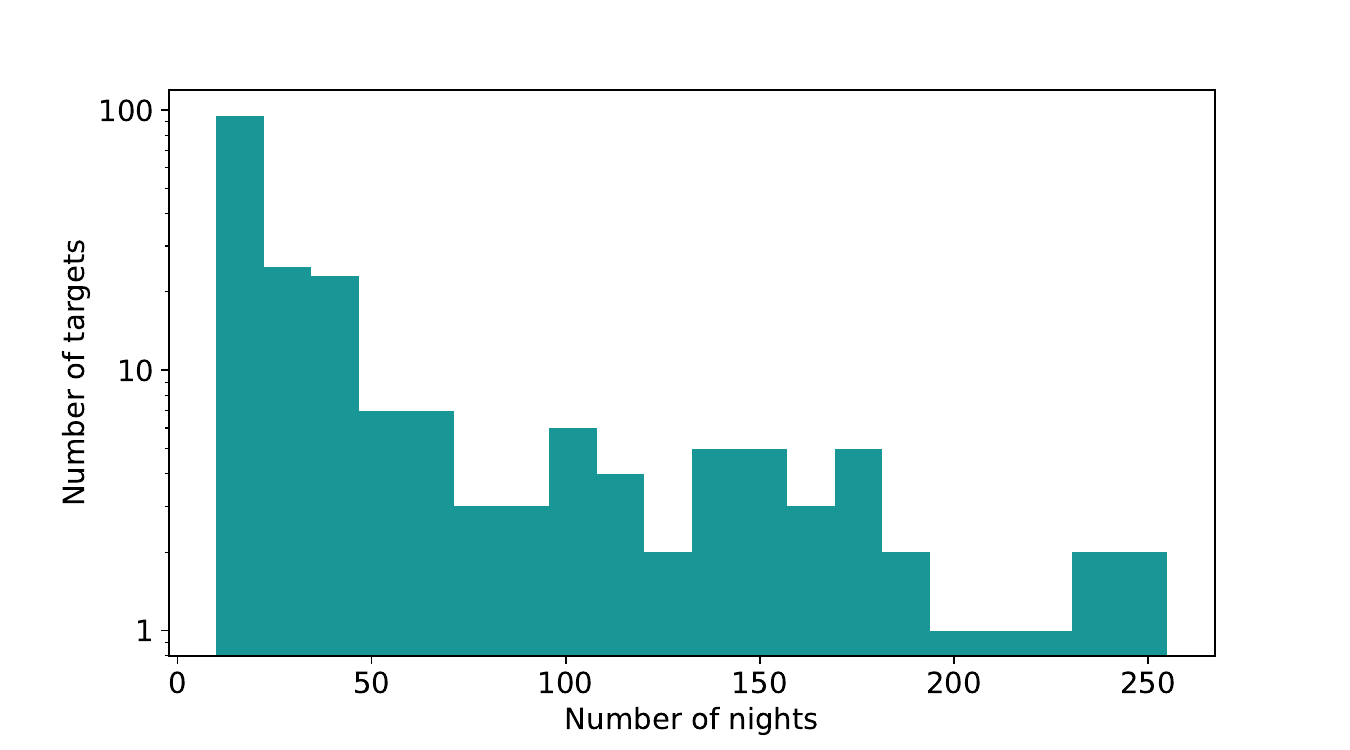}
    \caption{Distribution of observed nights.}
\label{mes}
\end{figure}

After these different selection steps, our final sample contained 200 stars, representing 10513 spectra.
Figure~\ref{mes} shows the distribution of the number of nights per star.
Some targets were no longer observed after only a few measurements, which explains the steep shape of the distribution.
At this stage of the study, RV time series were binned per night and considered corrected from outliers.

\section{Long-term analysis}\label{secAnalysis1}

The first step of the analysis is a study of the long-term trends in the 200 time series.
To automatically identify the long-term pattern dominance, we tested four different models labelled from 0 to 3:
\begin{itemize}
\item (0) a constant model;
\item (1) a linear trend;
\item (2) a second-order polynomial, called quadratic model;
\item (3) a sinusoidal fit with a period of the same order as that of the temporal coverage of the series.
\end{itemize}

We use statistical tests to determine whether a long-term correction should be applied before searching for short-term periodic signatures.


\subsection{Correction of the secular acceleration}\label{subsecCorr}

Perspective or secular acceleration \citep{vandekamp77} is a geometric effect that results in a variation in the radial component of velocity \citep[see e.g. Fig. 3 in][]{zechmeister2009}, measured for the first time by \cite{kuerster2003low} on the Barnard star.
We applied a secular acceleration ($\dot{v}$) correction on RV time series as

\begin{equation}
\dot{v} = \frac{dv}{dt} = \frac{\mu_{RA}^2 + \mu_{DEC}^2}{\pi}
\label{dvdt}
,\end{equation}where $\mu_{RA}$ and $\mu_{DEC}$ are the right ascension and declination components of the proper motion, and $\pi$ is the parallax.

The uncertainties associated with the Gaia measurements of proper motions and parallaxes are much smaller (by a factor of 10) than the uncertainties previously used in \cite{zechmeister2009} and \cite{bonfils2013harps}, and derived from \citet[][]{van2007validation}.
Consequently, the values of the acceleration correction terms are of the same order as in these previous works, but the associated uncertainties are now negligible compared to the photon noise.

\subsection{Excessive RV dispersion: Limits of the constant model}\label{subsecExcess}

\begin{figure}
\includegraphics[width=\linewidth]{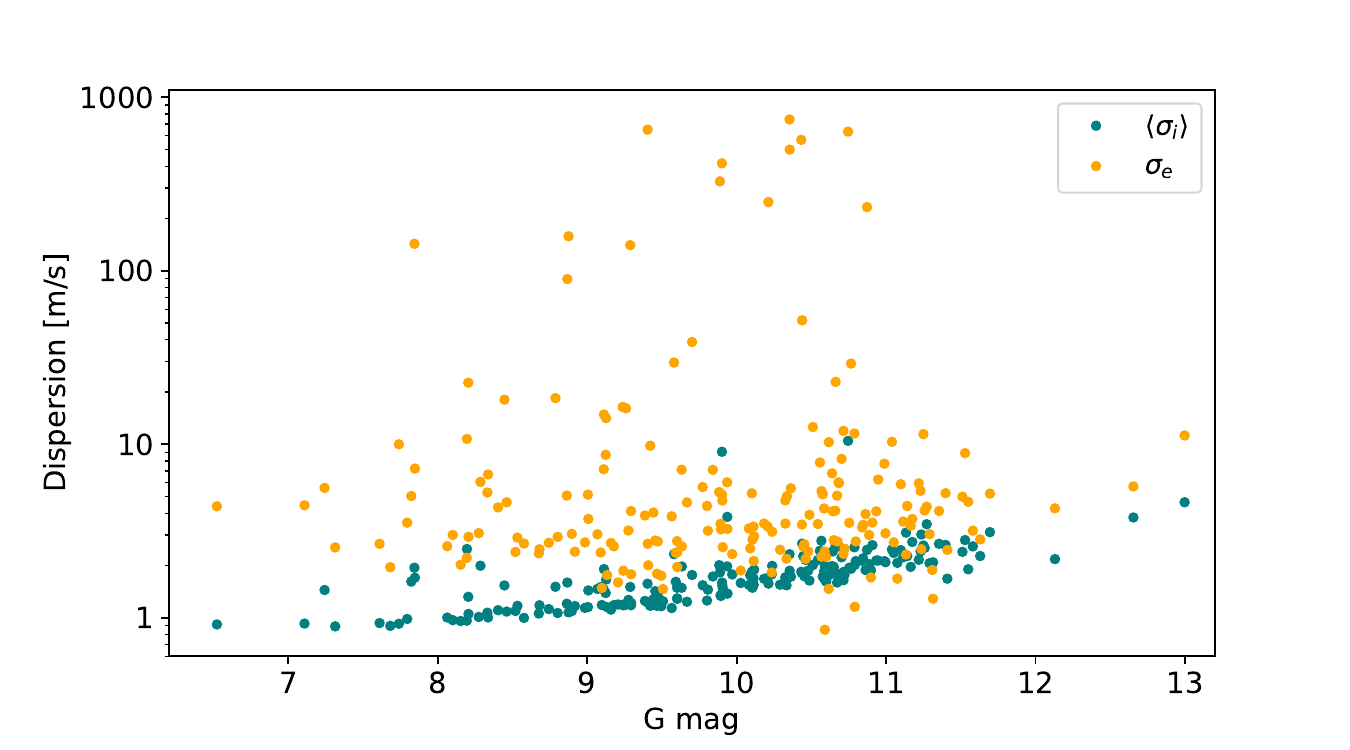}
\caption{Averaged internal error $\langle \sigma_{i} \rangle$: $\sigma_{i}$ (green) and RV dispersion $\sigma_{e}$ (orange) expressed in m/s vs G-band magnitude.}
\label{dispG}
\end{figure}

In this first step we measure the variability of the RV times series as in \cite{bonfils2013harps} to determine whether the constant model is sufficient to explain the dispersion.
Fig.~\ref{dispG} shows, for each star, the averaged internal error ($\langle \sigma_{i} \rangle$, $\sigma_{i}$, which is the quadratic sum of \textit{svrad}, in green) and the measured RV dispersion ($\sigma_{e}$, in orange) as a function of G-band magnitude.

For the majority of the stars in our sample, the RV series show an excessive dispersion that cannot be explained by internal noise alone, and we observe $\langle\sigma_{i}\rangle \leq \sigma_{e}$ in most cases, as in \cite{bonfils2013harps}.
To characterise this excessive dispersion, we performed a chi-squared test, with its $\chi^2_0$ value and probability P($\chi^2_0$) between the dispersion and the constant model.
A low value of P($\chi^2_0$) highlights that the model tested is not sufficient to explain the dispersion measured, even taking uncertainties into account.

Excessive variability can also be expressed as in \cite{cumming1999lick}, \cite{cumming2008keck}, \cite{zechmeister2009}, and \cite{bonfils2013harps}, using the definition of $Fvalue_0$ and its probability P($Fvalue_0$):
\begin{equation}
Fvalue_{0} = \frac{\sigma_{e}^2}{\langle\sigma_{i}\rangle^2} .
\label{fvalue}
\end{equation}

\begin{figure}
\includegraphics[width=\linewidth]{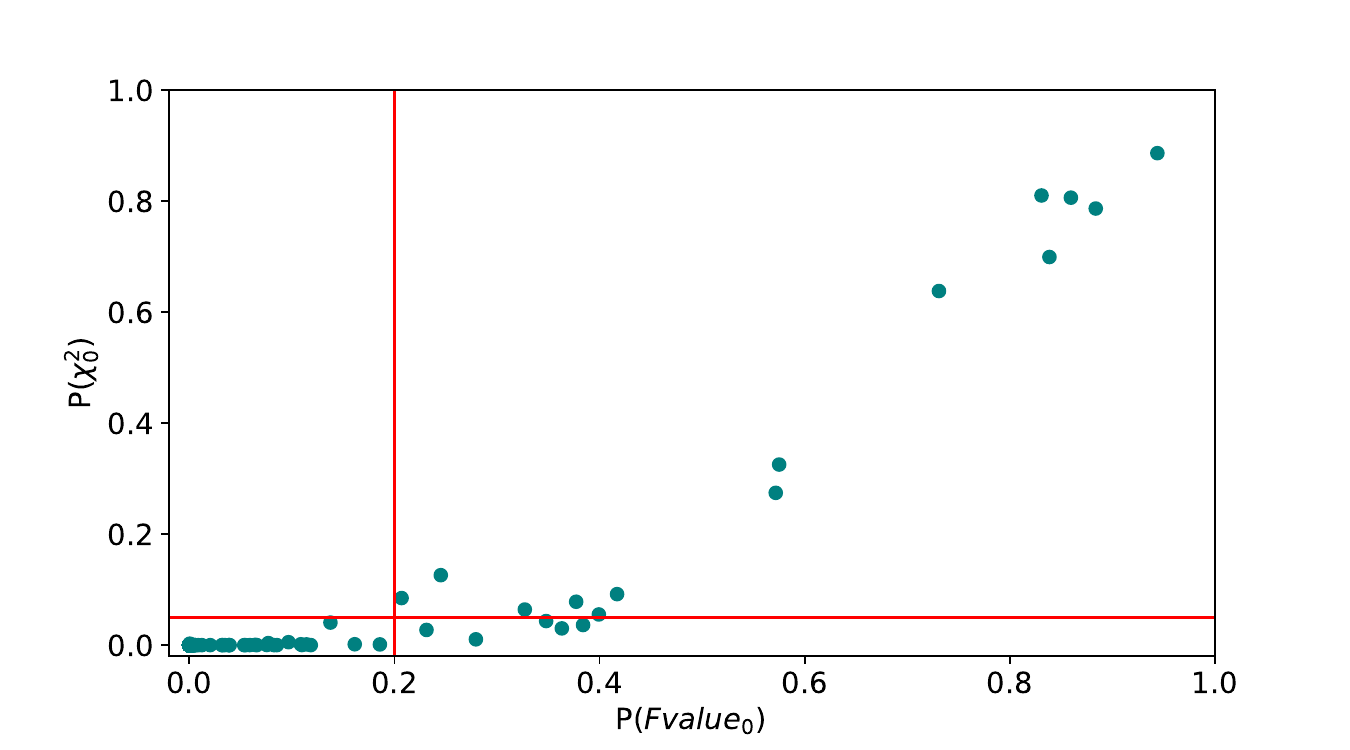}
\caption{P($\chi^2_{0}$) vs P($Fvalue_{0}$) for the constant model. The red vertical (resp. horizontal) line corresponds to the 0.2 (resp. 0.05) threshold adopted for P($Fvalue_0$) (resp. $P(\chi^2_0)$).}
\label{pFchi}
\end{figure}

As explained in the previous works cited, a high $Fvalue_0$ (and low P($Fvalue_0$)) indicates a dispersion that cannot be explained by internal errors.
These two complementary tests measure variability in two slightly different ways: $Fvalue_0$ evaluates average dispersion, while the impact of outliers is reduced in the $\chi^2$ test.
An outlier with very high uncertainties will have less impact on the rest of the dataset in the $\chi^2$ computation than in the $Fvalue_0$ computation.
Consequently, P($\chi^2_0$) is lower than P($Fvalue_0$), as announced by \citet{zechmeister2009}.
In our study, since the series are cleaned of points with aberrant uncertainties, P($\chi^2_0$) is only slightly lower than P($Fvalue_0$) as shown in Fig.~\ref{pFchi}.

\begin{figure}
\includegraphics[width=0.99\linewidth]{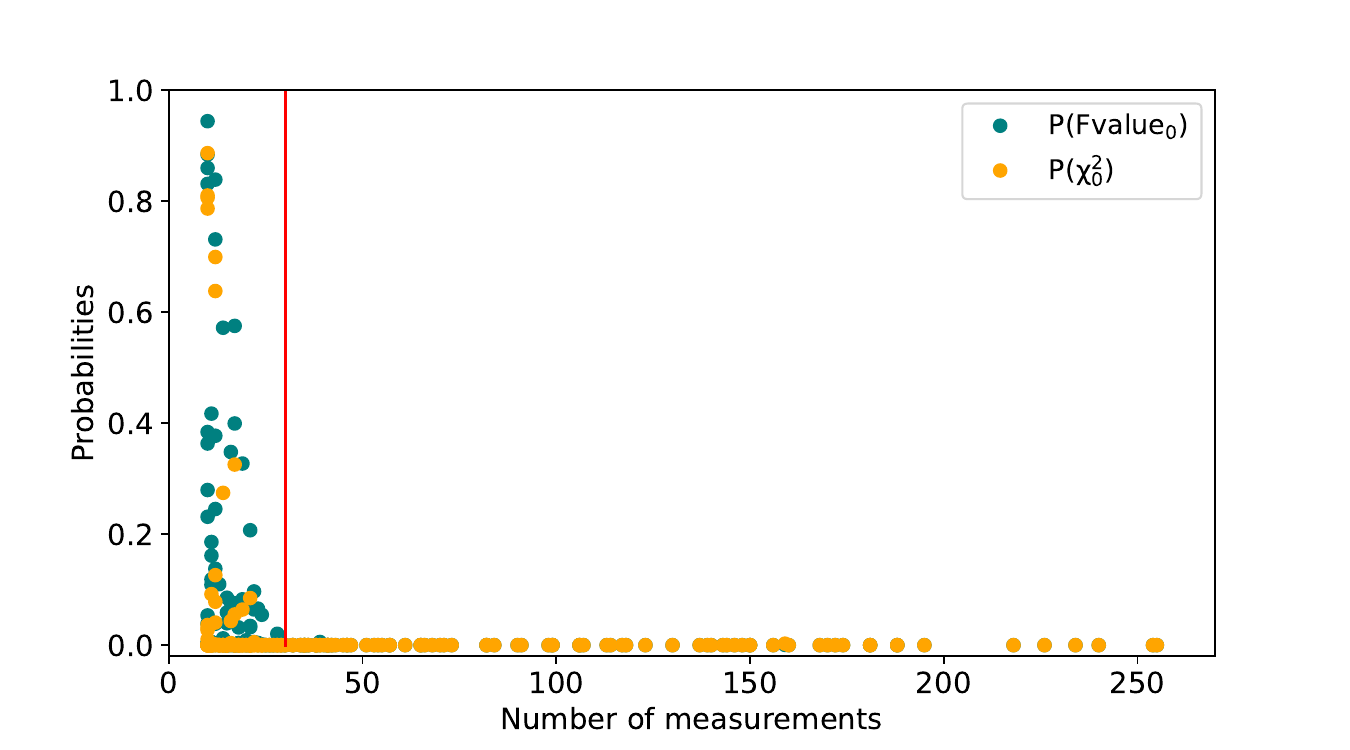}
\caption{P($Fvalue_0$) (orange) and $P(\chi^{2}_0)$ (green) as a function of the number of measurements for each star. The probabilities fall to zero after the threshold of 30 measurements (highlighted by the red vertical line).}
\label{probcst}
\end{figure}

To consider that a constant model cannot explain the dispersion of the dataset, we adopt the same threshold as \citet{zechmeister2009} and \cite{bonfils2013harps}: $P(\chi^2_0)<0.05$.
As shown in Fig.~\ref{pFchi}, this corresponds statistically to an equivalent threshold of $P(Fvalue_0)<0.2$.
Figure~\ref{probcst} shows P($\chi^2_0$) and P($Fvalue_0$) for our whole sample as a function of the number of measurements.
All the values of these tests and the parameters used are listed in Table~\ref{tabtot}.
The probability that the dispersion measured in each data series is explained solely by internal errors associated with each measurement falls to zero for all time series with more than 30 measurements.

The question is whether this is due to an underestimation of the errors or to the presence of a real signal in each RV time series.
The first hypothesis is simply verified by simulating the uncertainties required to obtain a uniform distribution of P($\rm \chi^{2}$) and P(Fvalue).
We obtain simulated uncertainties that are ten times greater than those calculated in Sect.~\ref{subsec31}, which is inconsistent with the characterisation of the precision obtained with HARPS.
Finally, we conclude that a signal (planetary system and/or activity impact) is present in almost all RV time series.
This result is consistent with the occurrence rate of $0.9^{+0.04}_{-0.03}$ planets (R < 4~R$\rm _{\oplus}$ and P < 50 days) per small star established on the Kepler missions \citep[e.g.][]{dressing2013,dressing2015}, but also with the $1.32^{+0.33}_{-0.32}$ rocky planets (M < 10~M$\rm _{\oplus}$ and P<100days) obtained on the RV survey of CARMENES \citep{sabotta2021}.


\subsection{Long-term signals}\label{stat}

Following the approach of \cite{zechmeister2009}, we begin the characterisation of excessive dispersion by searching for long-term variations in the time series.


\subsubsection{Fitting method and statistical test}

Fitting method:
The models tested in this section are fitted with a least-squares optimisation algorithm from Limited memory Broyden-Fletcher-Goldfarb-Shanno (L-BFGS-B) \citep{byrd1995limited,zhu1997algorithm}.
We use the diagonal terms of the covariance matrix associated with the parameters fitted as uncertainties, considering them uncorrelated.\\

\noindent
Reduced chi-squared ($\chi^2_*$):\label{chisec}
The $\chi^2$-probability is used to check whether the proposed long-term model is sufficient to explain the time series.
Since we expect this to rarely be the case, we instead compare the improvements brought by the models with each other.
As in \cite{mignon23a}, we use a test on the ratio of the reduced $\chi^2$ values of two models defined as
\begin{equation}
F_{red\:j/k} = \frac{\chi^2_{k*} - \chi^2_{j*}}{\chi^2_{k*}} 
\label{fred}
,\end{equation}where $\chi^2_{j*}$ (respectively $\chi^2_{k*}$) is the reduced chi-squared value of model j (resp. k). 
This ratio is close to 1 when model j brings a major improvement compared to model k.
Conversely, a negative value or a value close to 0 indicates that the improvement is either negligible or nonexistent.\\


\noindent
F test:\label{ftestsec}
We use also the F test to verify the improvement made by the model tested, as presented in \cite{cumming1999lick} and \cite{cumming2008keck}:
\begin{equation}
F_{j/k} = (d_{free}). \frac{\chi^2_{k} - \chi^2_{j}}{ \chi^2_{j}} 
\label{ftest}
.\end{equation}
Here $d_{free}$ is the degree of freedom of model j, and $\chi^2_{k}$ and $\chi^2_{j}$ are respectively the chi-squared values of models k and j.
We compute its probability as in \cite{bonfils2013harps} (constructed on Beta functions) to express the significance of the improvement. 
A very low probability expresses a significant improvement and a probability near 100\% expresses a non-significant improvement.


\subsubsection{Linear model}

We automatically examine if the RV time series of each star is dominated by a long-term linear trend pattern (identified by the index 1).
This signal can be caused by the presence of a massive companion with a large separation with a poorly covered orbit.
We consider that this model dominates the RV time series when the following criteria are met:
\begin{itemize}
\item the ratio $\rm F_{\rm red\:1/0}$ is higher than the ratio of the other models (including the two other types of models described below $\rm F_{red\:2/0}$ and $\rm F_{red\:3/0}$), and is above our 0.2 threshold;
\item the value of the P($F_{1/0}$) is under the threshold of 0.05;
\item the ratio $\rm F_{red\:2/1}$ is negative.
\end{itemize}

\begin{figure*}
 \centering
 \includegraphics[width=0.99\textwidth]{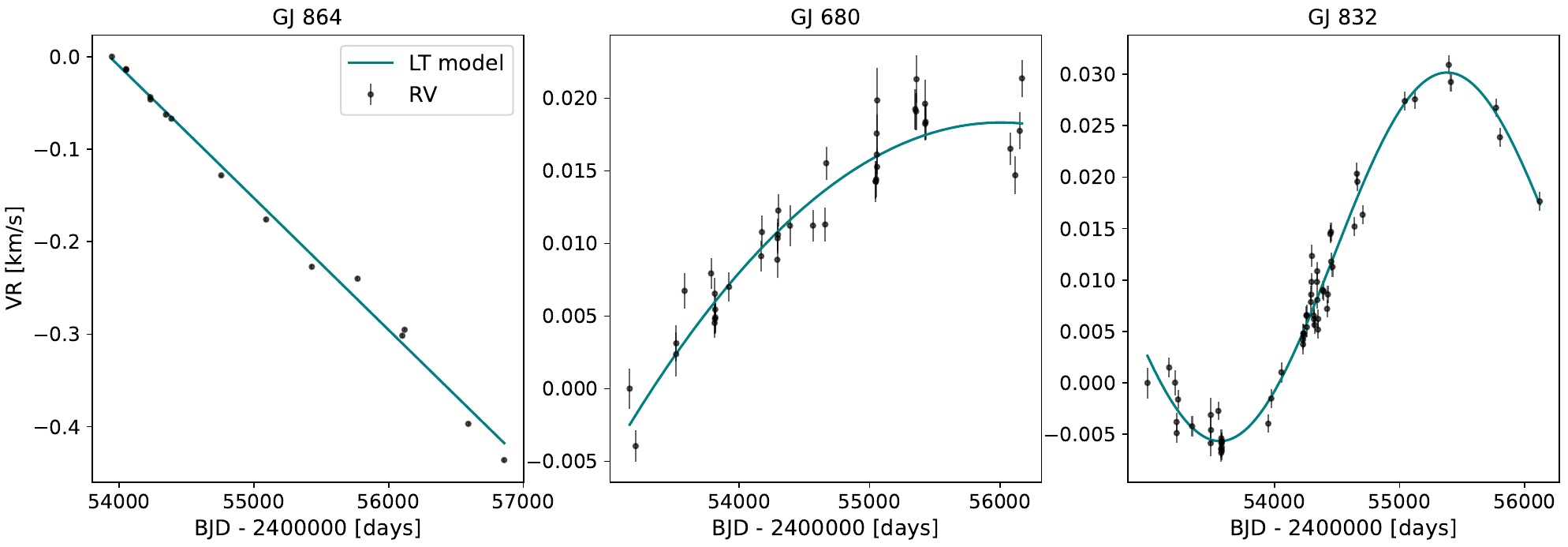}
 \caption{Example of detection of dominant long-term signal: GJ~864 \textit{(left panel)} as an example of linear trend, GJ~680 \textit{(middle panel)} for the quadratic trend, and GJ~832 \textit{(right panel)} for the sinusoidal function.}
 \label{fig:exemple-vr-LT}
\end{figure*}

These criteria guarantee the improvement (and its significance) provided by the linear model compared with the other models tested (described below).
The 16 targets whose RV time series is dominated by a linear trend are listed in Table~\ref{tablepente}.
This slope is subtracted before proceeding to the frequency analysis in the next section.
Figure~\ref{fig:exemple-vr-LT} shows the example of GJ~864.

\subsubsection{Quadratic model}\label{subsecPoly}

At the same time as the trend search, we also test a second-degree polynomial model (identified by the index 2).
Such a signal may reflect the presence of an intermediate-period companion that is better covered than in the case of the trend, or it may be caused by a variation in stellar activity on a scale of several years.
A second-degree polynomial is considered to be dominant in the RV time series when these criteria are met:
\begin{itemize}
\item the ratio $\rm F_{red\:2/0}$ is higher than the ratio of the other models ($\rm F_{red\:1/0}$ and $\rm F_{red\:3/0}$), and is above the 0.2 threshold;
\item the value of the P($F_{2/0}$) and P($F_{2/1}$) are under the threshold of 0.05;
\item the ratio $\rm F_{red\:2/1}$ is positive.
\end{itemize}

Once again these criteria guarantee the significant improvement provided by the quadratic model compared with the others.
This corresponds to the 29 stars listed in Table~\ref{tablepente}.
The fitted second-degree polynomial is subtracted before proceeding to the frequency analysis in the next section.
Figure~\ref{fig:exemple-vr-LT} shows the example of GJ~680.


\subsubsection{Sinusoidal model}\label{subsecsinus}

To complete the systematic analysis of the long-term variations, we test one last model: a sinusoidal function with and without a linear trend (identified by the index 3).
Its period is of the order of the duration of the RV time series (shorter periods are identified during frequency analysis), and its amplitude is of the order of the peak-to-peak RV variation of the time series.
We chose a sinusoidal function instead of a cubic trend in order to detect a wider range of long-term variations and minimise the impact of a bad sampling on the model chosen.
We note that this step does not represent a definitive diagnosis of these systems, which are analysed in Sect. \ref{candidate} through a Keplerian analysis.
Given the sporadic sampling of RV time series, testing the sinusoidal model helps avoid incorrect automated linear or quadratic trend corrections.
In particular, an RV series dominated by the signature of the presence of a massive companion can be misinterpreted as a simple trend in cases of very poor sampling, resulting in an incomplete phase coverage.
The choice for this model 3, between a sinusoidal function or the sum of a linear trend and a sinusoid, is made by selecting the solution giving the smallest error computed in using the covariance matrix.
The sinusoidal model is considered as dominant in the RV time series when these criteria are met: 
\begin{itemize}
\item the ratio $\rm F_{red\:3/0}$ is higher than the ratio of the other models ($\rm F_{red\:1/0}$ and $\rm F_{red\:2/0}$), and is above the threshold of 0.2;
\item the value of  P($F_{3/0}$), P($F_{3/1}$), and P($F_{3/2}$) are under the threshold of 0.05.
\end{itemize}

Following these criteria, the RV time series of 12 stars are dominated by a sinusoidal pattern (listed in Table~\ref{sin}) and Fig.~\ref{fig:exemple-vr-LT} shows the example of GJ~832.
The signal identified corresponds to a stellar or massive planet companion already discovered for seven of them (see references in Table~\ref{sin}).
We finally highlight five new massive candidates and/or binary systems described in Sect. \ref{candidate} of this paper. 
The targets flagged with an asterisk in Table~\ref{sin} are those dominated by the sum of a sinusoidal function and a linear trend.
These time series were corrected for linear trend before performing the analysis described in the next section.

\begin{table}
\caption{Sinusoidal model}
\label{sin}
\begin{center}
\renewcommand{\footnoterule}{} 
  \begin{tabular}{ l l | c | c |c }
  \hline
  Star && Period& K & Ref \\ 
   && (d) & (m/s) & \\\hline
  GJ~179* &b & 2288 & 22.10 & \cite{howard2010california} \\
  GJ~317 &b & 762 & 76.24 & \cite{johnson2007new} \\
  GJ~676A* &b & 1056 & 166.47 & \cite{forveille2011harps} \\
  GJ~832 &b & 3687 & 17.91 & \cite{bailey2008jupiter} \\
  GJ~849* &b & 1924 & 29.90 & \cite{butler2006long} \\
  GJ~3307& & 3539 & 970.41 & this paper \\
  GJ~4001& & 2672 & 571.92 & this paper \\
  GJ~4254*& & 1453 & 4595.92 & this paper \\
  GJ~4303& & 1267 & 47.92 & this paper \\
  GJ~9425* &c & 1800 & 31.02 & \cite{moutou2011harps} \\
   &b & 700 && \cite{moutou2011harps} \\
  GJ~9482* &b & 6700 & 75.52 & \cite{segransan2011harps} \\
  GJ~9588*& & 620 & 540.65 & this paper \\
  \hline
\end{tabular}
\end{center}
\tablefoot{Targets dominated by a sinusoidal function. The last column gives the companion discovery reference, including the five in this work. Targets flagged with an asterisk correspond to those dominated by a linear trend in addition to the sinusoidal function.}
\end{table}


\subsection{RV offset from HARPS fibre change}\label{fibers}

As discussed in Sect.~\ref{fibre}, the change of HARPS fibres (in May 2015) induces an offset in the RV time series, which is different for each star and depends on the spectral type \cite[][]{curto2015, trifonov2020public}.
As a result, the RV series of the 37 stars observed sufficiently before and after the change of fibres consist of two series extracted from two templates that can be considered as coming from two distinct instruments (HARPS03 and HARPS15).
For these stars, we fit an additional free parameter at the date of the change of fibres to each model tested.
The values of this parameter are listed in Table~\ref{off} with the corresponding model.

\subsection{Conclusion of the long-term analysis}

Among the 200 RV time series in our sample, 57 (28.5\%) are dominated by a long-term trend.
Among these, we corrected 49 series; we subtracted a linear function from 20 series (including 4 where it was necessary to model a sinusoid to properly identify the trend), and subtracted a quadratic function from 29 others.
RV time series dominated by a sinusoidal function are individually examined in the next section.

After these corrections, we recomputed the $Fvalue_0$ ratio to estimate the excessive dispersion after subtracting the long-term pattern.
Only 5\% of the RV time series have an $Fvalue_0 < 3$ (10 out of 200).
This indicates that even after subtracting the long-term trend, the RV time series still exhibit excessive variability, certainly due to shorter-term periodic or stochastic signals.


\section{Frequency analysis}\label{secAnalysis2}

Once the RV time series are corrected for the long-term signal, identified in the previous section, we analyse the higher-frequency signals corresponding to periodic or quasi-periodic variability (Keplerian or activity signals).
Our aim here is not to optimise the identification of individual planetary systems, but to carry out a homogeneous search that can be used for subsequent statistical analysis.
The signal detections we report in this section are compared with previously published systems.

\subsection{Method}

\subsubsection{Generalised Lomb-Scargle periodogram}

For each star we compute a generalised Lomb-Scargle periodogram (GLS), proposed by \cite{zechmeister2009generalised} and based on \cite{lomb1976least} and \cite{scargle1982studies}.
For a given period, the GLS gives the generalised power corresponding to the lowest $\chi^2$ obtained with a sinusoidal function at that period.
We compute the GLS from the minimum period of 1.6 days up to the maximum period of twice the time coverage of the RV time series.
We summarise the main steps in our procedure as follows: 

\begin{itemize}
  \item We compute the GLS periodogram of the observation window on a synthetic data series with a regular sampling of one point per night, with an arbitrary value of 1 for each observed night and 0 for the others; we reject any peak in the GLS of the original series if its period corresponds to that of a peak of the window function (and within an interval corresponding to the width of this peak).
  \item Once a peak is identified in the GLS, we reject aliases caused by observation frequencies.
  \item The false alarm probability (hereafter FAP) is computed using a classical bootstrap method: we generate 1000 time series with dates similar to those of the original series, but we assign each of them a random RV value from the set of RV values in the original series; we then compute the GLS for each of these arranged series. The resulting periodograms are then sorted in ascending order of the power of the main peak, and the FAP level of the original series corresponds to its position in this ranking; for example, a FAP level of 1\% corresponds to a ranking in the tenth position.
\end{itemize}

 
\subsubsection{Keplerian analysis}

We apply an iterative method to detect systems with multiple signals.
For each star, as described above, we compute the GLS periodogram, determine the FAP level of its maximum peak, and verify that it is not caused by the sampling (window function or alias).
If the FAP level of the peak is below the 1\% threshold, we fit a Keplerian signal to the corresponding period by using a L-BFGS-B minimisation method.
We use the procedure in the set of functions proposed in the Python script of DACE\footnote{https://dace.unige.ch/tutorials/?tutorialId=3} by \citet{delisle2016}. 

After subtracting the fitted Keplerian signal, we repeat the procedure on the residuals: computing a new GLS, checking for sampling effects, and determining the FAP level.
If the FAP level is again below the 1\% threshold, we fit a new model with an additional Keplerian to the original series.
We stop this iterative process when there are no significant peaks left in the GLS of the residuals, if the Keplerian fitting fails to converge, or if the RMS of the residuals is below the internal noise parameter $\langle\sigma_{i}\rangle$.


\subsection{Results}\label{results}

We obtain a total of 120 significant periodic signals, resulting in 108 Keplerian models (12 fittings did not converge) across the time series of 52 stars.

\subsubsection{Previously known planetary systems}\label{biblio}

We identify the periodic signals corresponding to previously published planetary systems by cross-referencing them with the NASA Exoplanet Archive.\footnote{https://exoplanetarchive.ipac.caltech.edu/index.html}
In Table~\ref{tableobjetsconnus} we list the 51 signals identified among the 108 peaks.

\begin{table*}
\begin{center}

\caption{\label{tableobjetsconnus}Planetary systems previously discovered and recovered.}
\renewcommand{\footnoterule}{} 
\begin{tabular}{ r | c| c| c| c|c}

Star & FAP & Period (d) & K (m/s) & Correction & Discovery reference \\
\hline

GJ~54.1~c &0.0 & 3.0599$\pm$0.0002 & 1.91 & - & \cite{astudillo2017harps} \\
 d &0.0 & 4.6566$\pm$0.0003 & 1.69 & & \cite{astudillo2017harps} \\
 b & 0.0 & 2.02101$\pm$0.00005 & 1.61 & & \cite{astudillo2017harps} \\
GJ~163~b & 0.0 & 8.6300$\pm$0.0003 & 5.70& - & \cite{bonfils2013GJ163} \\
 c & 0.0 & 25.608$\pm$0.006 & 3.83& & \cite{bonfils2013GJ163} \\
 d & 0.0 & 642.15$\pm$1.65 & 3.40 & & \cite{bonfils2013GJ163}\\
GJ~176~b & 0.0 & 8.7754$\pm$0.0002 & 4.14 & - & \cite{forveille2009harps} \\
GJ~179~b & 0.0 & 2321$\pm$500 & 23.87 & - & \cite{howard2010california} \\
GJ~180~b & 0.0 & 17.124$\pm$0.006 & 4.26 & - & \cite{feng2020search} \\
GJ~221~c &0.0 & 125.44$\pm$0.09 & 7.27 & quadratic & \cite{curto2013harps} \\
 b &0.0 & 3.87$\pm$0.01 & 4.32 & & \cite{curto2013harps} \\
GJ~229~c &0.0 & 121.49$\pm$0.07 & 7.23 & quadratic & \cite{feng2020search} \\
 b & 0.0 & 528.17$\pm$6.7 & 1.26 & & \cite{feng2020search} \\
GJ~273~b & 0.0 & 18.661$\pm$0.004 & 1.44& - & \cite{astudillo2017harps0} \\
 c & 0.0 & 4.7230$\pm$0.0002 & 1.14& - & \cite{astudillo2017harps0} \\
GJ~393~b & 0.0 & 7.0223$\pm$0.0008 & 1.04 & - & \cite{amado2021} \\
GJ~433~b & 0.0 & 7.3721$\pm$0.0006 & 3.05 & - & \cite{delfosse2013} \\
GJ~447~b & 0.0 & 9.861$\pm$0.002 & 1.68& quadratic & \cite{bonfils2018temperate} \\
GJ~480~b & 0.002 & 9.548$\pm$0.005 & 5.61 & quadratic & \cite{feng2020search} \\
GJ~536~b & 0.0 & 8.7089$\pm$0.0004 & 3.59 & linear trend & \cite{mascareno2017super} \\
GJ~551~b & 0.001 & 11.187$\pm$0.001 & 1.42 & quadratic & \cite{anglada2016terrestrial} \\
GJ~581~b & 0.0 & 5.36856$\pm$0.00006 & 12.77& - & \cite{bonfils2005harps} \\
 c & 0.0 & 12.914$\pm$0.001 & 3.25 & - & \cite{udry2007harps} \\
 e & 0.0 & 3.1489$\pm$0.0001 & 1.66 & - & \cite{udry2007harps} \\
GJ~628~d & 0.0 & 217.914$\pm$0.305 & 2.48 & - & \cite{astudillo2017harps0} \\
c & 0.0 & 17.865$\pm$0.003 & 2.10 & - & \cite{astudillo2017harps0} \\
b & 0.0 & 4.8872$\pm$0.0003 & 1.63 & - & \cite{wright2016three} \\
GJ~667C~b &0.0 & 7.2005$\pm$0.0001 & 3.97 & linear trend & \cite{delfosse2013} \\
f &0.0 & 38.948$\pm$0.03 & 1.07 & - & \cite{anglada2013dynamically}\\
c & 0.0 & 28.151$\pm$0.005 & 1.77 & - & \cite{delfosse2013} \\
GJ~674~b & 0.0 & 4.69507$\pm$0.00002 & 8.75 & - & \cite{bonfils2007harps} \\
GJ~676A~b & 0.0 & 1039.3$\pm$0.6 & - & linear trend & \cite{forveille2011harps} \\
GJ~752A~b & 0.0 & 106.34$\pm$0.08 & 2.81 & - & \cite{kaminski2018carmenes} \\
GJ~832~b & 0.0 & 3684$\pm$97 & 17.39 & sinusoidal & \cite{bailey2008jupiter} \\
GJ~849~b & 0.0 & 1796$\pm$3 & 20.8 & sinusoidal & \cite{butler2006long} \\
GJ~876~b & 0.0 & 61.0320$\pm$0.0003 & 219.01 & linear trend & \cite{delfosse1998closest} \\
c & 0.0 & 30.2310$\pm$0.0002 & 82.11 & & \cite{marcy2001pair} \\
d & 0.001 & 1.93789$\pm$0.00001 & 5.90 & & \cite{rivera20057} \\
GJ~1132~b & 0.0 & 1.6287$\pm$0.0002 & 3.13 & - & \cite{Berta-Thompson2015} \\
c & 0.0 & 8.933$\pm$0.006 & 2.49 & - & \cite{bonfils2018radial} \\
d & 0.0 & 177$\pm$2 & 3.3 & - & \cite{bonfils2018radial} \\
GJ~3053~b & 0.0 & 24.84$\pm$0.03 & 5.36 & - & \cite{dittmann2017temperate} \\
c & 0.007 & 3.7816$\pm$0.0008 & 2.07 & - & \cite{ment2019second} \\
GJ~3293~b & 0.0 & 30.611$\pm$0.006 & 14.10 & - & \cite{astudillo2017harps0} \\
c & 0.0 & 123.4$\pm$0.1 & 24.44 & - & \cite{astudillo2017harps0} \\
d & 0.0 & 48.16$\pm$0.03 & 5.33 & - & \cite{astudillo2017harps0} \\
GJ~3323~b & 0.002 & 5.3551$\pm$0.0007 & 2.55 & - & \cite{astudillo2017harps0} \\
c & 0.001 & 40.36$\pm$0.07 & 1.37 & - & \cite{astudillo2017harps0} \\
GJ~3341~b & 0.0 & 14.208$\pm$0.005 & 2.83 & - & \cite{astudillo2015harps} \\
GJ~3634~b & 0.0 & 2.6458$\pm$0.0002 & 5.48 & quadratic & \cite{bonfils2011harps} \\
GJ~9018~b & 0.0 & 15.79$\pm$0.02 & 4.64 & - & \cite{tuomi2014bayesian} \\
\hline
\end{tabular}
\tablefoot{The 0.0 FAP value corresponds to a value under the 10$^{-3}$ threshold.}

\end{center}
\end{table*}

In this table, the periods in Col. 3 represent the values returned by our Keplerian fit (rather than the published ones) with the corresponding FAP level of the peak in the GLS given in Col. 2.
Since the FAP level corresponds to its ranking, a value of 0 indicates that none of the 1000 permuted series yielded a peak as strong as that of the original series, implying a true FAP level below 10$\rm ^{-3}$.
The first column indicates the name of the recovered planet, and the last column provides the reference for its discovery.
For multi-planetary systems the signals are listed in order of the successive iterations, and are indicated by the corresponding letter of the planet in the literature.
The semi-amplitudes (K) of the fitted Keplerian signals are provided in Col. 4, and we also specify the long-term model applied to data before the Keplerian analysis (Col. 5).
Several published planets are not retrieved by our analysis; according to the NASA Exoplanet database, there are 78 published planets orbiting the 200 M dwarfs in our sample. 
The reasons are diverse and are discussed in detail in Sect.~\ref{discus}.
We summarise them as follows:
\begin{itemize}
  \item when the discovery is based on an accumulation of data from other instruments (HIRES and/or CARMENES);
  \item when the period is shorter than the lower limit of 1.6 days used here;
  \item when a detailed analysis of the impact of stellar activity is required (Gaussian process).

\end{itemize}

In addition, two long-period planets, GJ~317~c \citep{johnson2007new} and GJ~676A~c \citep{sahlmann2016mass}, are not recovered because their orbital periods are longer than twice the duration of our data.

\subsubsection{Stellar activity}\label{act}

M dwarfs are often active stars \citep{hawley1991, delfosse1998rotation, morin2008large, newton2016, astudillo2017magnetic}, and their variability impacts RVs. It is therefore possible to confuse stellar activity signal with a planetary signal.

The surface heterogeneities of the star (spot and plage contrasts, magnetic inhibition of surface convection) shift and/or deform stellar lines.
Due to stellar rotation, stellar activity can thus add a quasi-periodic signal at the rotation period (or one of its harmonics) in the RV time series \citep[see e.g.][]{queloz2001, robertson2014b}.
If the rotation period is known or identified, this signal is often modelled and subtracted using a Gaussian process regression \citep{haywood2014, rajpaul2015, aigrain2022}.
However, most of the rotation periods of the M dwarfs in our sample are unknown, and performing such a process is therefore outside the scope of this study.

However, it is well known that the activity signal on M dwarfs can remain coherent over a long period and can create a peak in the periodograms.
A periodic function can then model it, ultimately leading to false positive detections such as GJ~832 c \citep{wittenmyer2014gj} and GJ~9066 b \citep{feng2020search}, both recently ruled out.

Therefore, we compare the periods of the 57 unidentified signals with the stellar rotation periods published in the literature (determined through photometric or chromospheric emission variability).
An RV signal detected at a period close to the published rotation period of the star (within 10\%) is therefore attributed to a rotational signature.
In addition, we perform a frequency analysis of three chromospheric proxies (the H- and K-bands of CaII, the sodium doublet, and the H$\rm _{\alpha}$ line) from each spectrum series in our sample.
Finally, an RV signal detected at a period close (within 10\%) to a periodic signal identified in the chromospheric indices is also conservatively attributed to a rotational signature.
Out of the 57 remaining unidentified signals in the RV data, 35 are attributed to the stellar rotation signature of 23 stars, listed and detailed in Table~\ref{tableprotconnus}.

\subsubsection{Non-identified signals}

After these two steps 22 unidentified periodic signals remain. We attribute them to the following different origins:
\begin{itemize}
  \item an unpublished or unidentified stellar rotation, requiring a more complete study, as was the case for GJ~581 \citep{robertson2014b};
  \item a long-term magnetic variability, of the order of a few hundred days or more, leading to a long-term RV variation (see analysis in our companion paper \citealt{mignon23a});
  \item an unidentified planetary or binary system.

\end{itemize}
These remaining 22 signals are listed in Table~\ref{tablenonidentifie}, where we also indicate the dominant long-term model found in the previous section.

\begin{table}
\renewcommand{\footnoterule}{} 
\caption{Periodic signal in RV times series whose origin is not yet identified (discussed individually in Sect.~\ref{candidate}).}
\label{tablenonidentifie}
\begin{center}
\begin{tabular}{ l | c| c| c| c}
Star & FAP & Period (d) & K (m/s) & Correction \\
\hline
GJ~300 & 0.0074 & 8.33 & 7.14 & quadratic \\
GJ~317 & < 10$^{-9}$ & 314.40 & 38.95 & - \\
 & 9 10$^{-9}$ & 236.46 & 21.65 & - \\
GJ~361 & 2.4 10$^{-8}$ & 28.94 & 4.94 & -\\
GJ~393 & 4.4 10$^{-6}$ & 775.94 & 1.43 & - \\
& 7.3 10$^{-5}$ & 51.9 & 1.73 & - \\
GJ~569A &0.00073 & 14.38 & 11.17 & linear \\
GJ~588 & 0.0021 & 21.02 & 2.5 & - \\
& 0.0067 & 24.19 & 2.3 & - \\
GJ~654 & 0.00019 & 260.83 & 1.81 & - \\
GJ~699 & 6.3 10$^{-8}$ & 373.09 & 2.5 & - \\
GJ~739 & 0.00066 & 138.73 & 4.06 & linear \\
GJ~754 & 0.00061 & 250.43 & 3.3 & - \\
& 0.0031 & 16.99 & 1.9 & - \\
GJ~846 & 0.0049 & 10.80 & 4.5 & - \\
GJ~880 & 0.0081 & 40.17 & 2.32 & quadratic \\
GJ~3307 & 7.4 10$^{-6}$ & 3539.87 & 1271.02 & - \\
GJ~4001 & 1.4 10$^{-6}$ & 4274.92 & 608.94 & - \\
GJ~4254 & < 10$^{-9}$ & 1453.68 & 3393.67& linear \\
GJ~4303 & 1.8 10$^{-9}$ & 1267.26 & 45.06 & - \\
& 0.0012 & 17.63 & 11.1 & - \\
GJ~9588 & < 10$^{-9}$ & 558.72 & - & linear \\
\hline

\end{tabular}
\end{center}
\end{table}

\section{Candidates}\label{candidate}

In this section we focus on the remaining 22 Keplerian signals as well as the massive candidates detected with the sinusoidal function in the long-term analysis.
The spectral types of the M dwarfs discussed below are taken from \citep{gaidos2014}.
The RV time series derived from the template-matching method of the respective stars are imported into the DACE\footnote{https://dace.unige.ch/tutorials/?tutorialId=3} platform, where we use the Keplerian analysis of \cite{delisle2016} and the analytical FAP computation routines from \cite{baluev2008}.
The fitted parameters on the platform are listed in Table~\ref{tablenonidentifie}.

\subsection{Stellar or brown dwarf companions} \label{massive companions}
With a temporal coverage of nearly 20 years, HARPS is an ideal instrument for detecting, monitoring, and completing the orbits of long-period companions.
In this section we provide detailed information on the high-amplitude RV signals found during the long-term and Keplerian analyses.
For the specific analysis of these massive companions, we utilise a software called \texttt{orvara}\xspace \citep{Brandt2021_orvara}, which is a Markov chain Monte Carlo (MCMC) programme designed to fit Keplerian orbits to various combinations of proper motion anomalies, absolute and relative astrometry, and radial velocities.
It uses the Hipparcos-Gaia Catalog of Accelerations (HGCA, \cite{Brandt2021_HGCA}), which relies on a combination of the Hipparcos and Gaia EDR3 catalogues.
It has been cross-checked and adjusted to address any discrepancies and ensure that all proper motions in the Gaia EDR3 are accurately aligned;
this enables us to obtain precise dynamical masses and orbital elements.
We are using a modified \texttt{orvara}\xspace version\footnote{https://github.com/nicochunger/orvara/tree/period-prior} that incorporates extra priors on orbital periods and semi-major axes, along with primary mass and stellar jitter priors from the original code, to enhance accuracy and improve orbital element consistency, especially for intermediate-separation binaries.

\subsubsection{GJ~3307}

\begin{figure}
\includegraphics[width=0.99\linewidth]{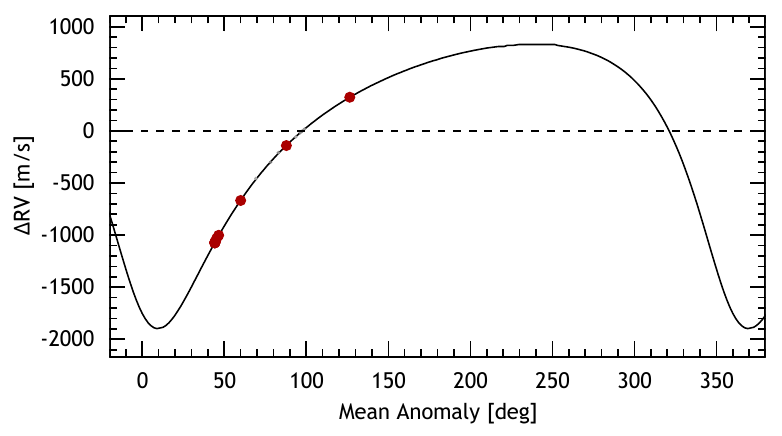}
\caption{GJ~3307 RV phase-folding}
\label{gj3307rv}
\end{figure}

\begin{figure}
\includegraphics[width=0.99\linewidth]{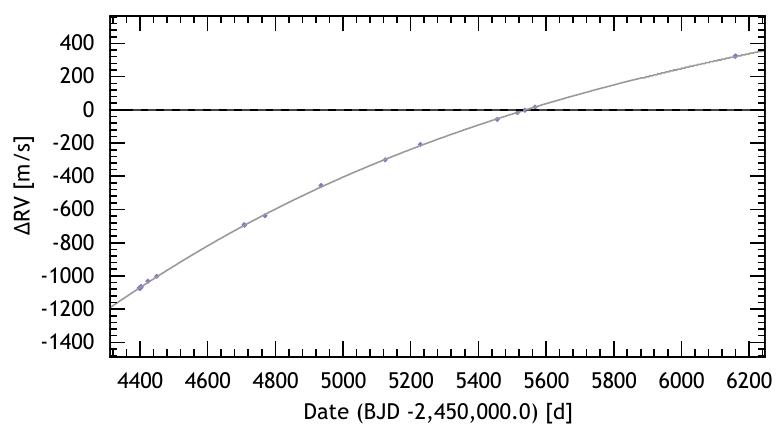}
\caption{GJ~3307 RV time series and best fitting orbit model.}
\label{gj3307}
\end{figure}

GJ~3307 is an M2V dwarf at a distance of $\sim$ 23~pc, one of the most massive stars in our sample (0.63 $\rm \pm$ 0.01 M$\rm _{\odot}$).
Our modelling of its RV time series converges to a large-amplitude sinusoidal function (peak-to-peak variation of 1.6~km/s).
By using the estimated period from Sect.~\ref{subsecsinus} as a prior to the Keplerian fit on DACE, we obtain a minimum period of 8396 $\rm \pm$ 52300 days and a minimum mass of 0.091 M$\rm _{\odot}$ (see Fig.~\ref{gj3307rv}).
The orbit is not well constrained by the 20 HARPS data points taken between October 2007 and September 2012 
(a time coverage of 1800 days), but the fitted signal can be subtracted from the RV series shown in Fig.~\ref{gj3307}, and the periodogram of the residuals shows no significant peak.
Therefore, the companion is likely a very low-mass star with a typical orbital period of about 20 years or longer, and should be resolvable with high angular resolution imaging.
The FWHM of the CCF for this star remains stable over time, indicating that this system is not an SB2 and that the contrast in luminosity between the two components must be significant.
Since this star has not been observed by Hipparcos, we are not able to use orvara to combine the astrometry and the RV to extract the system's inclination and the true companion mass.

\subsubsection{GJ~4001} 

GJ~4001 (HIP~84652) is classified as M1V and is in the top 10\% of the most massive stars in our sample, with a determined mass of 0.66$\pm$0.03 M$_{\odot}$, also classified as a K7V in previous studies \citep{Stephenson1986}.
Once again, the FWHM of the CCF remains stable over time, indicating that GJ4001 is not an SB2 system.
We use the period estimated in Sect.~\ref{subsecsinus} as a prior in the Keplerian fitting on DACE, and obtain a minimum projected mass 42.49 M$_{\rm J}$ with a period of 7378 $\pm$ 84 days and a high eccentricity (e = 0.717 $\pm$ 0.005).
Since measurements were taken over a 4115-day time span, slightly more than half of the period is covered, but fortunately the passage to periastron was well monitored.
With a separation of 6.6$\pm${0.17}~au at a distance of 22.5~pc \citep[Gaia DR2][]{gaia2018vizier}, the maximum apparent separation of the system is around 290 $mas$, making it an ideal target for high angular resolution observations and also for astrometric detection with Gaia.
\citet{Kervella2019}, in studying perturbations in proper motion using HIPPARCOS and Gaia DR2 measurements, found a possible astrometric signal for this system corresponding to a companion of $\sim$54M$\rm _{J}$ at a separation of 1.28 au at the time of the Gaia DR2 observation windows.
Using \texttt{orvara}\xspace, we obtain an inclination of 8.6 degrees, indicating that the system is seen from the pole and suggesting a fairly massive companion of 362~$\rm M_{J}$.

\begin{figure}
\includegraphics[width=0.99\linewidth]{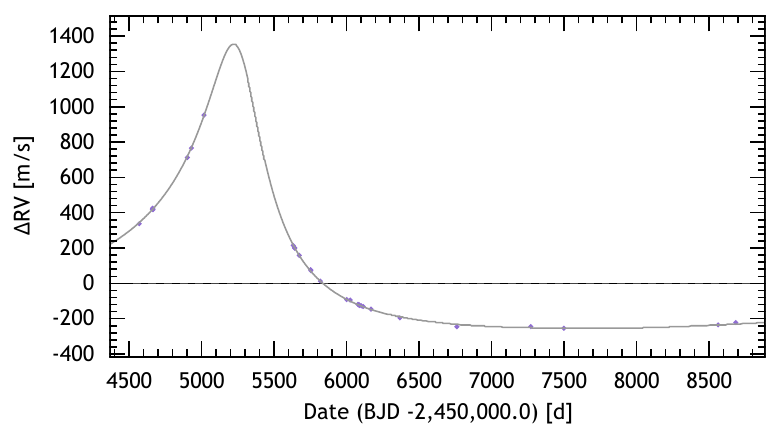}
\caption{GJ~4001 HARPS RV measurements and best fitting orbit model.}
\label{GJ4001vr}
\end{figure}

\subsubsection{GJ~4254}

\begin{figure}
\includegraphics[width=0.99\linewidth]{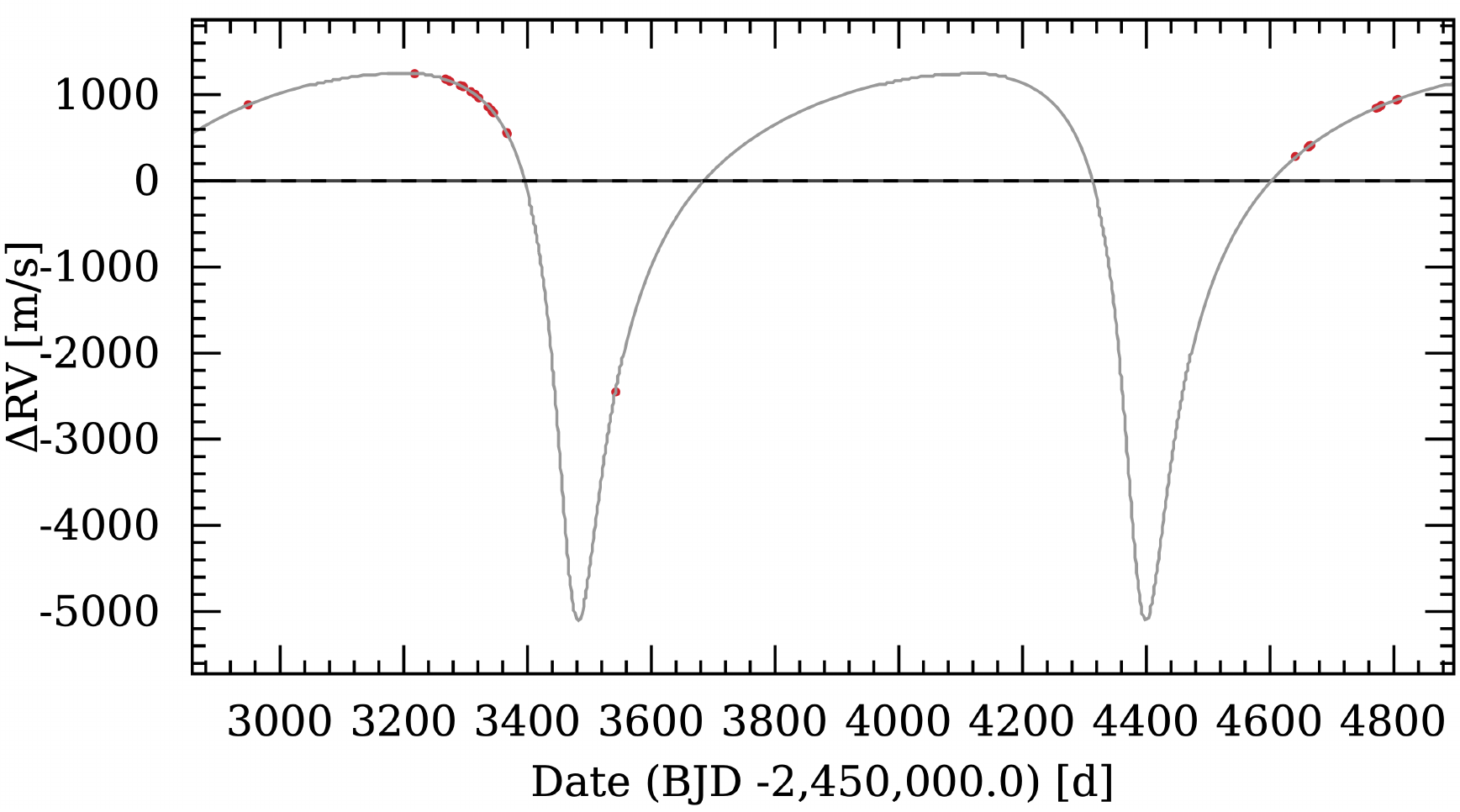}
\caption{GJ~4254 HARPS RV measurements and best fitting orbit for the first Keplerian model.}
\label{GJ4254vr}
\end{figure}

GJ~4254 (HIP~109084) is a massive M0V star (0.79 $\pm$0.02 M$\rm _{\odot}$) in our sample, also classified as a K7V in \cite{bidelman1985}.
The 30 HARPS measurements, shown in Fig.~\ref{GJ4254vr}, exhibit a peak-to-peak variation of 3.7 km/s.
Using the DACE platform, the Keplerian fit converges to a period of 917.4 $\rm \pm$ 0.8 days and a semi-amplitude of $ \rm K = 3.18 \rm \pm 0.08 km/s$, corresponding to a companion mass of m.$\sin{i}$ = 0.1 M$\rm _{\odot}$.
Unfortunately, for this system the periastron passage has not been sufficiently monitored to constrain the eccentricity (0.65 in the solution presented in Fig.~\ref{GJ4254vr}).
The true amplitude of the signal is therefore strongly underestimated in our fit.
This source would require additional measurements, ideally at a time when the RV separation is greatest, to identify the two individual components in the CCF.

\subsubsection{GJ~4303} 

GJ~4303 (HIP~113201AB) is a M1 system with a mass of 0.54 $\pm$0.01 M$\rm _{\odot}$, at a distance of 23.8~pc.
The system has been monitored using direct imaging with SPHERE, astrometry with HIPPARCOS and RV measurements with HARPS by \citet{biller2022} in recent years.
They detected a low-mass stellar companion of 0.145 M$\rm _{\odot}$ with a period of 33 years.
In this study we use 73 of the 76 public RV measurements of HARPS, excluding the last 3 measurements taken after the change of fibres as they do not allow the construction of a proper template.
We fit the orbit using priors from our long-term signal analysis.
We obtain a period of 11709 $\pm$ 2030 days, with an eccentricity of 0.66 $\pm$ 0.05, and a companion mass of m.$\sin{i} \sim$ 0.07 M$\rm _{\odot}$, both consistent with \citet{biller2022}.
Additionally, the periodogram of the residuals shows two significant peaks at 8.8 and 17.6 days, once again consistent with the stellar rotation published in \citet{biller2022}.

\subsubsection{GJ~9482}

GJ~9482 is an M0V star (0.68 $\pm$ 0.01 M$\rm _{\odot}$) with a massive planet of m.$\sin{i}~=~9\pm6~M\rm _{J}$ and a period of 17337$\rm \pm$15512 days, detected in RV by \citet{segransan2011harps}.
For this star, the dominant long-term model is a sinusoidal function, with a fitted period of 6700 days and an offset value of -27.1 +/- 15.7 m/s, as this target was monitored both before (41 nights) and after (13 nights) the fibre change.
However, the Keplerian fit performed during the iterative search for signals did not converge.
Using the DACE platform, we obtain a good solution (with a residual dispersion of 3.39 m/s) that constrains the period and mass much better than in \citet{segransan2011harps}, thanks to the addition of 13 new data points (2015.08-2019.08).
We obtain a companion mass of m.$\sin{i}=4.58\pm 0.09~M\rm _{J}$ and a orbital period of 3659$\rm \pm$4 days.
However, in this solution the RV offset at the time of the fibre change is 39.2 $\rm \pm$ 0.6 m/s.
The correction of the time series with an incorrect offset value explains why the automatic Keplerian fitting did not converge.
This poor estimation of the offset is due to the high eccentricity of the orbit (e = 0.660 +/- 0.004), modelled at this step as a sinusoidal function.
This case, with a high-eccentricity orbit, highlights a limitation of our systematic study of long-term signals which do not converge, and so must be treated individually.
Using \texttt{orvara}\xspace, the inclination is not well constrained (i = ${90}_{-21}^{+24}$), but it is sufficient to confirm that the system is observed close to edge-on, giving a maximum mass of the planet of 4.6~$\rm M_{J}$.

\begin{figure}
\includegraphics[width=0.99\linewidth]{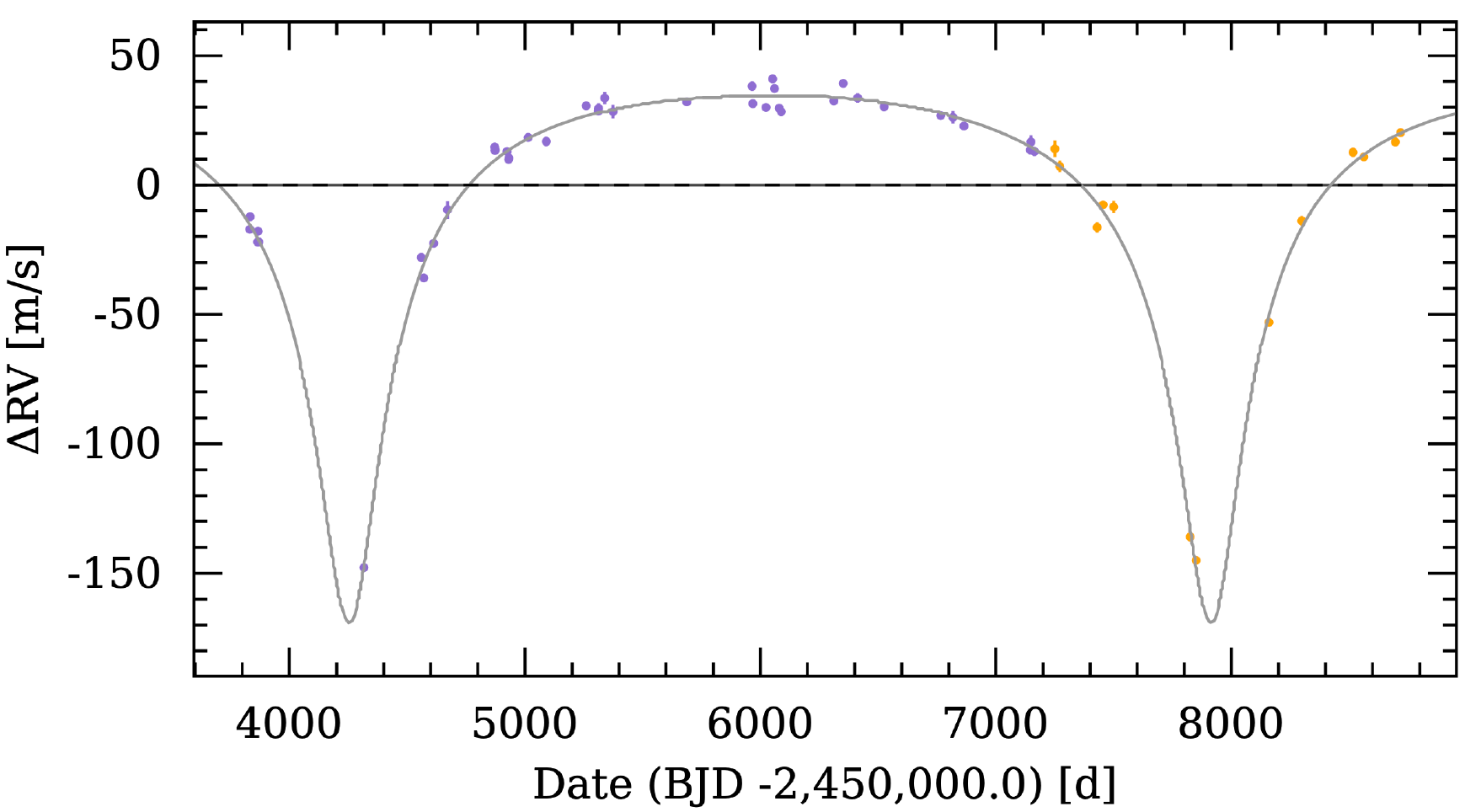}
\caption{GJ~9482 HARPS 54 RV measurements and best fitting orbit model. The 41 RV measurements done before the change are in purple and the 13 after the change are in orange.}
\label{gj9482vr}
\end{figure}

\subsubsection{GJ~9588} 

GJ~9588 is an M1 star with a mass of 0.44 $\pm$ 0.03 M$\rm _{\odot}$ and a companion detected through direct imaging, with a spectral type of M9 ($\pm$2) and an estimated mass between 0.075 and 0.080 M$\rm _{\odot}$ \citep{schneider2011}.
We measure the signature of this companion in the RV time series as a linear trend of -7.43 $\pm$ 2.94 m/s/yr, consistent with the reflex motion due to the presence of an object close to the hydrogen-burning limit at a separation of 120 au.
However, this is not the dominant signal in RV time series.
We detect a clear signal of a new companion with a period of 618.54$\pm$0.11 days and a minimum projected mass of 34 M$_J$.
Using \texttt{orvara}\xspace, we obtain an inclination of 54 degrees and a true mass of 44~$\rm M_{J}$.
The FWHM of the CCF remains stable over time, indicating that this system is not an SB2.
Therefore, GJ~9588 is a very interesting M-dwarf system with two companions, one in the brown dwarf mass range and the second at the boundary between a star and a brown dwarf, both being at very different orbital periods.
As illustrated, for example, by Fig.~10 in \citet{artigau2021}, a brown dwarf companion to an M dwarf with an orbital period of $\sim$2 yr is a rare object.
This system is of interest from the perspective of formation models.

\begin{figure}
\includegraphics[width=0.99\linewidth]{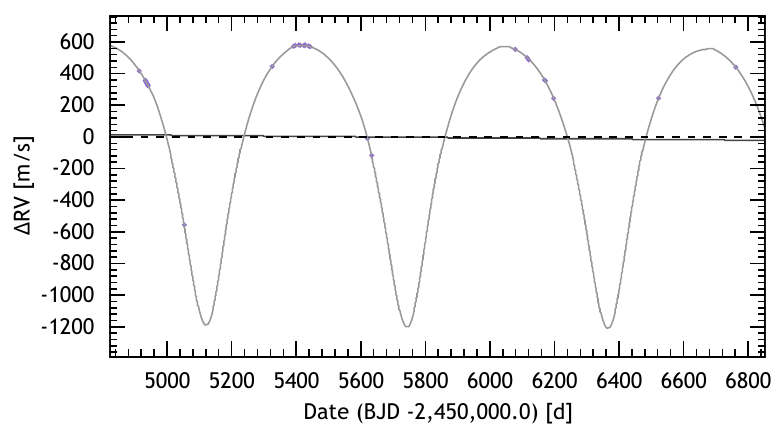}
\caption{GJ~9588 HARPS 28 RV measurements done before the change of fibres and best fitting orbit model.}
\label{gj9588vr}
\end{figure}

\subsubsection{Stars with RV drift} 

Table~\ref{tablepente} lists the stars for which we detect RV drift (linear or quadratic).
The sources of these drifts can be companions and/or long-term activity signal (magnetic cycle), making these stars good candidates for future monitoring.
Focusing on the 14 stars with the highest amplitude of RV variation (50 m/s peak-to-peak), the majority of them are well-known binaries (e.g. GJ~234 or GJ~660), but some do not have a known companion at the moment and need to be investigated in the future.

\subsection{Planetary candidates}

Beyond the stellar or brown dwarf companion signatures presented and discussed, we point out 16 previously unidentified periodic or quasi-periodic signals in this section.
This is only a preliminary analysis as the scope of this article is to make a homogeneous analysis of a large sample.
For several of these sources we identify planetary candidates for which additional measurements would be crucial to strengthen the significance of the signal.
It is worth noting that several of the stars listed below are part of the NIRPS GTO programme \citep{bouchy2017}, which should allow conclusions to be drawn about the status of some of these planetary candidates in the near future.
For other stars, a detailed analysis of the activity signal via Gaussian process regression or any other method would be important, but is outside the scope of our analysis.

\subsubsection{GJ~300}

This is a very low-mass star (M$_*$=0.27$\pm$0.09 M$\rm _{\odot}$) of M5V spectral type, 
observed on 39 nights between December 2003 and April 2010 with HARPS.
This target was poorly monitored during the first and last seasons, with only four measurements before 2008 and five measurements after 2009 (see Fig.~\ref{gj300complet}).
Therefore, we focus on the 30 measurements taken between October 2008 and April 2009 (see Fig.~\ref{gj300zoom}).
This RV time series exhibits a high dispersion (rms of 5.07 m/s), and its periodogram shows a significant maximum peak at 8.33 days with a FAP level about 0.1\% (see Fig.~\ref{gj300periodogrammcomplet} and Fig.~\ref{gj300periodogrammzoom}).
Using the DACE platform, we fit a Keplerian signal at 8.347$\pm$0.005 days, with a high eccentricity (0.64$\pm$0.07) and a semi-amplitude of 7.1$\pm$0.8 m/s, corresponding to a planet of approximately 7.3 M$\rm _{\oplus}$ (presented in Fig.~\ref{gj300phasefolding}).
The dispersion of the residuals is about 2.14 m/s and the reduced $\chi^2$ is below 3.
When we subtract this $\sim$8.33-day signal from the complete RV tine series of 39 nights, the periodogram of the residuals shows a maximum peak at 299 days.
This long-term signal is consistent with the dominant quadratic pattern found in Sect.~\ref{secAnalysis1}.
According to the mean activity level $\log{R'_{HK}}$ (-5.3066), the derived rotation period is about 67 days, which does not match with the 8.33-day RV signal.
Furthermore, the emission in the three activity indicators (used in Sect.~\ref{act}) is not consistent with a saturated regime of the chromosphere, typical of a fast rotator with a rotation period of less than ten days \citep{delfosse1998new,reiners2009,astudillo2017magnetic}.
This $\sim$8-day signal is therefore a good indication of the presence of a low-mass planet orbiting GJ~300.
However, we note that the high eccentricity required to fit this signal is not consistent with the orbital period of less than ten days; such a planet should be on a circularised orbit due to tidal effects.
This may indicate that the small number of measurements does not yet allow convergence to the correct orbital solution.
If we impose a circular orbit, the fit is of much lower quality (dispersion of the residuals of 3.37 m/s).
An alternative solution is to consider two planets in 2:1 resonance on circular orbits, which results in a very good fit (dispersion of the residuals of 2.43 m/s).
In any case, this source deserves further observations to lower the FAP of the signal and to study the long-term signal.

\subsubsection{GJ~317}
GJ~317 is an M4V-type star with a mass of 0.42$\pm$0.02 M$\rm _{\odot}$, known to host two confirmed planets in orbit.
GJ~317~b is a planet with a mass of 1.853$\pm$0.037 M$_J$ and an orbital period of 695.36$\pm$0.70 days \citep{rosenthal2021}, discovered using the RV method \citep{johnson2007new}.
GJ~317~c is a less constrained object with a mass of 1.673$^{0.078}_{-0.076}$ M$_J$ and an orbital period of 7500$^{1500}_{-720}$ days \citep{rosenthal2021}.
GJ~317 was observed on 99 nights from January 2010 to June 2018 with HARPS, and is therefore affected by the fibre change.
We find a signal similar to that of GJ~317~b as a large sinusoidal function (see Table~\ref{sin}) and that of GJ~317~c as a linear trend.
However, during the periodic analysis, we also find two signals at 314.40 and 236.46 days.
These signals stem from an RV time series corrected with an erroneous offset value due to the difficulty of identifying the two signals during our long-term analysis, as in the case of GJ~9482.
Using the DACE platform, we adopt the period of \cite{rosenthal2021} for GJ~317 c as a prior to fit the two Keplerian signals (as presented in Fig.~\ref{gj317}).
Two peaks in the periodogram of the residuals reach the 10\% FAP level at periods of 50.9 and 59.4 days.
These periods are of the same order of magnitude as the estimated rotation period (63 days with the $\log{R'_{HK}}$), and are close to the significant peak we find in the periodogram of the S$_{NaD}$ index at 64 days.
Consequently, these signals are not considered new candidates, but signatures of the rotation.


\subsubsection{GJ~361}

GJ~361 is an M2V-type star with a mass of 0.484$\pm$0.022 M$\rm _{\odot}$, observed on 103 nights with HARPS from December 2007 to March 2013 (1919 days of coverage).
The periodogram of the RV time series shows two significant peaks with a FAP level well below 0.1\% at 28.91 and 26.82 days of period.
The last one is a 365-day alias of the first.
Using the DACE platform, we fit a signal with a period of 28.95 days and the periodogram of the residuals exhibits a significant peak at 24.98 days.
The 33-day rotation period derived from the mean activity level is very close to the signal in the RV periodogram.
Furthermore, we find a significant peak in the S$_{Ca}$ index periodogram at 23.96 days (close to the alias), and another one at 29.79 days in the S$_{NaD}$ index periodogram.
Therefore the two significant periodicities in the HARPS data are similar to that of the stellar rotation, and we do not propose any planetary candidate at these periods.

\subsubsection{GJ~393}

GJ~393 is an M2V-type star with a mass of 0.43$\pm$0.11 M$\rm _{\odot}$, observed for a total of 169 nights with HARPS from December 2003 to April 2016.
At the first step of the iterative search for periodic signals in the RV periodograms, we find a signal with a period of 7.02 days.
This particular signal is identified as a planetary signature by \cite{amado2021}, using velocities from HARPS, CARMENES, and HIRES, and photometric data.
In this study they subtract a model based on a Gaussian process with a 34-day period signal, attributed to rotational signature, and a longer-term signal, also attributed to activity effect.
Numerous significant peaks (with FAP level around or well below 0.1\%) are present in the periodogram of the residuals (see Fig.~\ref{gj361}) around 35, 51, 450, 550, 780, and 1000 days.
In addition, we find significant peaks in the periodograms of two activity indices (S$_{Ca}$ and S$_{H\alpha}$) at a period of $\sim$340 days, indicating that activity may be responsible for the longer-period RV signals.
Consequently, we consider it highly likely that the remaining signals, after subtracting the 7.02-day planetary signature, are not due to planets, but rather to the impact of complex stellar activity.

\subsubsection{GJ~569A}

GJ~569A is a young M2-type star with a mass of 0.49$\pm$0.04~M$\rm _{\odot}$, and an estimated age of 100 to 125 Myr \citep{simon2006}.
It is sparsely monitored with HARPS, with only 25 measurements taken between 2003 and 2013.
The periodogram exhibits a maximum significant peak at a period of 14.38 days (see Fig.~\ref{gj569res}) with a FAP level of 0.1\%.
Given its young age this star is highly active, and we estimate a 13-day rotation period from the average $\log{R'_{HK}}$ level.
Additionally, we find signals at 11.5, 15.8, and 23.1 days in the periodograms of the three activity indices. Due to the poor sampling of the time series and the high level of activity, we are unable to confirm the presence of a planetary candidate signature. The primary signal observed in the RV data is most likely attributable to the stellar activity. 

\subsubsection{GJ~588}
GJ~588 is an M2V-type star with a mass of 0.42$\pm$0.20 M$\rm _{\odot}$, observed for 64 nights with HARPS\footnote{GJ~588 and GJ~887 have also been intensively monitored in recent years as part of observational programmes focused on asteroseismology. Although the hundreds of data points from these programmes are now publicly available and would undoubtedly offer valuable insights into the stars' characteristics, we have not incorporated them into our RV database to keep the RV database consistent, as a dataset of several hundred measurements would compromise the homogeneity of both this study and the next, given their specific goals.} from 2003 to 2019, and is therefore affected by the change of fibres.
Using the DACE platform, we observe a significant dispersion in RVs around the fibre change date, with some measurements taken shortly after the change.
To ensure the homogeneity of the dataset, we exclude these specific points.
Subsequently, the study of the remaining dataset exhibits a significant peak at a period of 48 days, close to the rotation period of 51.2 days derived from the average level of activity.
Furthermore, we find several significant peaks in the analysis of the chromospheric series, at distinct periods (11, 27, 63, and 73 days), indicating a complex manifestation of activity in this star.
The presence of multiple significant peaks in the chromospheric series further supports the notion of a dynamically active star, exhibiting a rich variety of activity phenomena. Although the RV data exhibited a significant peak at 48 days, the proximity of this period to the derived rotation period reinforces the idea that this signal is associated with the stellar activity rather than a planetary signal.

\subsubsection{GJ~654}

GJ~654 is an M2V star (M$_*$ = 0.48$\pm$0.04~M$_{\odot}$) observed for 186 nights with HARPS from April 2008 to September 2016, which is therefore affected by the change of fibres.
Using the DACE platform, the fit of the 260.8-day signal found in the Keplerian analysis does not converge to any satisfactory solution.
Consequently, we only focus on four well-sampled observing seasons (165 data points) between April 2013 to September 2016.
After subtracting a linear drift, the periodogram of the RV times series exhibits a large peak at 50.50 days and a clear secondary at 15.35 days.
In the S$_{Ca}$ periodogram, we found two significant peaks at 48.2 and 51.4 days, and a significant peak at 50.4 days in the S$_{NaD}$ periodogram.
Therefore, we consider the first signal at 50.5 days to be attributed to the star's rotational signature, while the remaining signal at 15.35 days is a strong planetary candidate.
The fitted Keplerian function corresponds to a 5.11$\pm$0.4 M$_{\oplus}$ planetary mass.
The complete characterisation of the system goes beyond the scope of this paper and will be addressed in a future paper.

\subsubsection{GJ~699}

At just 1.8~pc from our Sun, Barnard's Star (GJ~699) is the second-closest system after Proxima Centaury - $\alpha$~Centauri \citep{bailer2018,gaia2021}.
\cite{ribas2018candidate} proposes the presence of a candidate super-Earth planet on a 233-day orbit.
The star has a long rotation period of approximately 145$\pm$15~days, determined by variations in chromospheric activity indicators by \cite{toledo2019}, and 137$\pm$5~days derived from the longitudinal magnetic field variation by \citet{fouque23}.
However, \cite{lubin2021} does not confirm this planet and proposes it to be an alias of the rotation period.
This conclusion is further supported by infrared RV measurements with SPIRou \citep{artigau22}, which strongly disfavour the existence of a 233-day planet.
The periodogam of our RV time series exhibits a significant peak at 273 days (i.e. twice the rotation period).
It is therefore likely that this signal is due to stellar activity.

\subsubsection{GJ~739}

GJ~739 is an M3V-type star with an estimated mass of 0.46$\pm$0.02~M$_{\odot}$, poorly monitored with HARPS with only 19 measurements taken from July 2008 to May 2012.
In our analysis we find a significant peak with a very low FAP level (well below 0.1\%) in the RV periodogram at 137.8 days. 
By fitting a Keplerian function on the DACE platform, we obtain a signal at 145.3$\pm$0.8 days, corresponding to a planet with a mass of 39$\pm$4 M$_{\oplus}$.
The rotation period derived from the level of activity is around 50 days, and we find a significant peak in the S$_{Ca}$ periodogram at 40.82 days and another significant peak at 47.9 days in the S$_{H\alpha}$ periodogram.
These findings suggest that the signal adjusted to about 138 days is unlikely to be a rotational signature.
However, to confirm this strong planet candidate, we require more than the current 19 measurements.

\subsubsection{GJ~754}

GJ~754 is an M4.5V-type star \citep{reid1995} with a mass of 0.18 $\pm$ 0.02~M$_{\odot}$, observed for a total of 170 nights between May 2004 and September 2016 with HARPS, which means it is affected by the change of fibres.
Upon analysing the entire dataset, we find two significant peaks in the RV periodogram at 249 and 17 days (in order of their power), with no long-term signal detected.
However, focusing on the well-monitored seasons (from 2012 to 2016), we find a significant peak at 77 days using the DACE platform (see Fig.~\ref{gj754}).
Using the activity proxies, we find seven significant peaks at different periods in the three chromospheric indexes (ranging from 12 to 1750 days), and the rotation period derived from the mean activity level is about 77 days.
In conclusion, this quiet mid-M dwarf exhibits a complex and evolving activity signal, which poses challenges in disentangling any potential planetary system from the RV time series.
A more in-depth analysis is required to determine if there is indeed a planetary system, which is beyond the scope of this article.

\subsubsection{GJ~846}

GJ~846 is an M1Vtype star with a mass of 0.61 $\pm$ 0.08~M$_{\odot}$, observed for a total of 51 nights from May 2004 to November 2010 with HARPS.
The RV periodogram exhibits a significant peak at 10.8 days with a FAP of around 0.1\%.
However, using the DACE platform for Keplerian fitting, the results are inconclusive, showing a high dispersion in the residuals and a high value of reduced $\chi^2$.
The rotation period derived from the average activity level is about 20 days, and we find significant peaks at 7.8, 10.7, 19.2, and 30.9 days in the periodograms of the three activity indices.
Previous studies identified rotation periods ranging from 26 to 31 days when determined by activity index variations \citep{suarez2015, suarez2017, DiezAlonso2019}, and from 10.7 to 11.01 days when determined from the magnetic field modulation \citep{hebrard2016modelling,fouque23}, highlighting the complexity of the activity signal.
In conclusion, the RV signal is consistent with one of the activity signals, leading us to consider that the RV time series is dominated by the short-term impact of activity.

\subsubsection{GJ~880}
GJ~880 is an M2V-type star with a mass of 0.58 $\pm$ 0.08~M$_{\odot}$, and was well monitored with HARPS over 137 nights from December 2003 to October 2016.
We find a significant peak at 40.05 days in the RV periodogram, with a FAP below 0.1\%, and another less significant peak at 37.2 days.
These signals are aliases of each other due to the one-year sampling.
In addition, we find a signal at 37.4 days in the S$_{NaD}$ periodogram, another at 37.0 days in the S$_{H\alpha}$ periodogram, and a third at 35.3 days in the S$_{Ca}$ periodogram.
\cite{suarez2015} and \cite{fouque23} both find a similar rotation period of $\sim$37~days using two complementary methods.
In conclusion, the signal at 37.2 days corresponds to the rotational signature, while the signal at 40.5 days is an alias of it.


\section{Limits and discussion}\label{discus}

The systematic analysis of 200 highly heterogeneous RV time series imposes the use of conservative methods, which entails certain limitations.
Here we discuss these choices and their consequences in terms of detection.

\subsection{Sampling}

\begin{figure}
\includegraphics[width=0.99\linewidth]{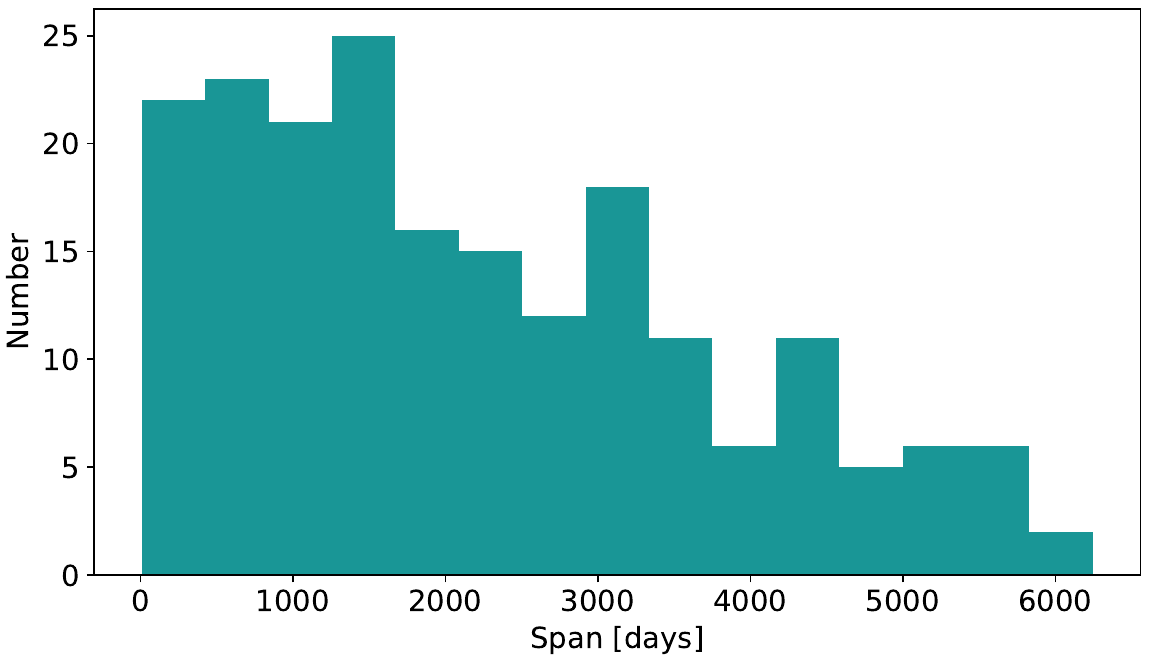}
\caption{Distribution of time coverage of the RV time series for the 200 stars in our sample.}
\label{span}
\end{figure}

Even if stars observed for fewer than 10 nights are excluded from the study, the sampling frequency remains the main bias in this analysis.
Stars that are well sampled over a short timescale allow the detection of short-term periodicity, and the better characterisation and subtraction of activity-induced signals from the rotation period.
Conversely, only stars monitored over a long timescale enable the analysis of long-period signals.
The histogram of the time interval between the first and last measurement for the 200 stars in our sample is presented in Fig.~\ref{span}.
It is evident that this span varies significantly, resulting in an uneven detectability.
This bias is taken into account in the second article of this analysis (Mignon et al. in prep.), where individual detection limits are computed in order to construct the overall occurrence frequency.

\subsection{Periodograms}

For this study we compute the periodogram proposed in \cite{zechmeister2009generalised}, previoulsy used in \cite{bonfils2013harps}, and adopt the FAP level of the highest peak as a conservative detection criterion for detecting a periodic signal.
Although more recent methods enable the better separation of Keplerian signals from correlated noise, which can be due to activity \citep{Delisle2020, Hara2022}.
They generally require a more accurate knowledge of the rotation period, which is not precisely known for all our targets.
Additionally, for certain M dwarfs in the sample, the effect of activity may manifest at different periods \citep{mignon23a}.

\subsection{Corrections of activity}
We correct 23 RV time series from 35 signatures of the stellar rotation using a Keplerian signal with the same phase across the entire dataset.
As the presence of activity-induced effects is a direct consequence of magnetic structures (mainly spots) on the surface, which have a finite lifetime, their behaviour is usually quasi-periodic rather then strictely periodic.
Our correction is therefore not optimal compared to a case-by-case study, but it is well-suited for our homogeneous and automated search.
However, this approach may hinder the detectection of some weak Keplerian signals.

\subsection{Missed planetary systems}

In this section we provide an explanation for the non-detection of some previously published planets, aiming to illustrate the limitations of conducting such a homogeneous study on a very large sample.
Additionally, this comparison allows us to identify the domains (e.g. mass, period, multiplicity) where our study remains robust. 
According to the NASA Exoplanet database, we identify 51 planets out of the 78 previoulsy published (65.4\%) within our sample of 200 M dwarfs in 2020--2021.
In terms of planetary systems, we identify 31 of the 40 previously discovered systems (75\%).

\subsubsection{System detected with additional measurements}

We limit our study to the analysis of RVs data from HARPS, to ensure a homogeneous dataset.
However, some of the planets we do not detect were originally discovered using datasets combined from other instruments.
GJ~191 hosts a two-planet system discovered by \cite{anglada2014two} using RV data from HARPs, HIRES, and PFS combined with photometry (ASAS 3).
GJ~357 \citep{luque2019planetary} hosts a three-planet system, which was revealed by combining transit and RV methods from TESS, CARMENES, HIRES, UVES, and HARPS.
The planetary systems orbiting GJ~422 and GJ~682 were detected by the RV method, but most of the data were obtained using UVES \citep{tuomi2014bayesian}.
GJ~740 and GJ~3779 are both single-planet systems detected by RVs, combining HARPS, HARPS-N, and CARMENES data \citep{toledo2021super} and HARPS and CARMENES data \citep{luque2018}, respectively.
By excluding these five systems, which were not detected mainly due to an insufficient number of measurements with HARPS, we recover over 88\% of the previously published systems within our sample.

\subsubsection{Very short-period planets}

We perform automatic Keplerian analysis for periods longer than 1.6 days to avoid confusion between aliases created by the one-day sampling and the actual period of the signal.
As a result, two planets with periods below this threshold were not detected: GJ~1214~b, which has a period of 1.58 days and was detected through the transit method \citep{carter2011transit}, and GJ~3543~b, which has a period of 1.12 days and was detected by the RV method \citep{astudillo2015harps}.
By lowering the threshold to 1.1 days, we do detect these two planets in the periodograms.
This confirms that they are only missed due to the choice of the tested period domain, and thus lie outside the scope of this study.
It is therefore possible that we might have missed unknown planets in our sample with orbital periods shorter than 1.6 days.
If we consider only systems with periods superior to 1.6~days, we recover 94\% of the previously published systems.

\subsubsection{The specific case of GJ~1061}

The case of GJ~1061 is complex.
This system includes three planet candidates detected from the HARPS data \citep{dreizler2020}, with minimum masses ranging from 1.4 to 1.8 Earth masses and RV signal semi-amplitudes between $K=1.8$ and $K=2.5$ m/s.
The RVs we obtain show an offset of about a dozen m/s over one observing season, without a clear explanation at this stage.
These data come from three different programmes, and the mask used in the CCF method is not the same for the various observer teams.
However, even if our template matching code uses this first CCF RV for its original iteration, its final result is independent of it.
Several other sources were initially analysed with different CCF masks, but they did not generate this type of offset.
This remains the only case where we do not detect the published signal without identifying the reason behind it.

\section{Conclusion}\label{secConclu}

This comprehensive analysis of RVs measured with HARPS on M dwarfs between 2003 and 2019 (200 stars with over ten measurements) significantly updates the set of detected periodic RV signals.
The investigation of long-term trends (linear, quadratic, or sinusoidal) reveals variability in 57 time series (28.5\% of our sample): 16 linear trends, 29 quadratic trends, and 12 sinusoidal functions.
Regarding the objects in the last category, we confirm seven detections of massive companions (very low-mass stars or brown dwarfs), identify four new companions of M dwarfs (ultra-cool dwarfs or brown dwarfs for GJ~3307, GJ~4001, GJ~4254, and GJ~9588), and significantly revise the parameters of a massive planet detected in 2011 (GJ~9482~b), providing better constraints on its orbit with a period of 3659$\pm$4 days and a mass of 4.5 M$_J$.
After subtracting the long-term variability, the analysis of the remaining 108 periodic signals leads to the recovery of 52 previously published planets and 34 known rotation period signatures.
The analysis of other previously unidentified periodic signals allows us to propose three planetary candidates in orbit around GJ~300 (7.3M$_{\oplus}$), GJ654 (5M$_{\oplus}$), and GJ739 (39M$_{\oplus}$), which will require additional measurements for confirmation.
For six other stars (GJ~361, GJ~393, GJ~569, GJ~754, GJ~846, and GJ~880), we suggest that the RV variation is more likely due to stellar activity.
This work serves as the foundation for a statistical study of the M dwarf population in the solar neighbourhood, from detection limits to occurrence statistics, which is the aim of the second article in this series.


\begin{acknowledgements} 
This research has made use of the VizieR catalogue access tool, CDS, Strasbourg, France. The original description of the VizieR service was published in A\&AS 143, 23
This work has been supported by a grant from LabEx OSUG@2020 (Investissements d'avenir - ANR10LABX56) 
This research has made use of the NASA Exoplanet Archive, which is operated by the California Institute of Technology, under contract with the National Aeronautics and Space Administration under the Exoplanet Exploration Program.
This research has made use of the Exoplanet Follow-up Observation Program (ExoFOP; DOI: 10.26134/ExoFOP5) website, which is operated by the California Institute of Technology, under contract with the National Aeronautics and Space Administration under the Exoplanet Exploration Program.
This work has been carried out within the framework of the NCCR PlanetS supported by the Swiss National Science Foundation.
NCS acknowledges funding by the European Union (ERC, FIERCE, 101052347). Views and opinions expressed are however those of the author(s) only and do not necessarily reflect those of the European Union or the European Research Council. Neither the European Union nor the granting authority can be held responsible for them. This work was supported by FCT - Fundação para a Ciência e a Tecnologia through national funds and by FEDER through COMPETE2020 - Programa Operacional Competitividade e Internacionalização by these grants: UIDB/04434/2020; UIDP/04434/2020.
\end{acknowledgements}


\bibliographystyle{aa}
\bibliography{library}


\begin{appendix}
\onecolumn

\section{Graphs}\label{append2}

\subsection{Planetary candidates}

\subsubsection{GJ~300}

\begin{figure}[h!]
   \centering
   \begin{subfigure}{0.49\textwidth}
     \centering
     \includegraphics[width=\textwidth]{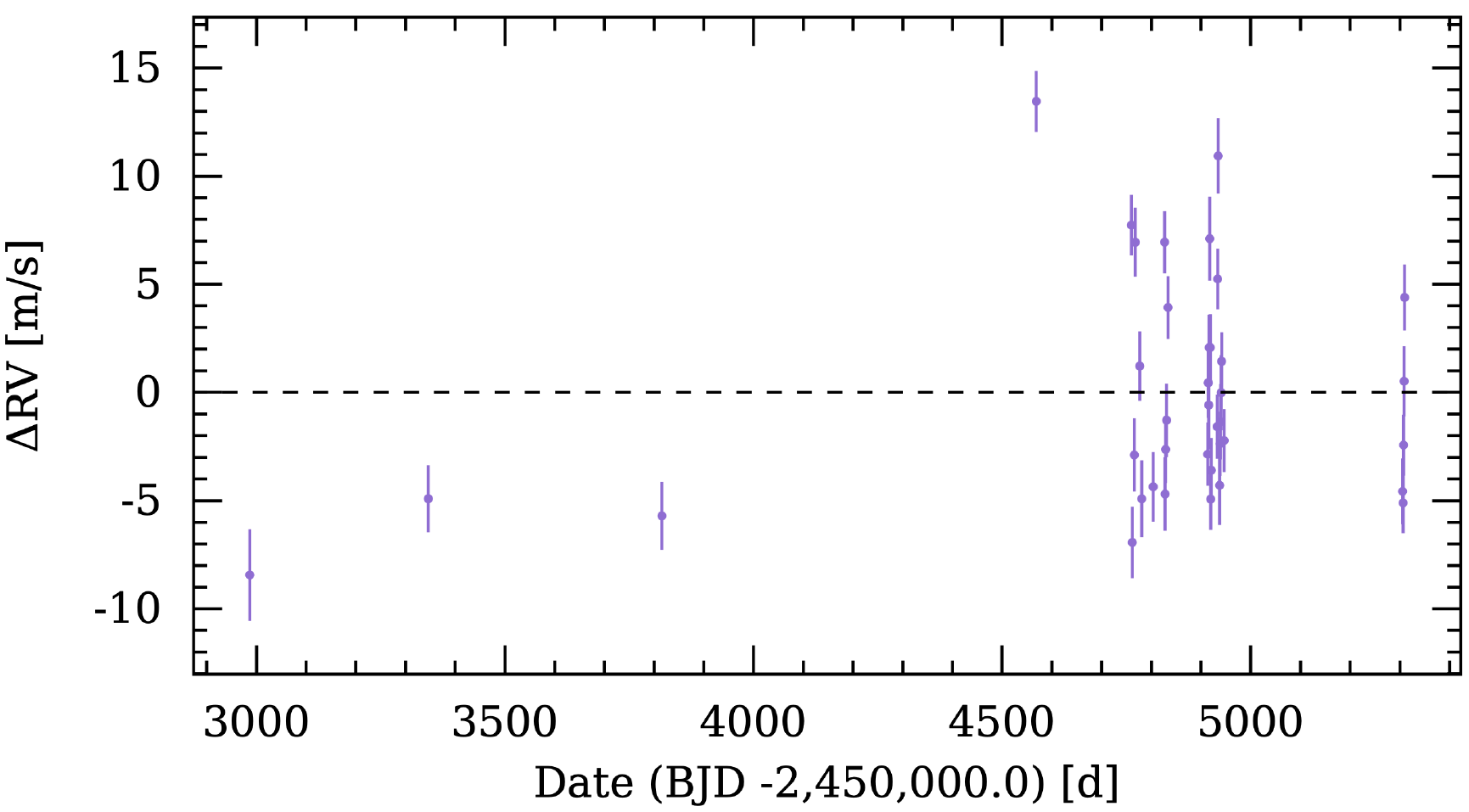}
   \end{subfigure}
   \hfill
   \begin{subfigure}{0.49\textwidth}
     \centering
     \includegraphics[width=\textwidth]{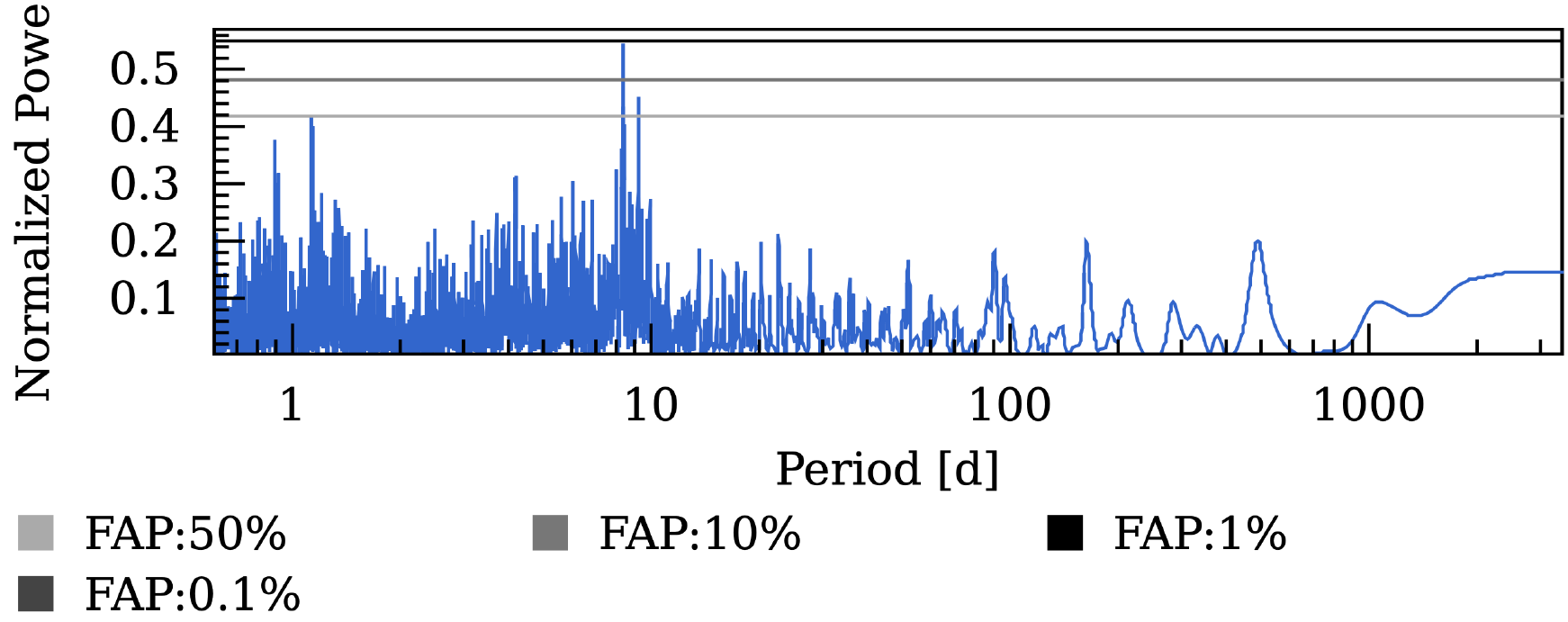}
   \end{subfigure}
   \hfill
   \caption{\label{gj300complet}RV time series and GLS periodogram of the complete RV times series of GJ~300 }
\end{figure}

\begin{figure}[h!]
  \centering
  \begin{subfigure}{0.49\textwidth}
    \centering
    \includegraphics[width=\textwidth]{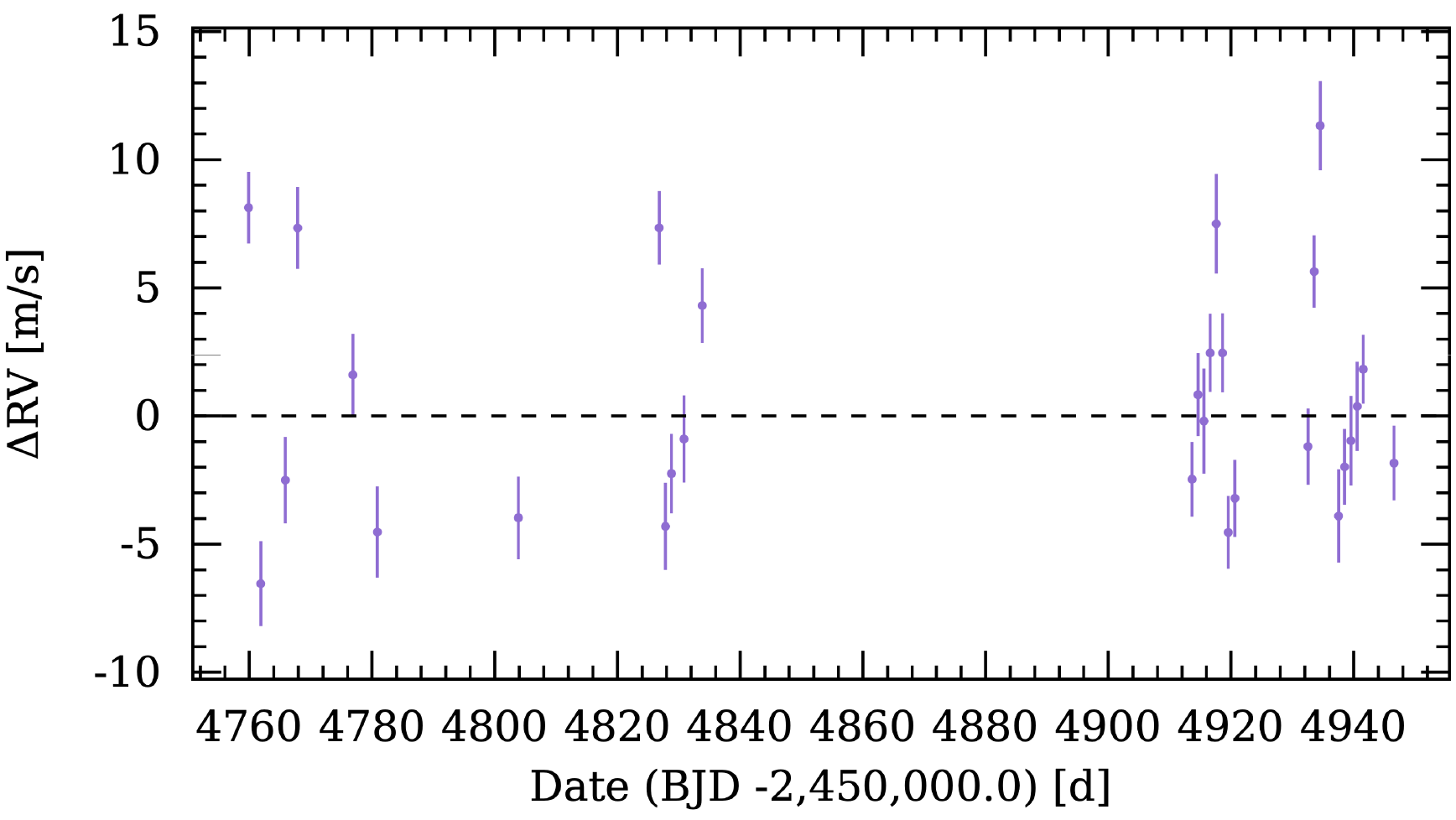}
  \end{subfigure}
  \hfill
  \begin{subfigure}{0.49\textwidth}
    \centering
    \includegraphics[width=\textwidth]{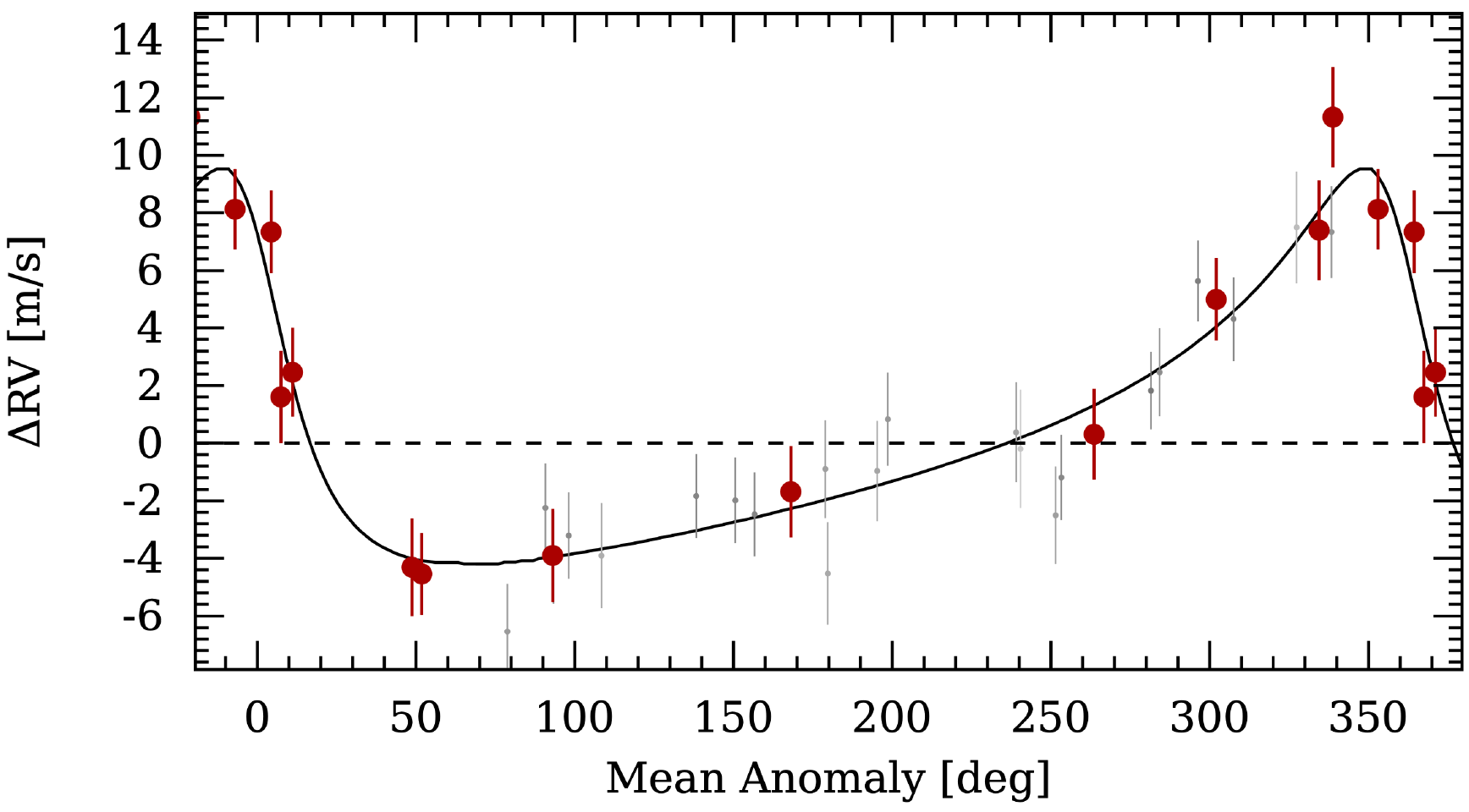}
  \end{subfigure}
  \hfill
  \begin{subfigure}{0.49\textwidth}
    \centering
    \includegraphics[width=\textwidth]{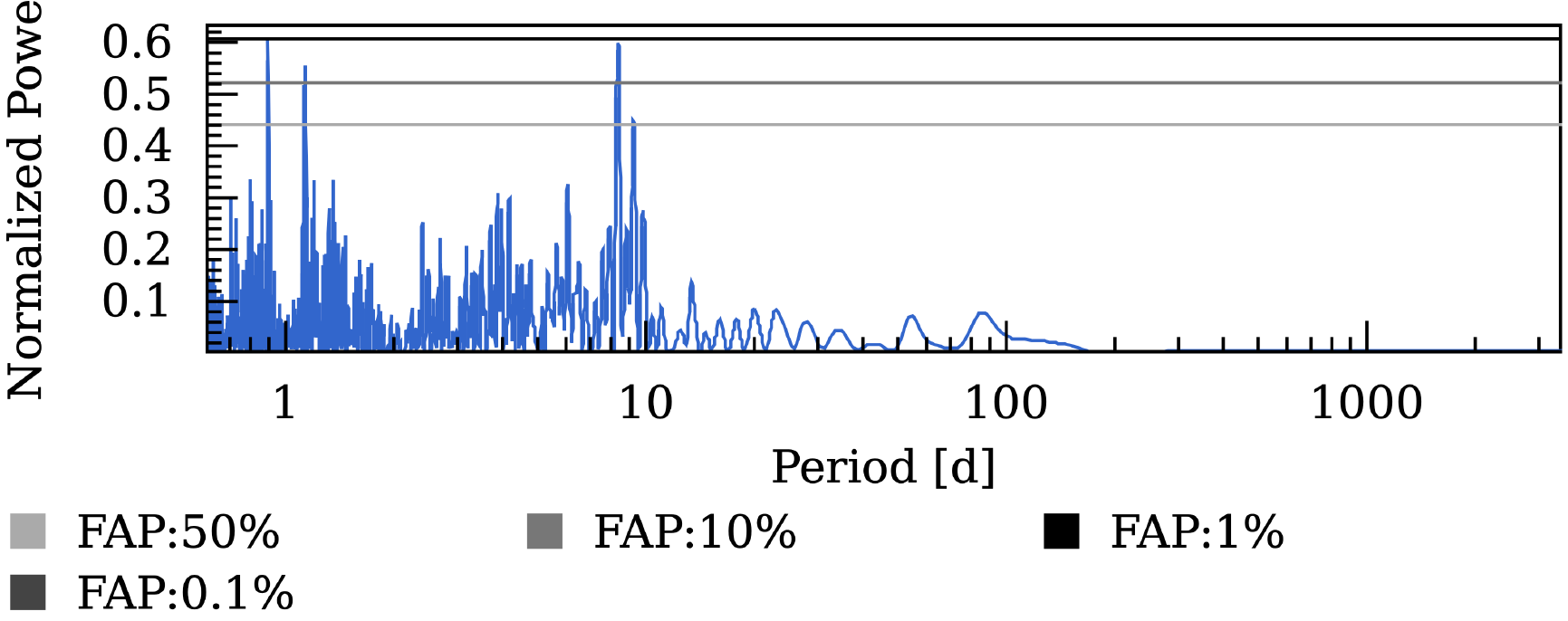}
  \end{subfigure}
  \hfill
  \begin{subfigure}{0.49\textwidth}
    \centering
    \includegraphics[width=\textwidth]{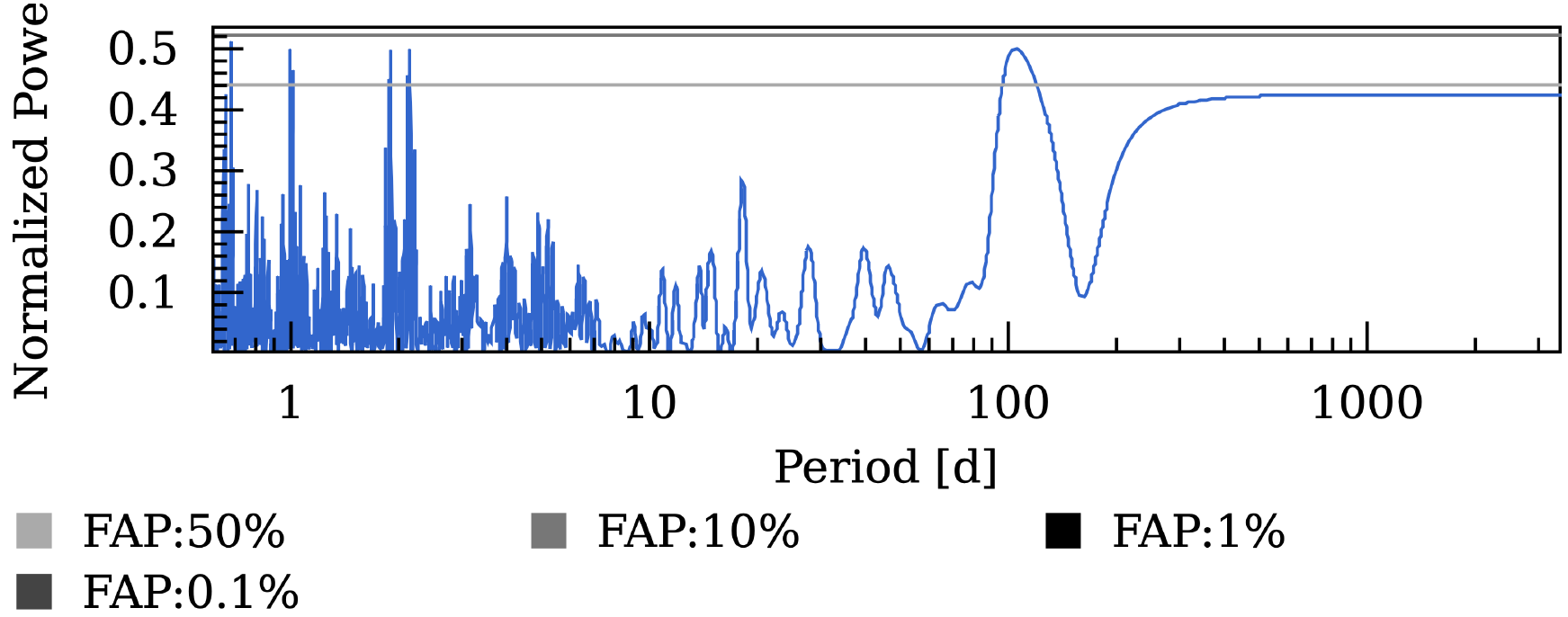}
  \end{subfigure}
  \hfill
  \caption{\label{gj300zoom}RV time series and GLS periodograms of the 28 nights between October 2008 and April 2009 (see text) of GJ~300 (left: original, right: residuals).}
\end{figure}

\newpage

\subsubsection{GJ~317}

\begin{figure}[h!]
  \centering
  \begin{subfigure}{0.49\textwidth}
    \centering
    \includegraphics[width=\textwidth]{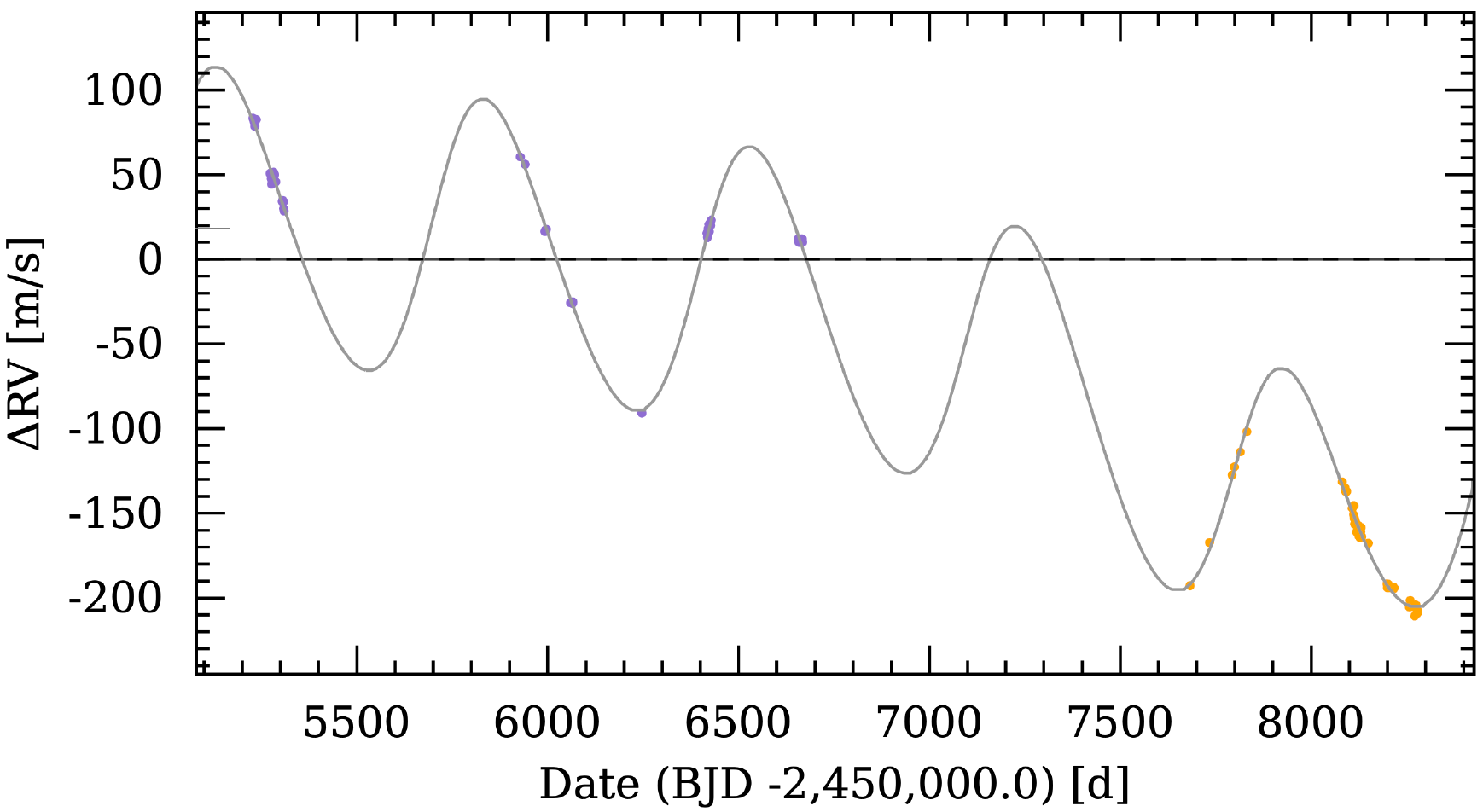}
  \end{subfigure}
  \hfill
  \begin{subfigure}{0.49\textwidth}
    \centering
    \includegraphics[width=\textwidth]{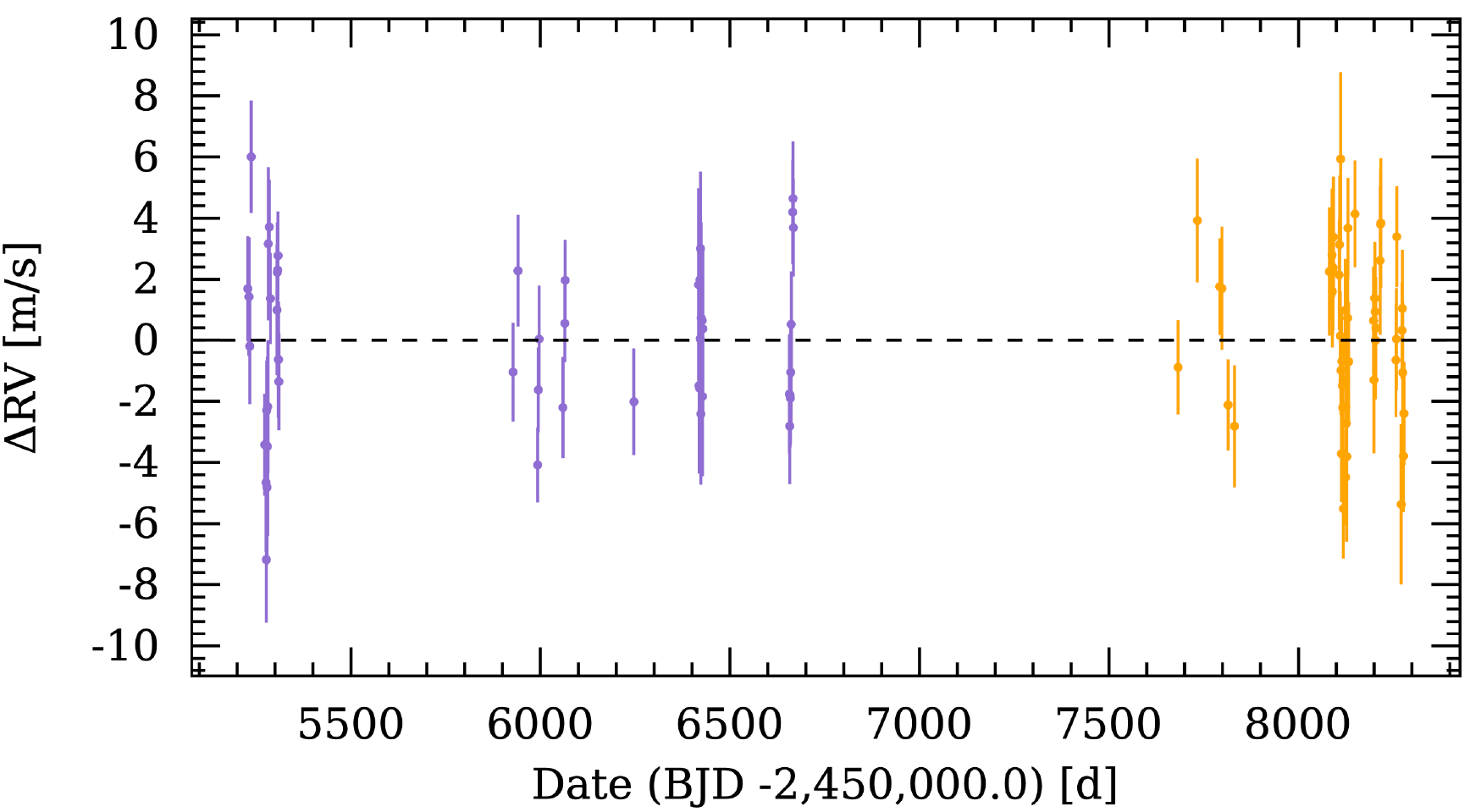}
  \end{subfigure}
  \hfill
  \begin{subfigure}{0.49\textwidth}
    \centering
    \includegraphics[width=\textwidth]{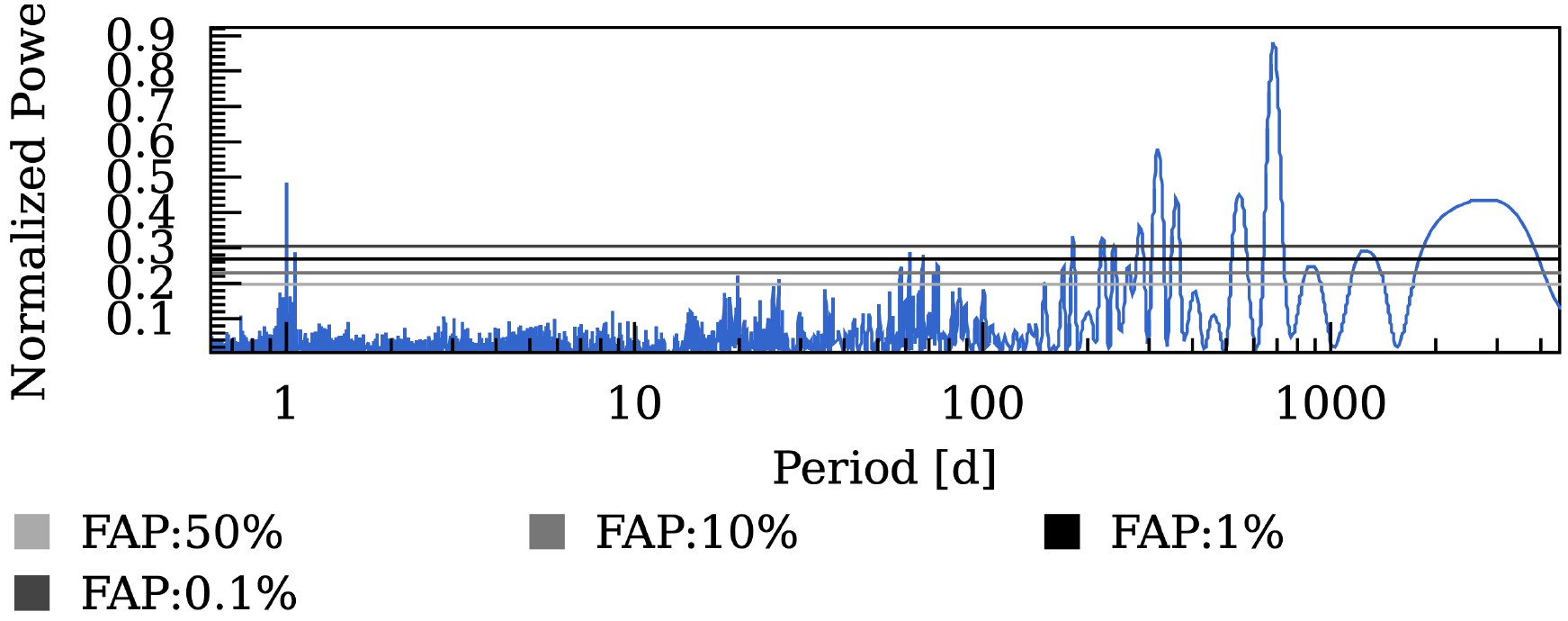}
  \end{subfigure}
  \hfill
  \begin{subfigure}{0.49\textwidth}
    \centering
    \includegraphics[width=\textwidth]{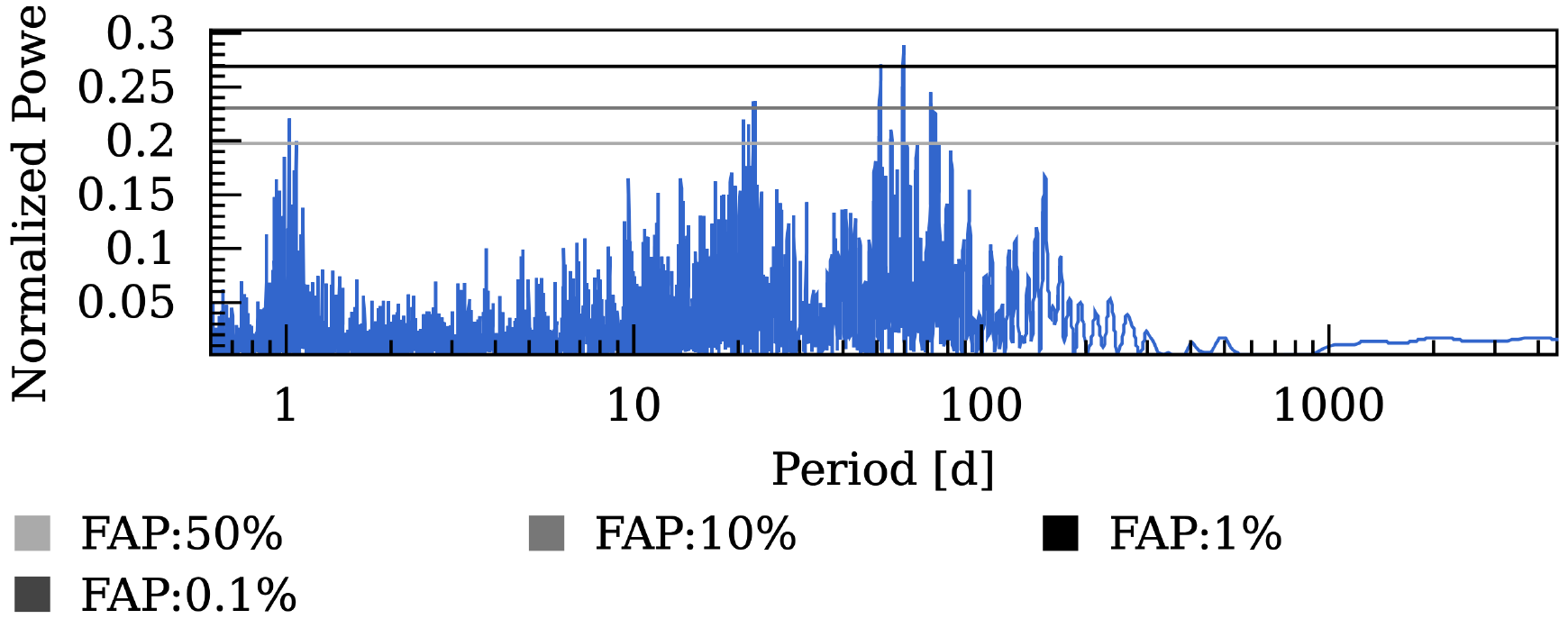}
  \end{subfigure}
  \hfill
  \caption{\label{gj317}RV time series and GLS periodograms of GJ~317 (left: original, right: residuals). The RV obtained before and after the fibre change is respectively plotted in blue and orange.}
\end{figure}

\subsubsection{GJ~361}

\begin{figure}[h!]
  \centering
  \begin{subfigure}{0.49\textwidth}
    \centering
    \includegraphics[width=\textwidth]{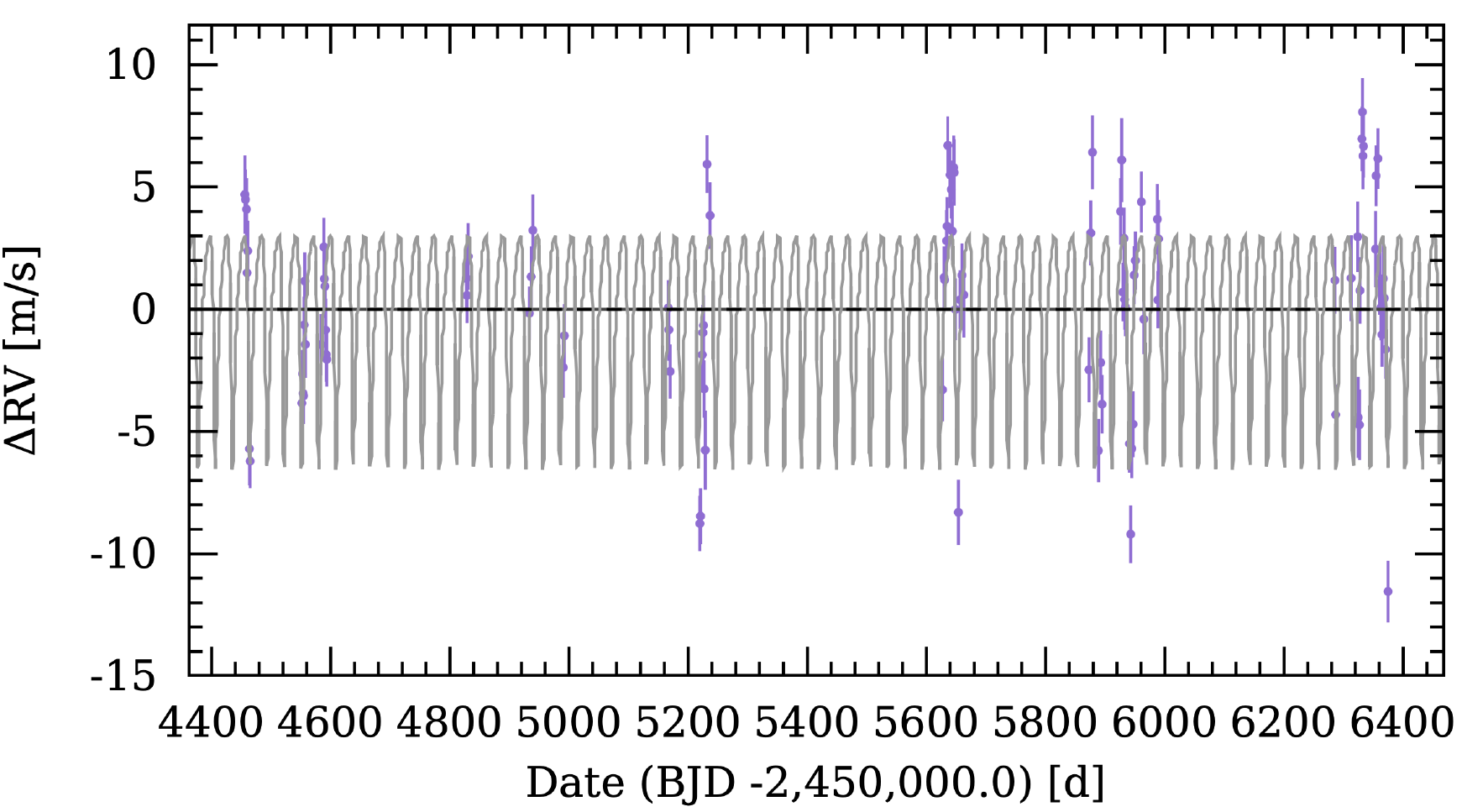}
  \end{subfigure}
  \hfill
  \begin{subfigure}{0.49\textwidth}
    \centering
    \includegraphics[width=\textwidth]{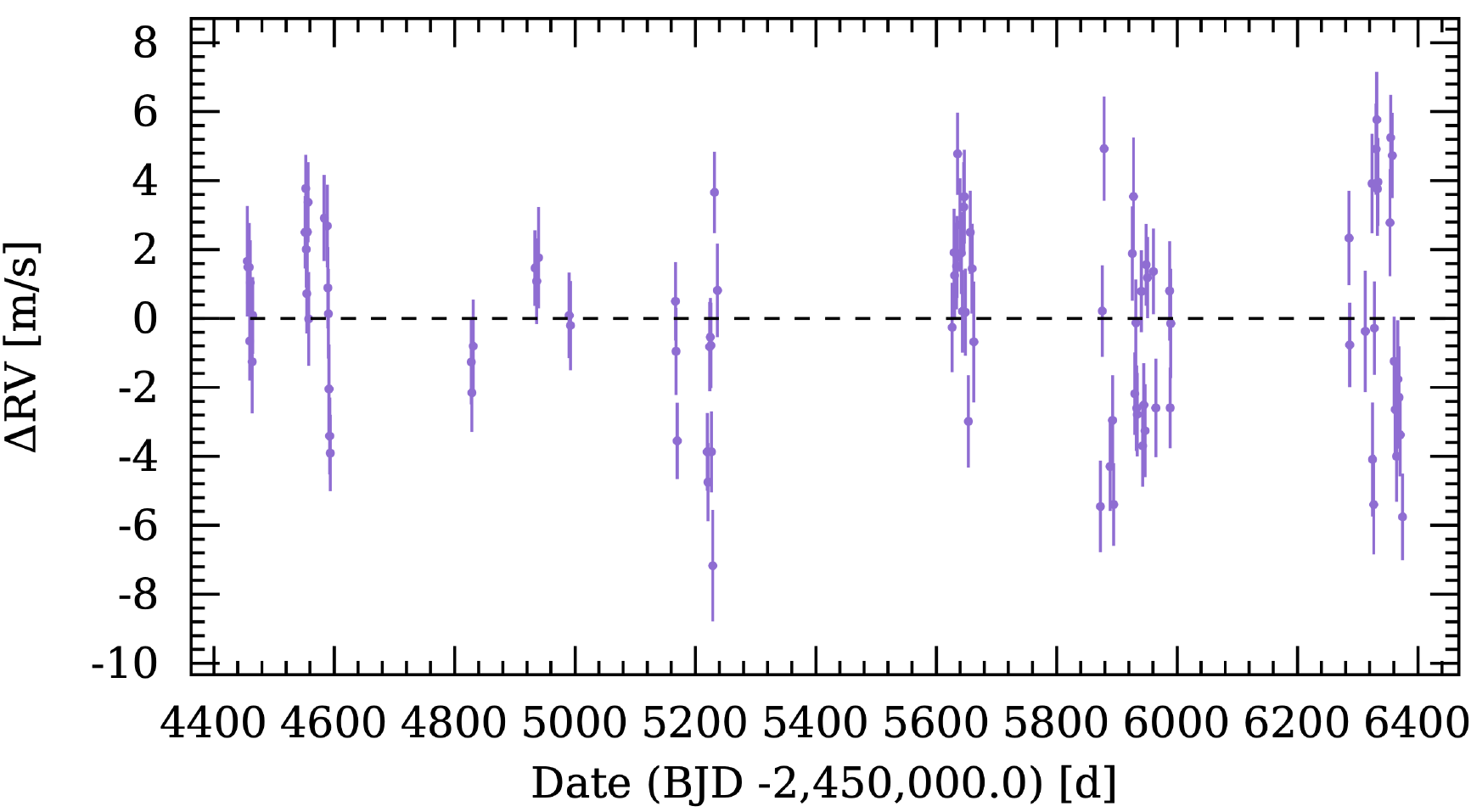}
  \end{subfigure}
  \hfill
  \begin{subfigure}{0.49\textwidth}
    \centering
    \includegraphics[width=\textwidth]{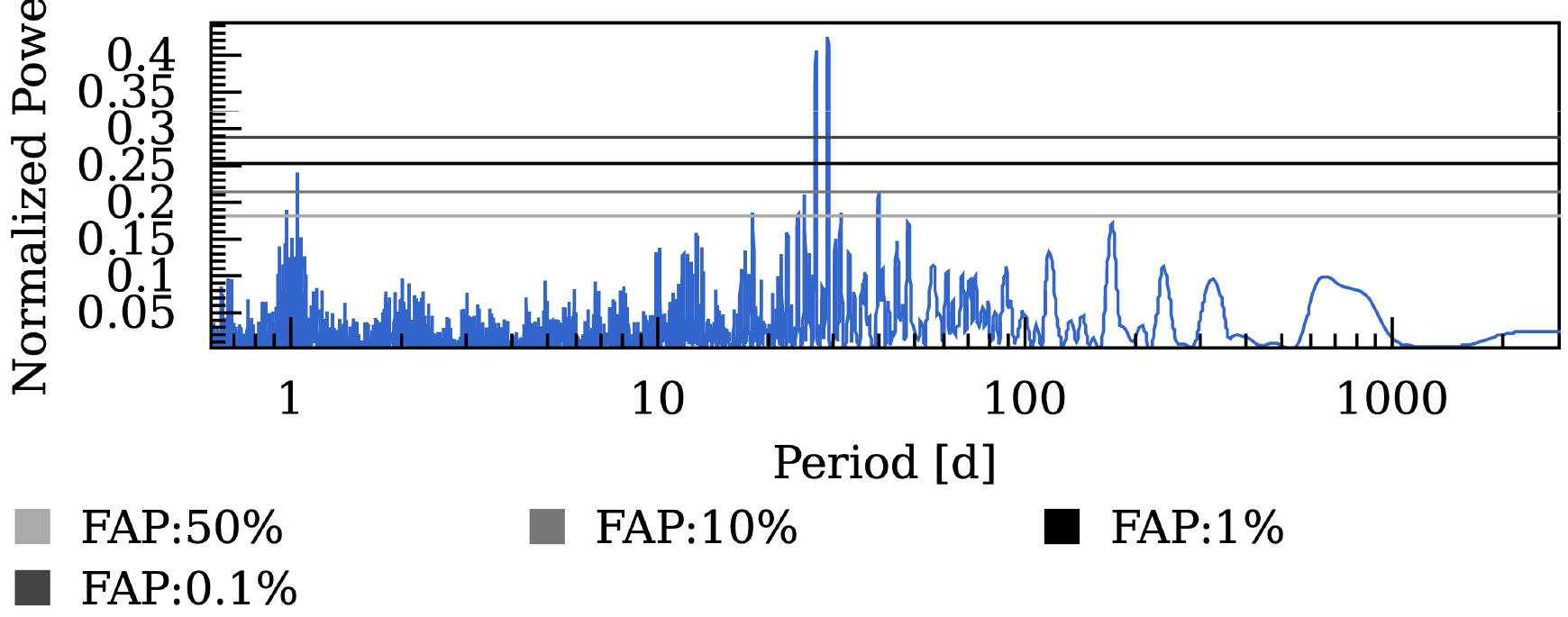}
  \end{subfigure}
  \hfill
  \begin{subfigure}{0.49\textwidth}
    \centering
    \includegraphics[width=\textwidth]{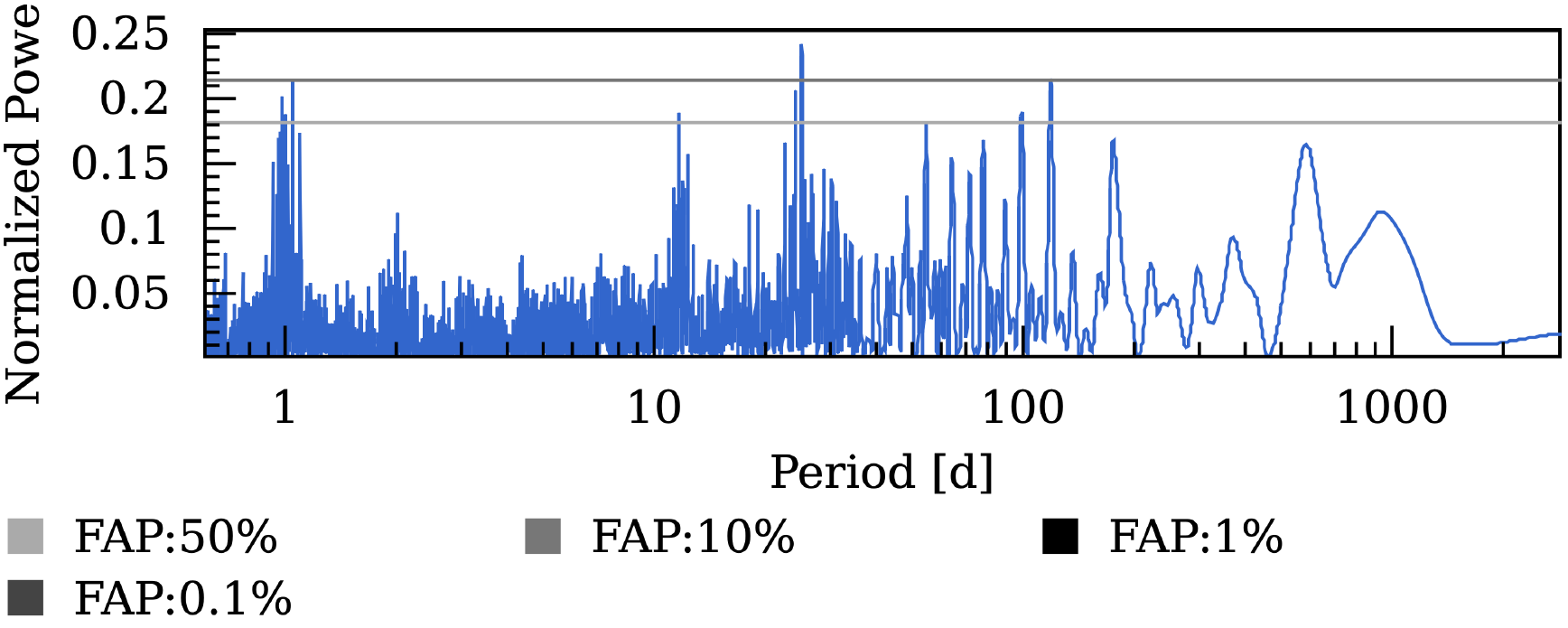}
  \end{subfigure}
  \hfill
  \caption{\label{gj361}RV time series with fitted signal and GLS periodograms of GJ~361 (left: original, right: residuals).}
\end{figure}

\newpage

\subsubsection{GJ~393}

\begin{figure}[h!]
  \centering
  \begin{subfigure}{0.49\textwidth}
    \centering
    \includegraphics[width=\textwidth]{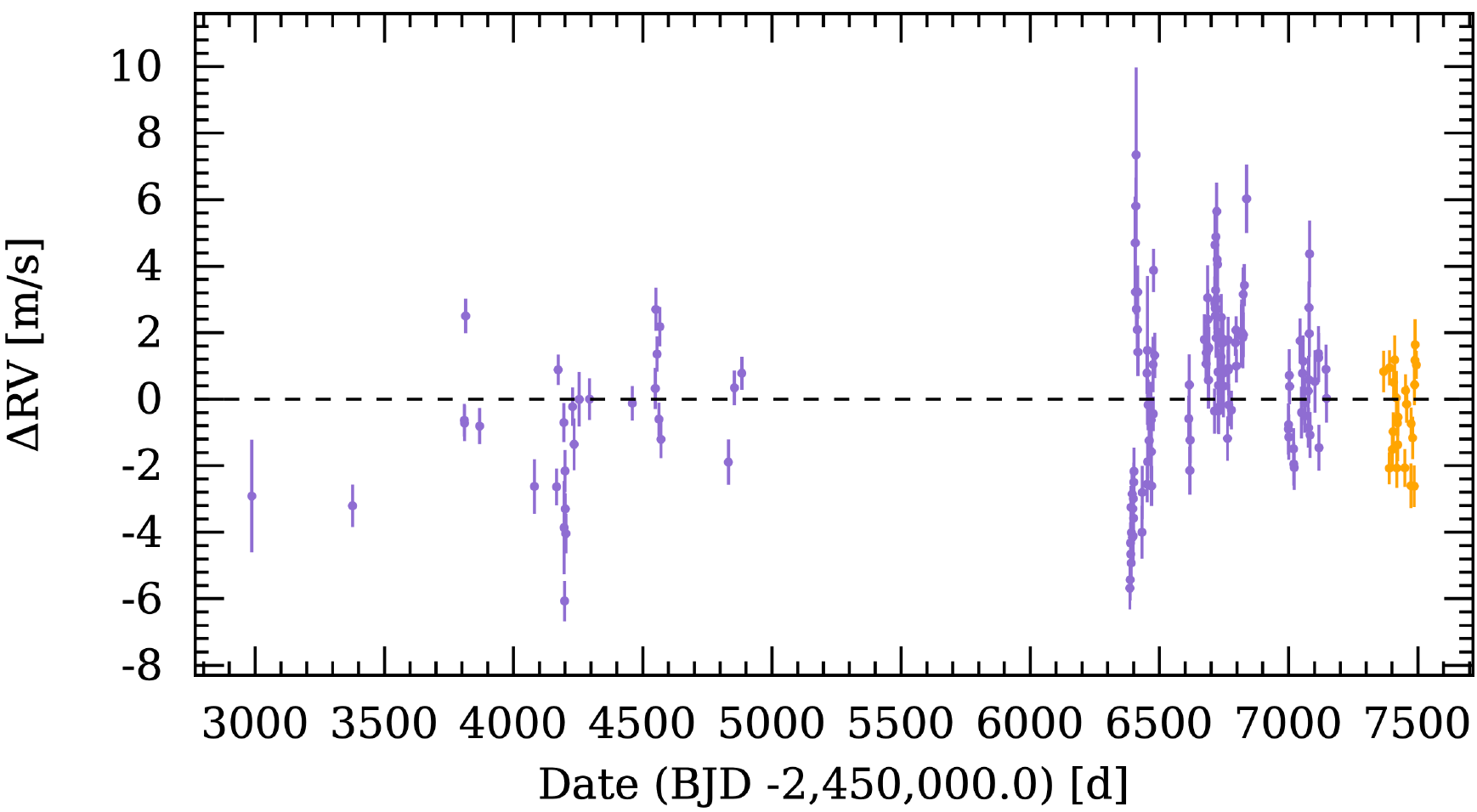}
  \end{subfigure}
  \hfill
  \begin{subfigure}{0.49\textwidth}
    \centering
    \includegraphics[width=\textwidth]{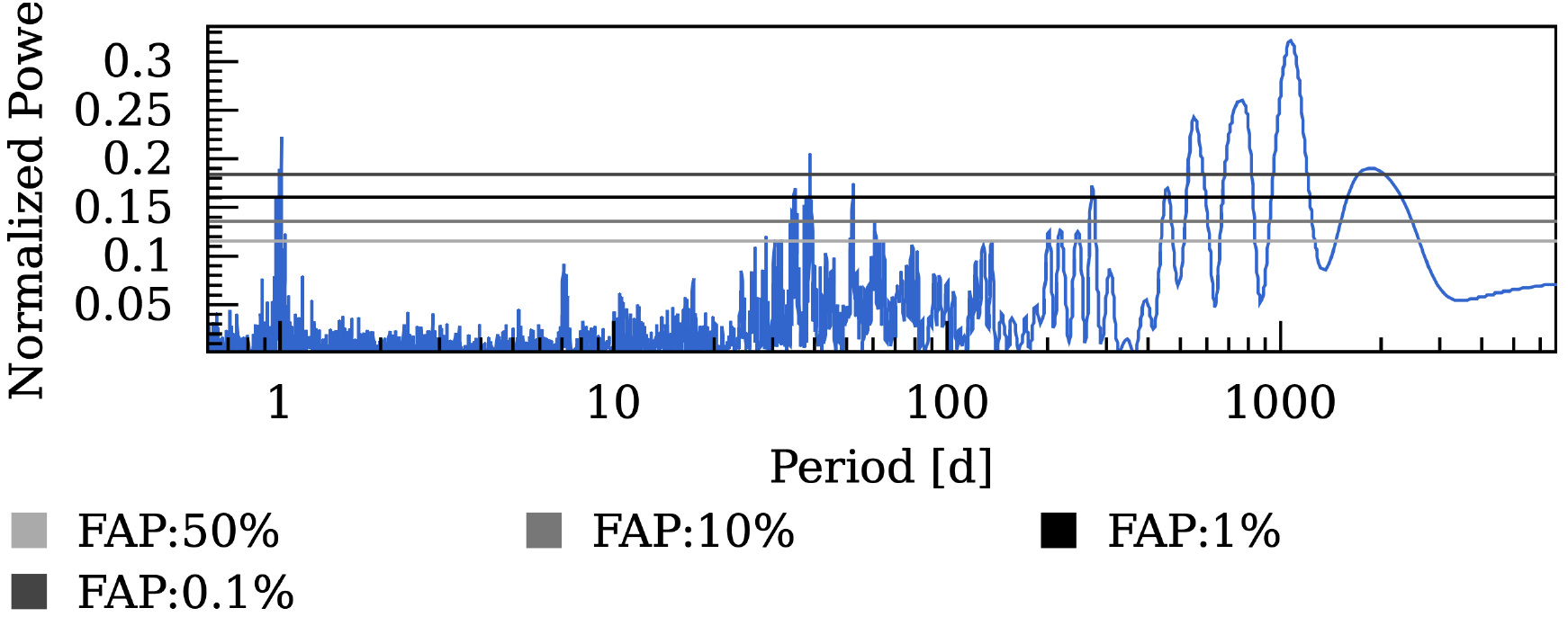}
  \end{subfigure}
  \hfill
  \caption{\label{gj393}RV time series and GLS periodogram of GJ~393 after the subtraction of the planetary signal at 7.02 days.}
\end{figure}

\subsubsection{GJ~569A}

\begin{figure}[h!]
  \centering
  \begin{subfigure}{0.49\textwidth}
    \centering
    \includegraphics[width=\textwidth]{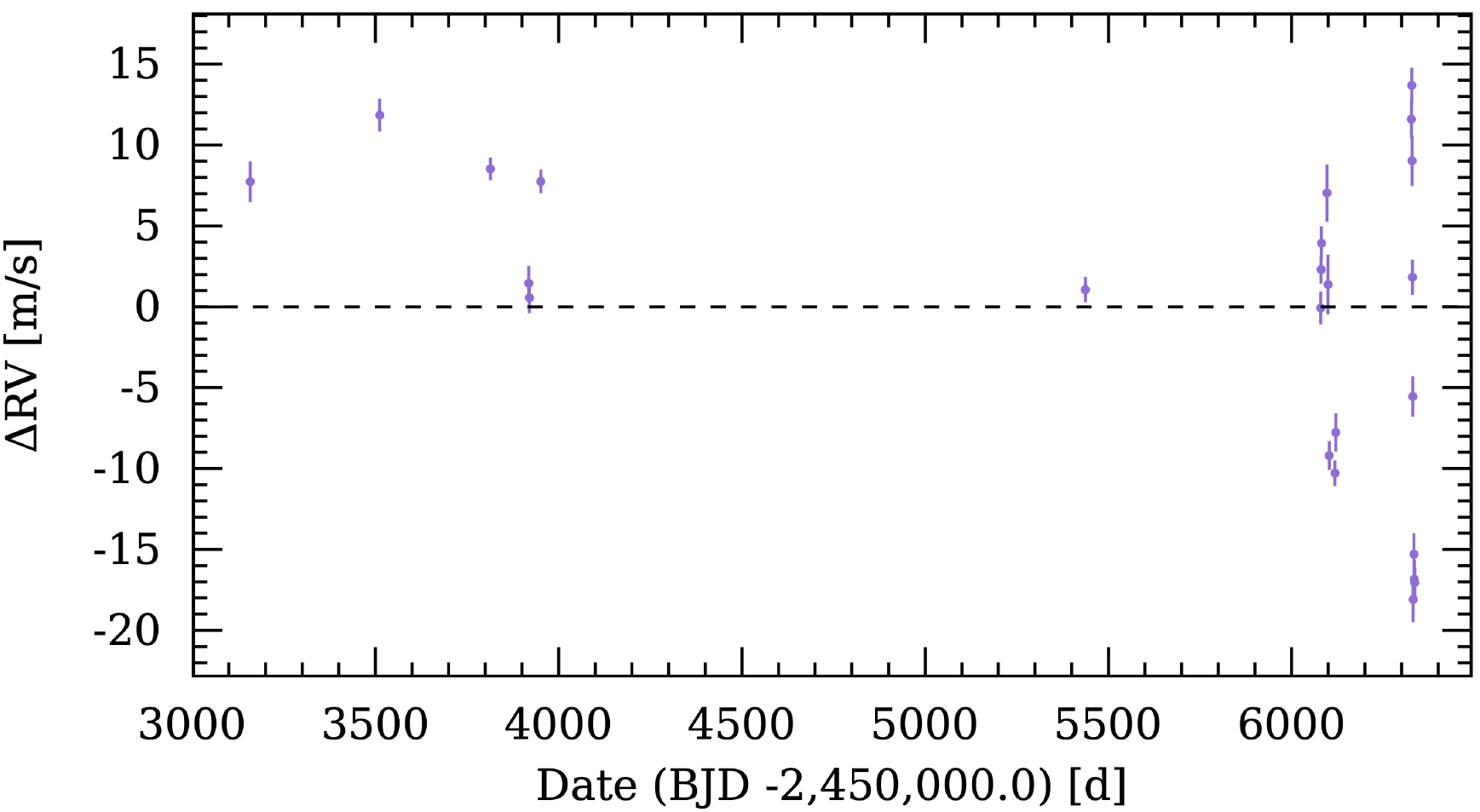}
  \end{subfigure}
  \hfill
  \begin{subfigure}{0.49\textwidth}
    \centering
    \includegraphics[width=\textwidth]{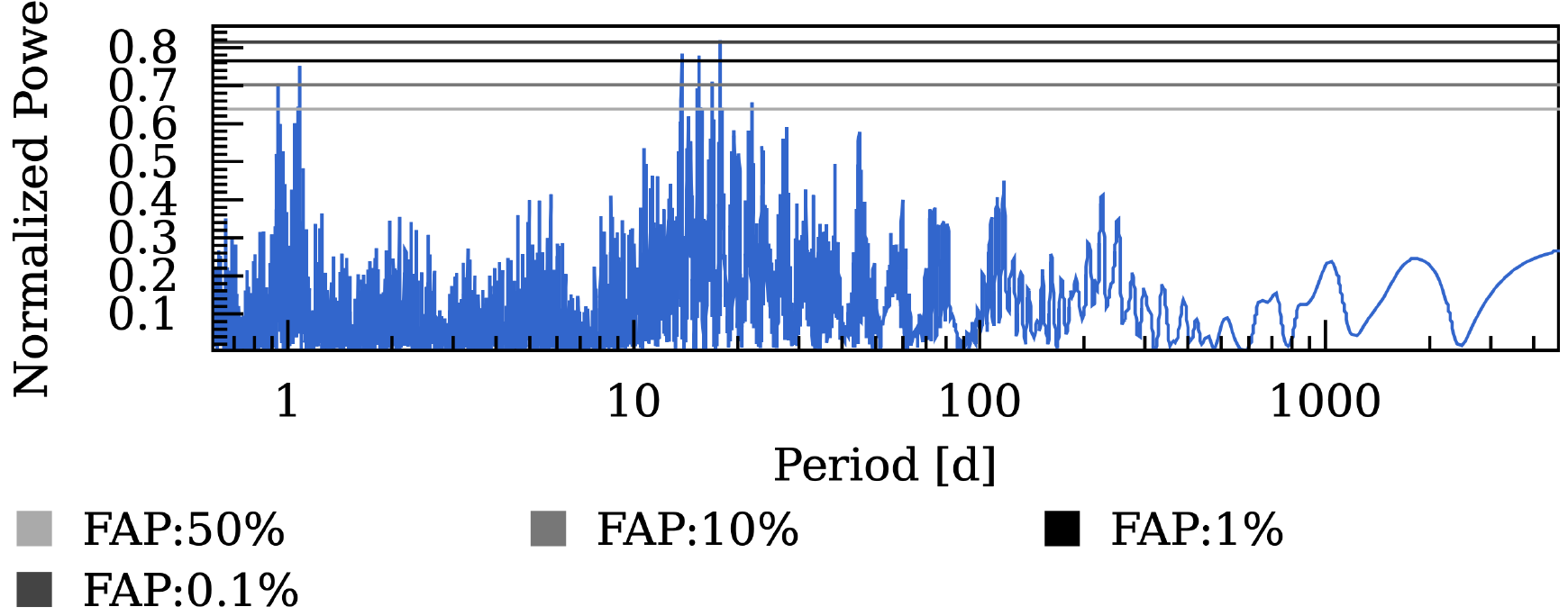}
  \end{subfigure}
  \hfill
  \caption{\label{gj569}RV time series and GLS periodogram of GJ~569A.}
\end{figure}

\newpage

\subsubsection{GJ~654}

\begin{figure}[h!]
  \centering
  \begin{subfigure}{0.49\textwidth}
    \centering
    \includegraphics[width=\textwidth]{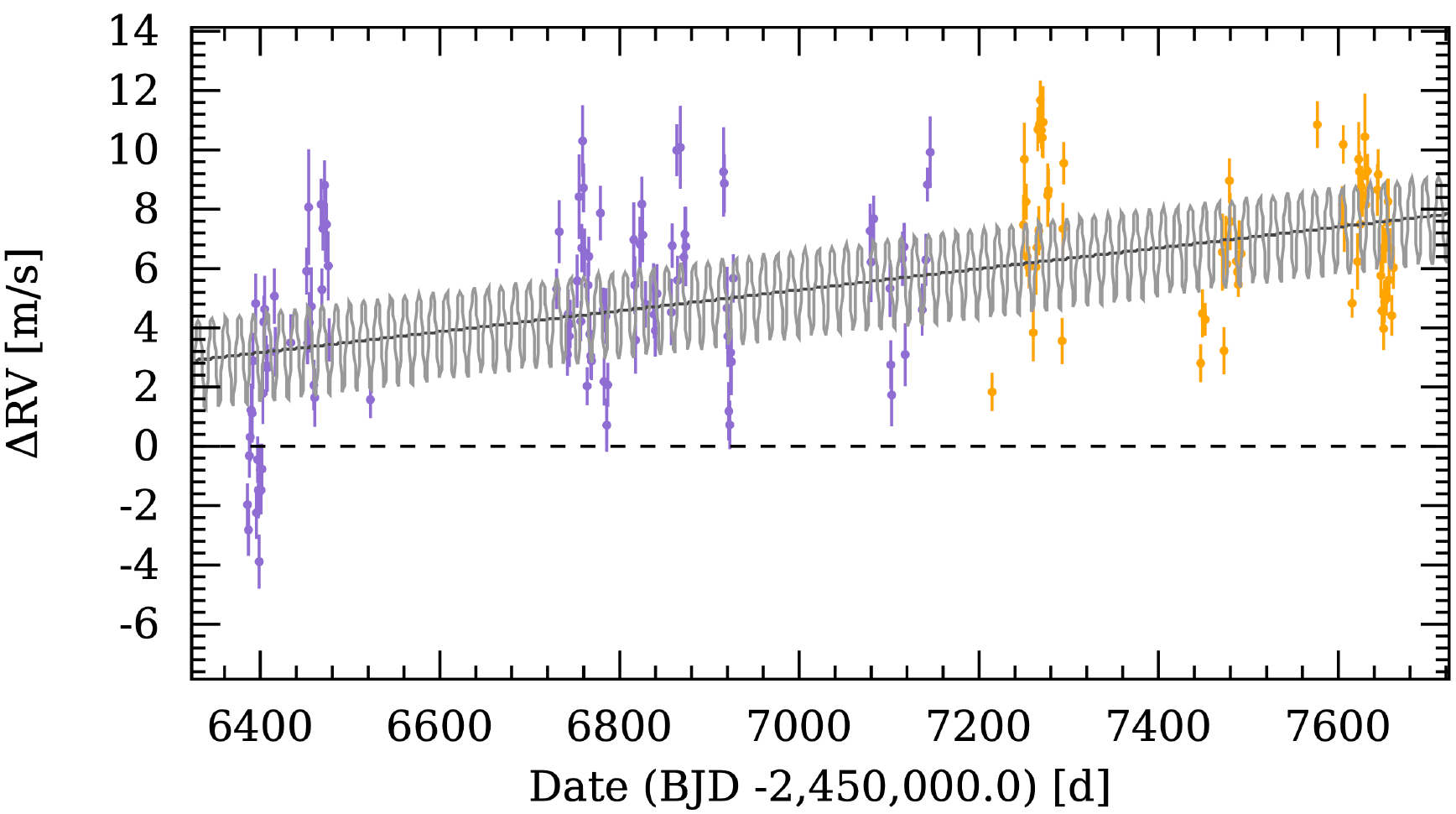}
  \end{subfigure}
  \hfill
  \begin{subfigure}{0.49\textwidth}
    \centering
    \includegraphics[width=\textwidth]{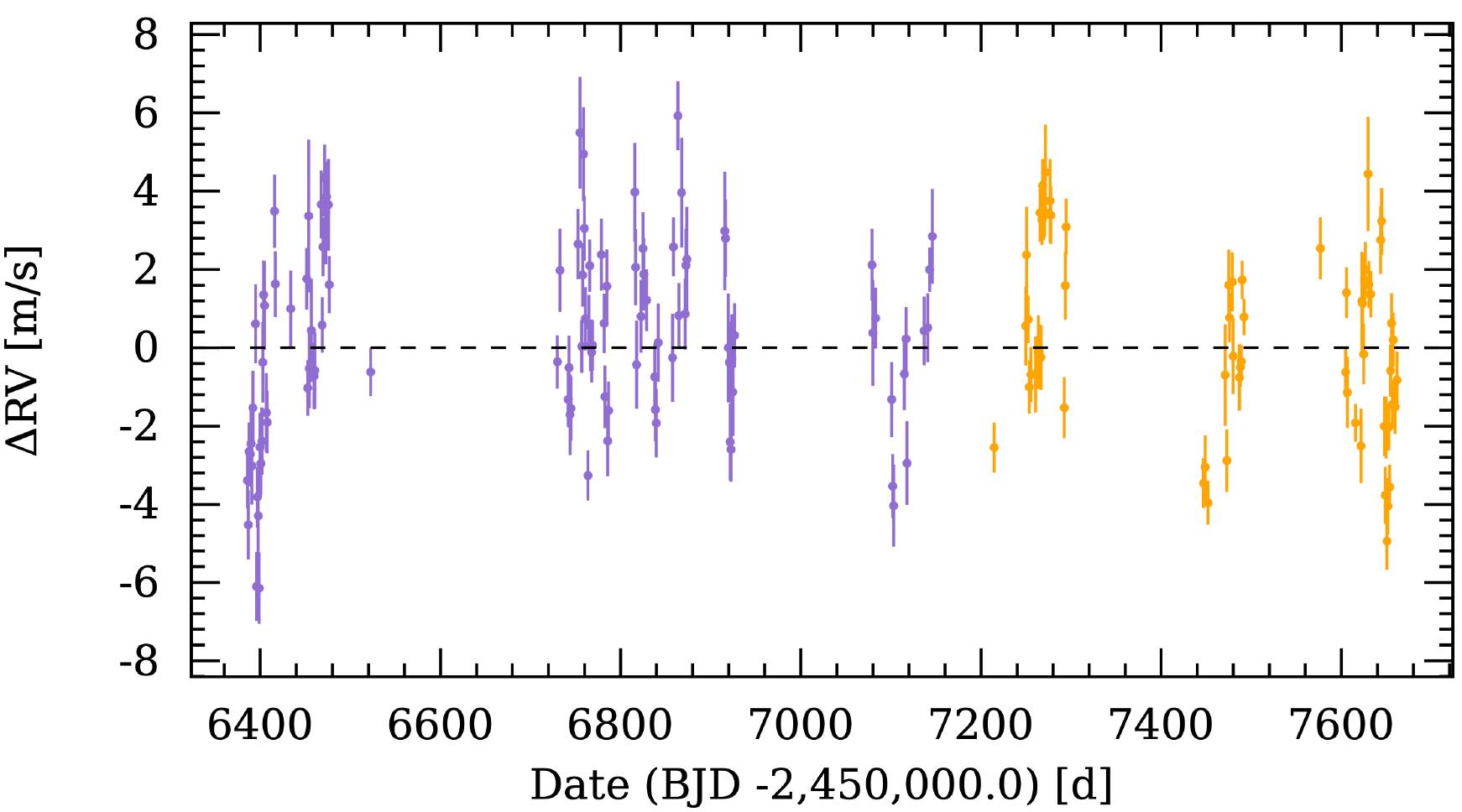}
  \end{subfigure}
  \hfill
  \begin{subfigure}{0.49\textwidth}
    \centering
    \includegraphics[width=\textwidth]{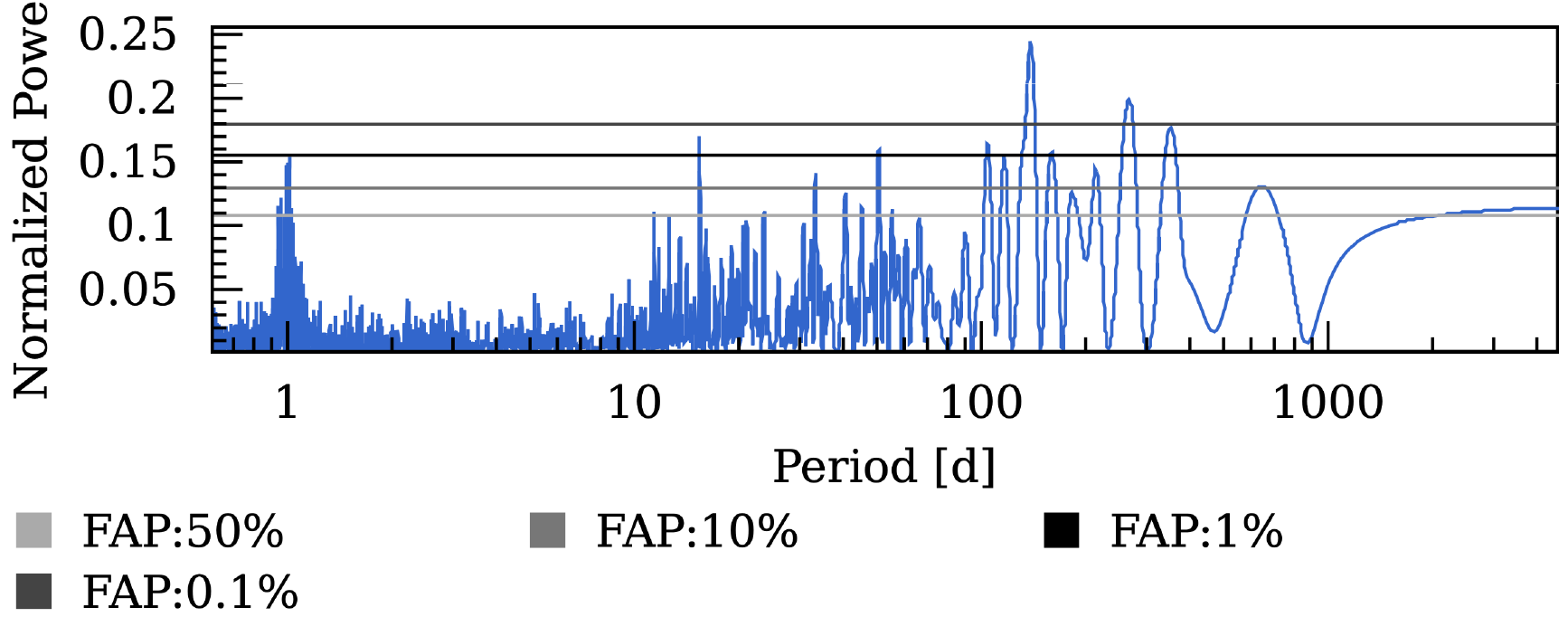}
  \end{subfigure}
  \hfill
  \begin{subfigure}{0.49\textwidth}
    \centering
    \includegraphics[width=\textwidth]{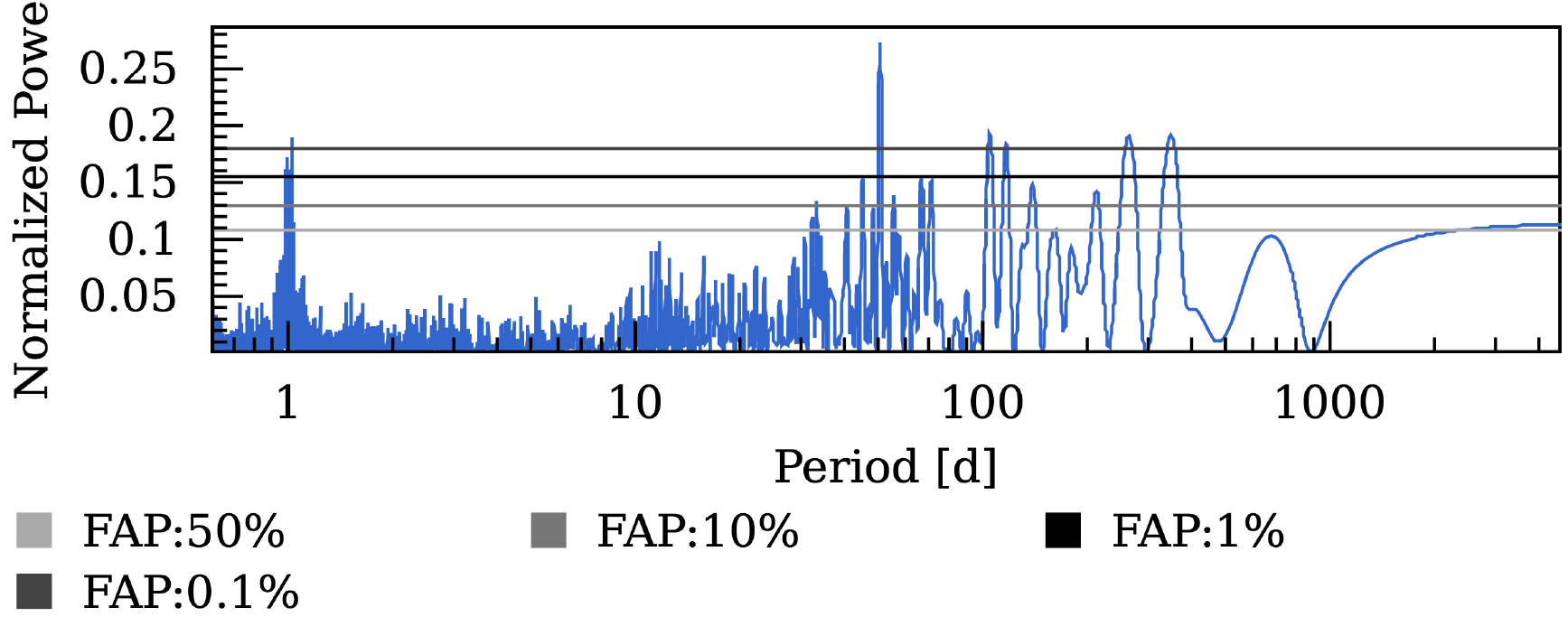}
  \end{subfigure}
  \hfill
    \begin{subfigure}{0.49\textwidth}
    \centering
    \includegraphics[width=\textwidth]{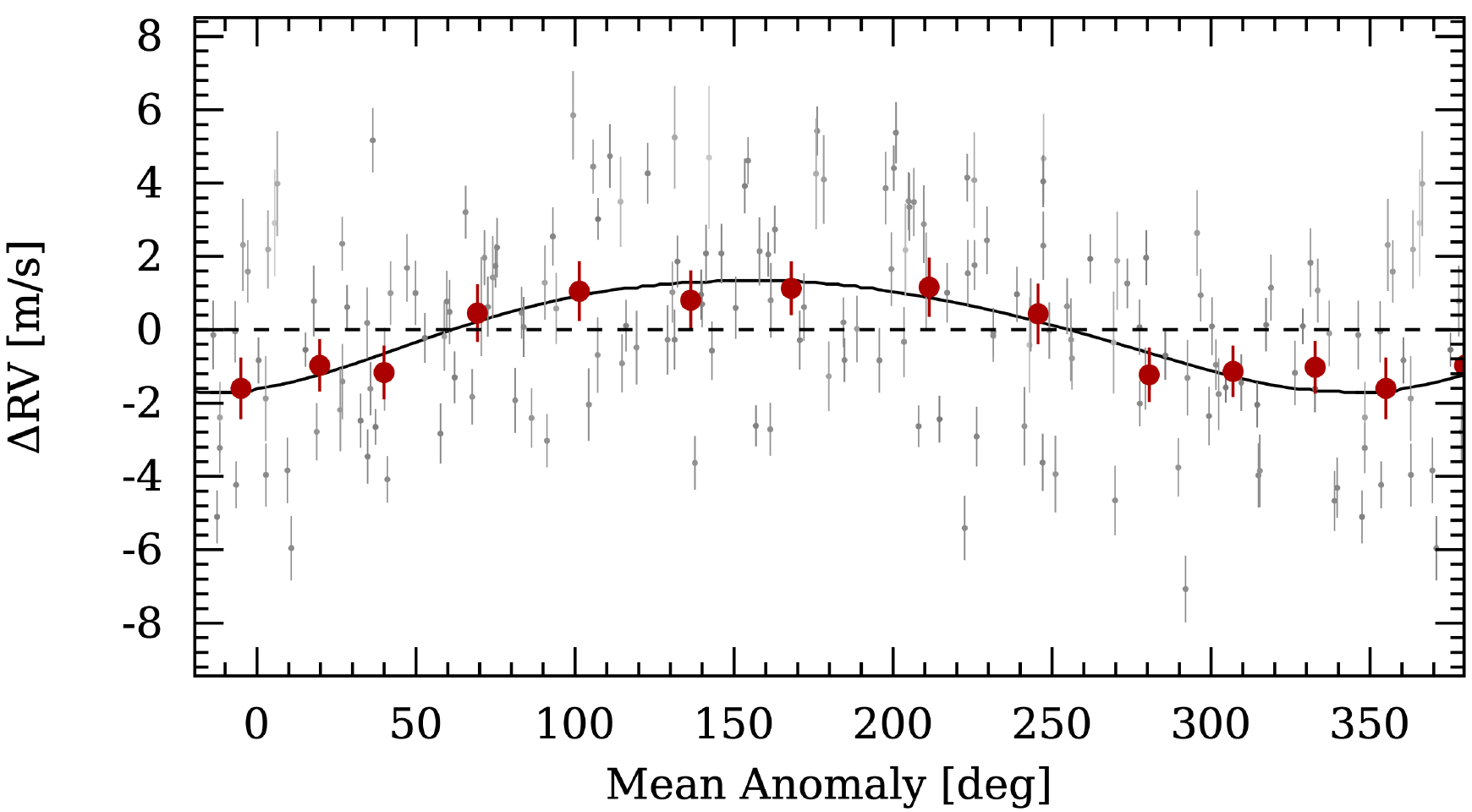}
  \end{subfigure}
  \hfill
  \caption{\label{gj654}RV time series and GLS of GJ~654. Upper left: RV time series zoomed in on the four well-monitored seasons with the trend and Keplerian signals fitted at 15.35 days. Upper right: RV time series of the residuals of the trend and the Keplerian signal, zoomed in on the four well-monitored seasons. Middle left: periodogram. Middle right: periodogram of the residuals of the trend and the Keplerian signal, zoomed in on the four well-monitored seasons. Bottom: signal fitted at 15.35 days.}
\end{figure}

\newpage

\subsubsection{GJ~739}

\begin{figure}[h!]
  \centering
  \begin{subfigure}{0.49\textwidth}
    \centering
    \includegraphics[width=\textwidth]{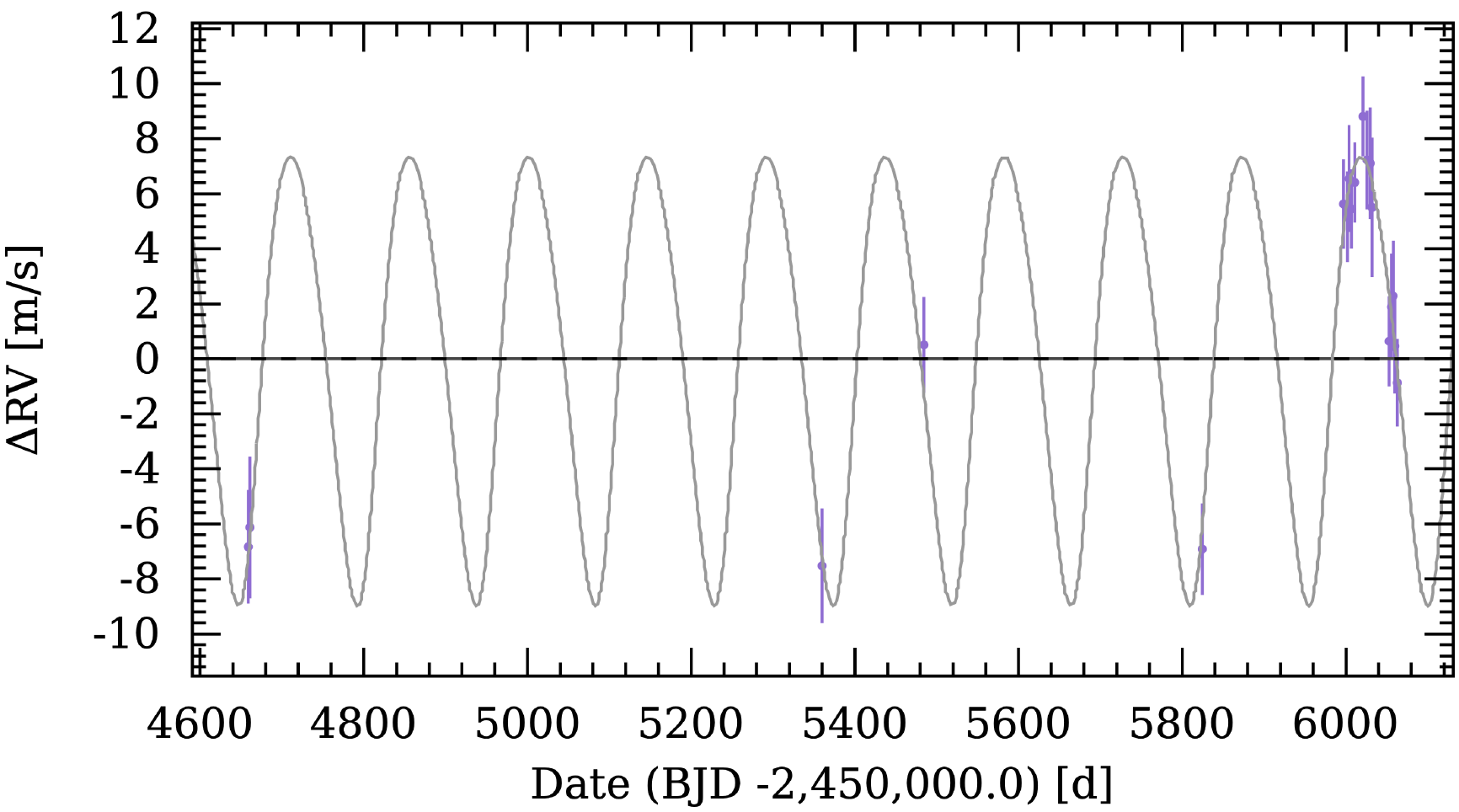}
  \end{subfigure}
  \hfill
  \begin{subfigure}{0.49\textwidth}
    \centering
    \includegraphics[width=\textwidth]{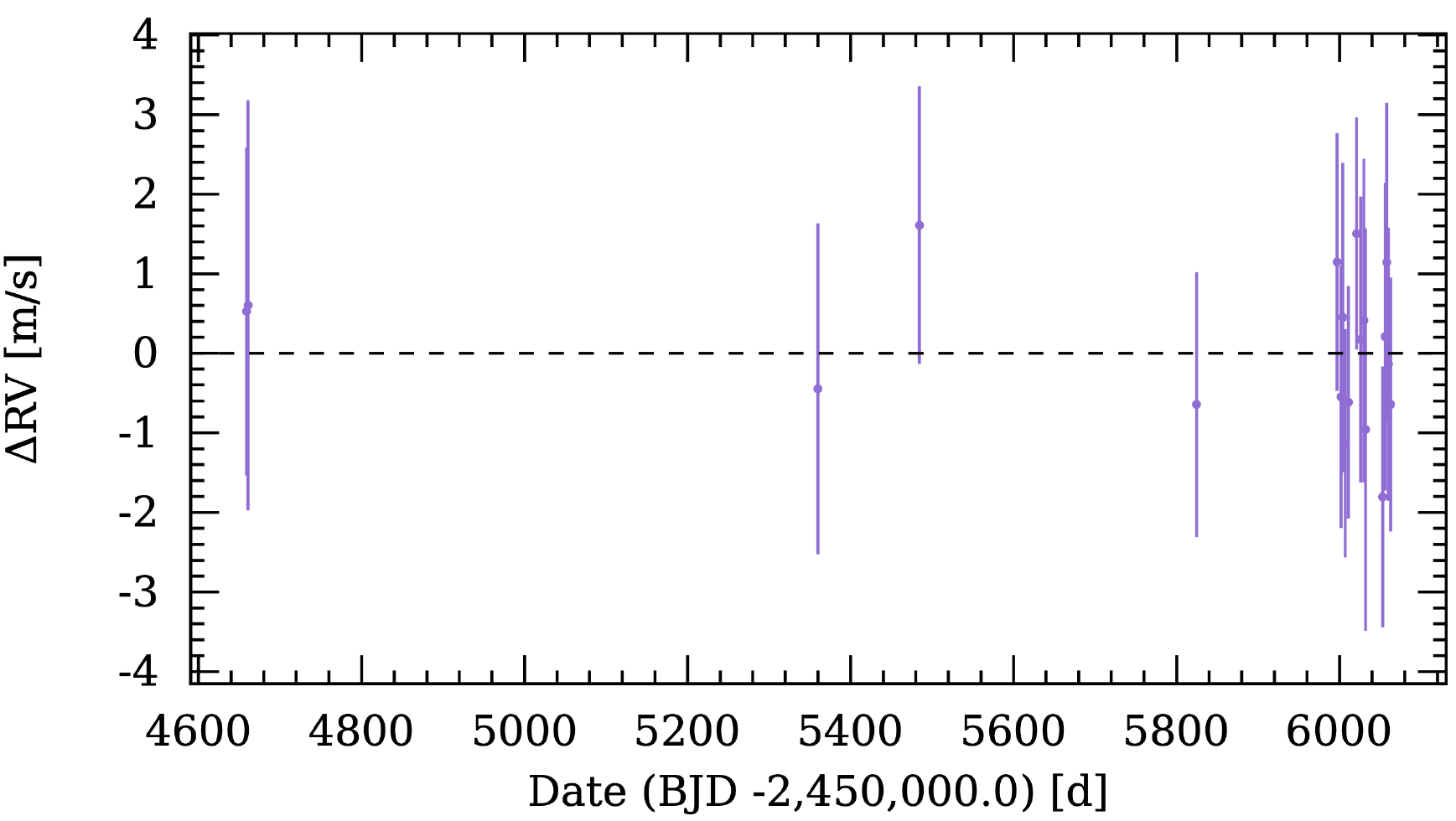}
  \end{subfigure}
  \hfill
  \begin{subfigure}{0.49\textwidth}
    \centering
    \includegraphics[width=\textwidth]{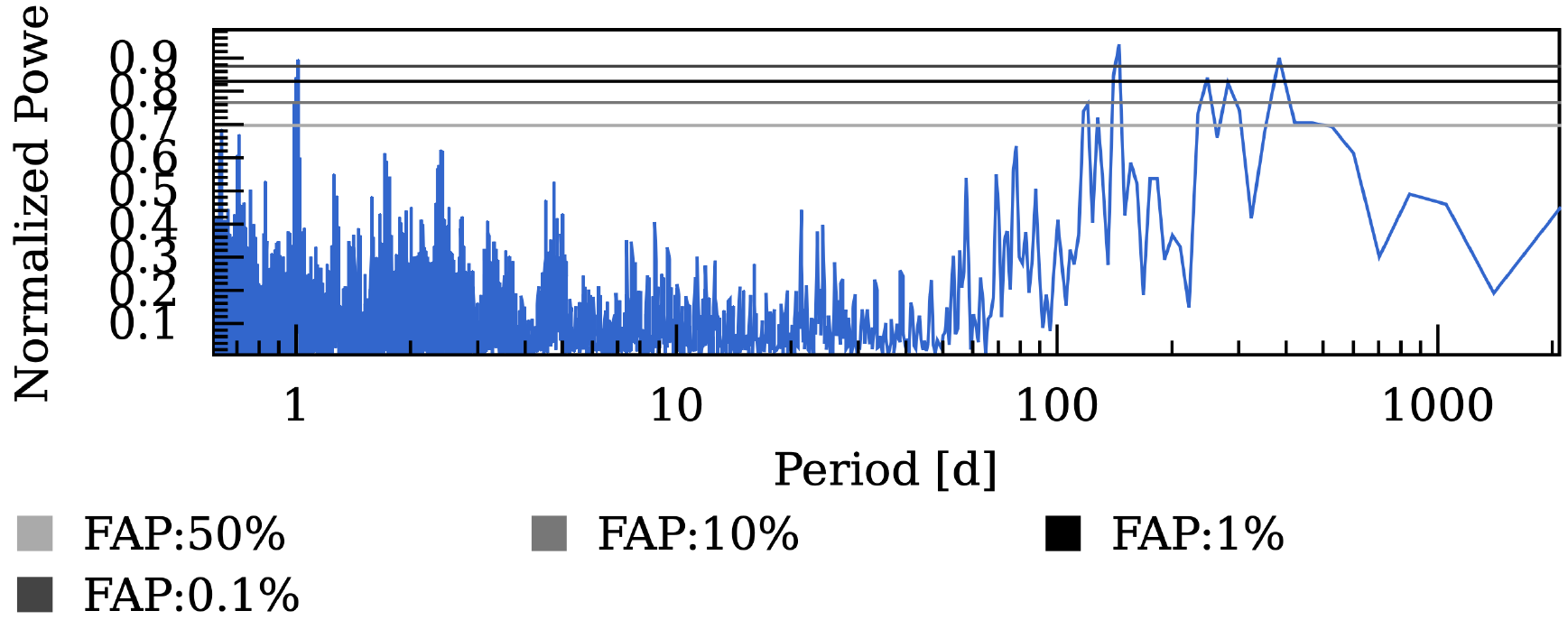}
  \end{subfigure}
  \hfill
  \begin{subfigure}{0.49\textwidth}
    \centering
    \includegraphics[width=\textwidth]{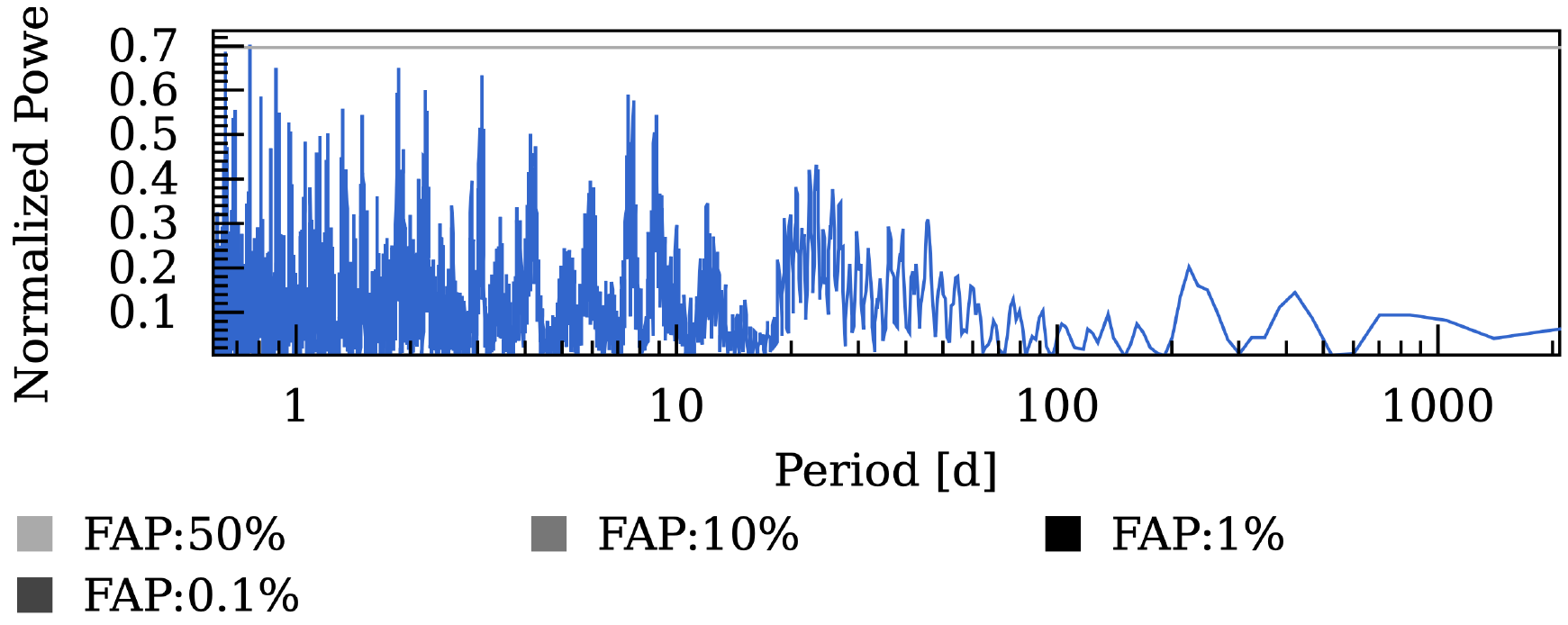}
  \end{subfigure}
  \hfill
    \begin{subfigure}{0.48\textwidth}
    \centering
    \includegraphics[width=\textwidth]{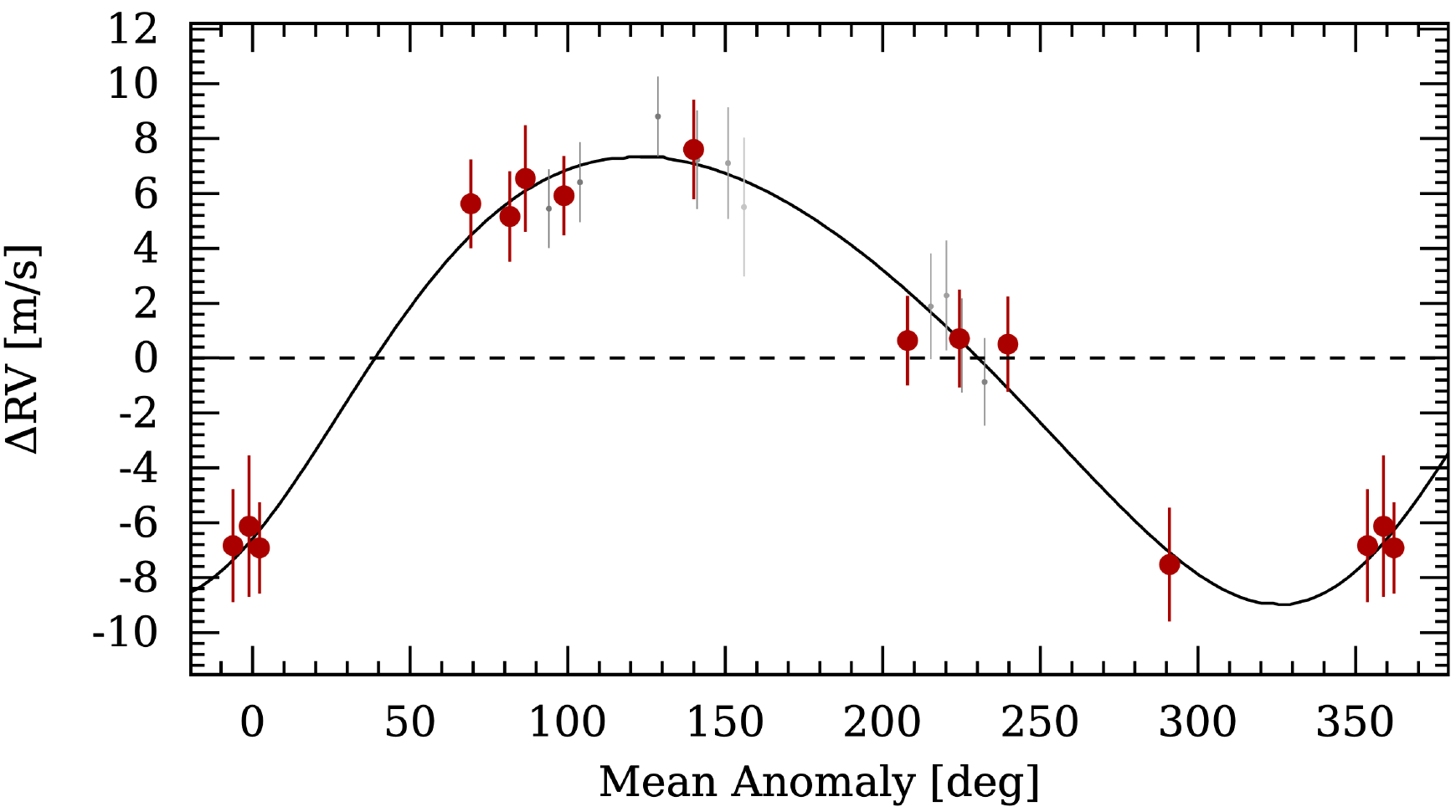}
  \end{subfigure}
  \hfill
  \caption{\label{gj739}RV time series and GLS periodogram of GJ~654. Upper left: RV time series with the Keplerian signal fitted. Upper right: residuals. Middle left: periodogram. Middle right: periodogram of residuals. Bottom: Signal fitted at 145.3 days.}
\end{figure}

\newpage

\subsubsection{GJ~754}

\begin{figure}[h!]
  \centering
  \begin{subfigure}{0.49\textwidth}
    \centering
    \includegraphics[width=\textwidth]{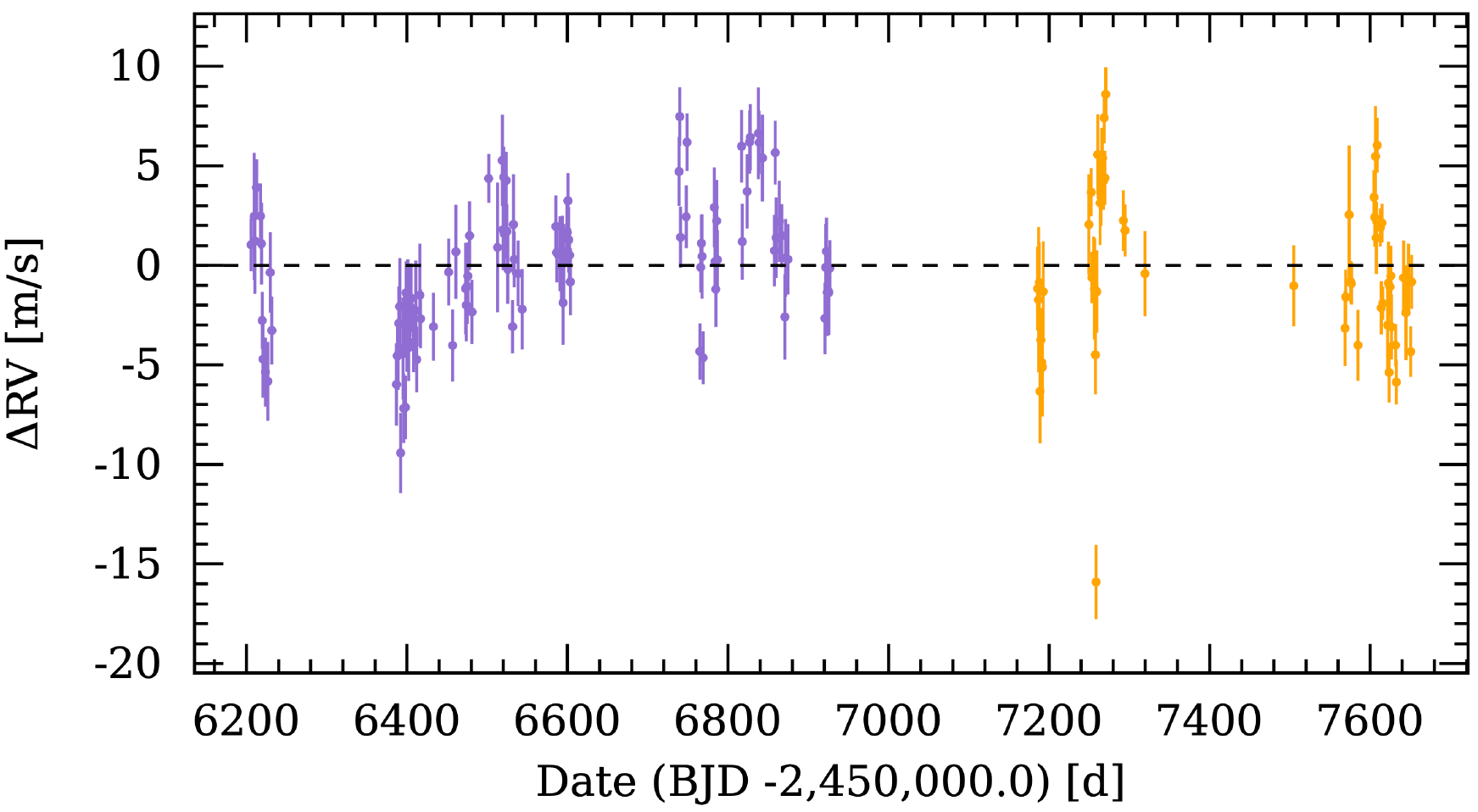}
  \end{subfigure}
  \hfill
  \begin{subfigure}{0.49\textwidth}
    \centering
    \includegraphics[width=\textwidth]{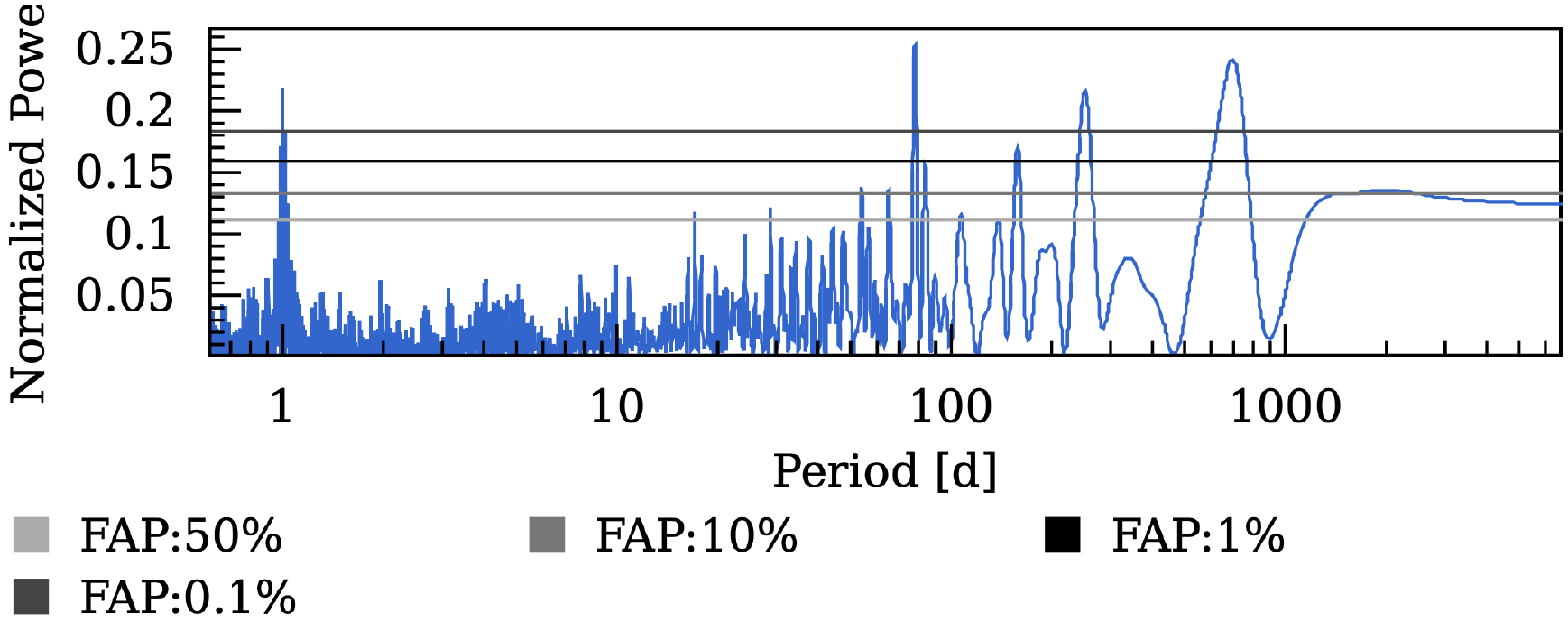}
  \end{subfigure}
  \hfill
  \caption{\label{gj754}RV time series and GLS periodogram of GJ~754 for the four well-monitored seasons (2012--2016).}
\end{figure}

\subsubsection{GJ~846}

\begin{figure}[h!]
  \centering
  \begin{subfigure}{0.49\textwidth}
    \centering
    \includegraphics[width=\textwidth]{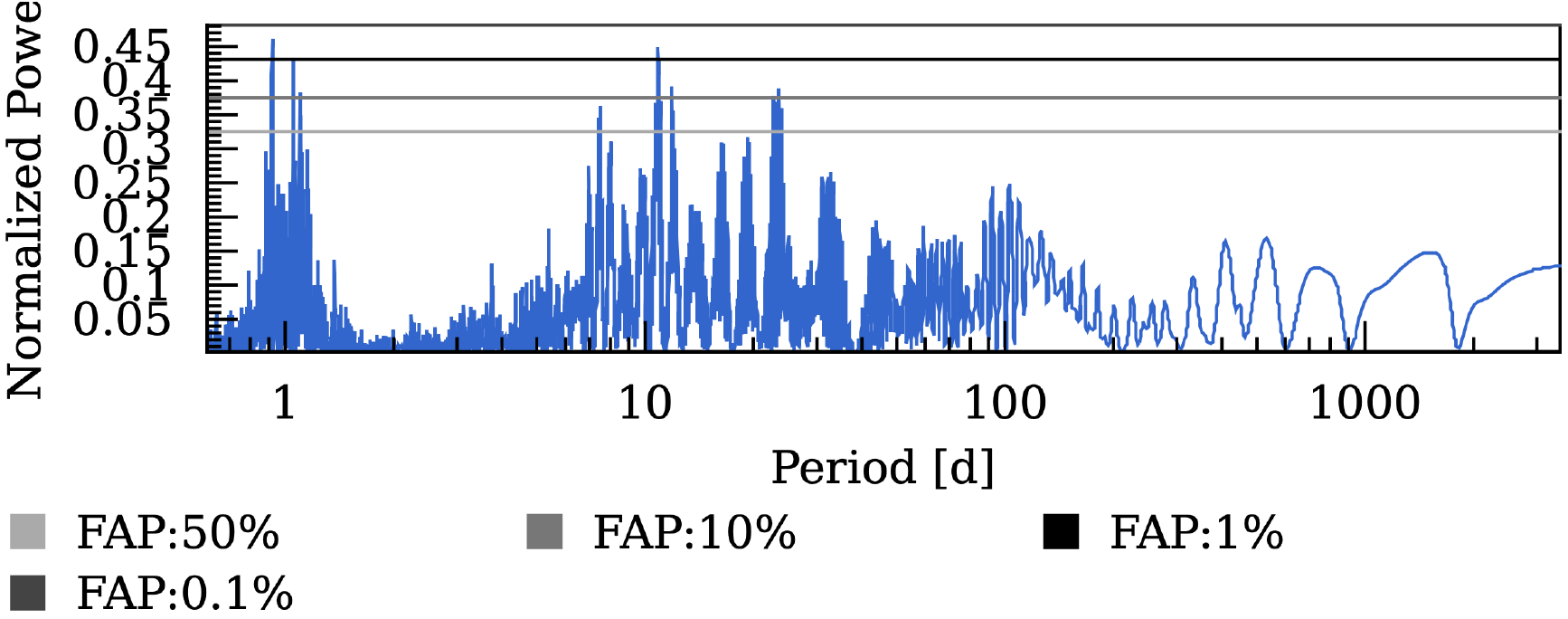}
  \end{subfigure}
  \hfill
    \begin{subfigure}{0.48\textwidth}
    \centering
    \includegraphics[width=\textwidth]{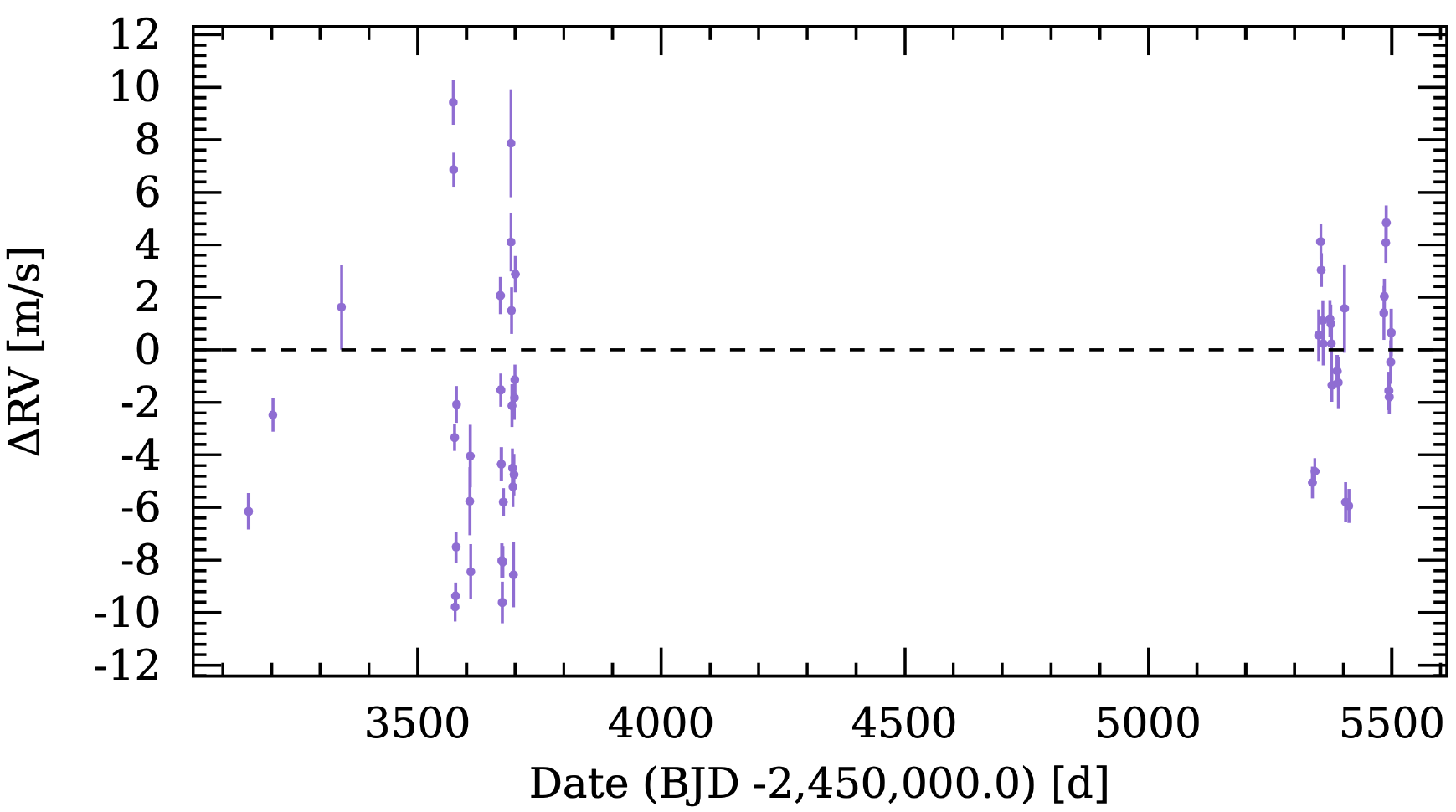}
  \end{subfigure}
  \hfill
  \caption{\label{gj846}RV time series and GLS periodogram of GJ~846 without any correction of long-term trend.}
\end{figure}

\newpage

\subsubsection{GJ~880}

\begin{figure}[h!]
  \centering
  \begin{subfigure}{0.49\textwidth}
    \centering
    \includegraphics[width=\textwidth]{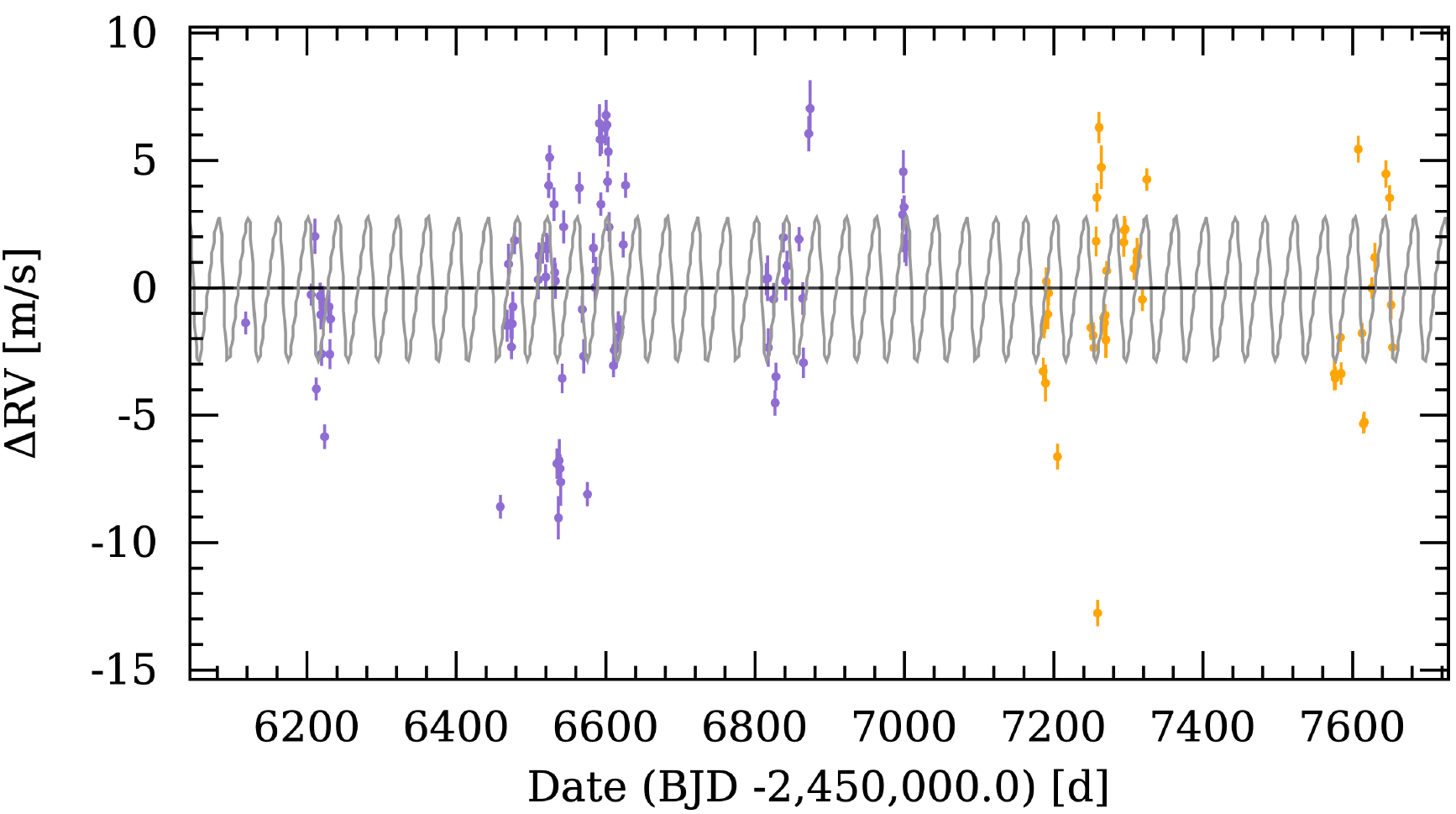}
  \end{subfigure}
  \hfill
  \begin{subfigure}{0.49\textwidth}
    \centering
    \includegraphics[width=\textwidth]{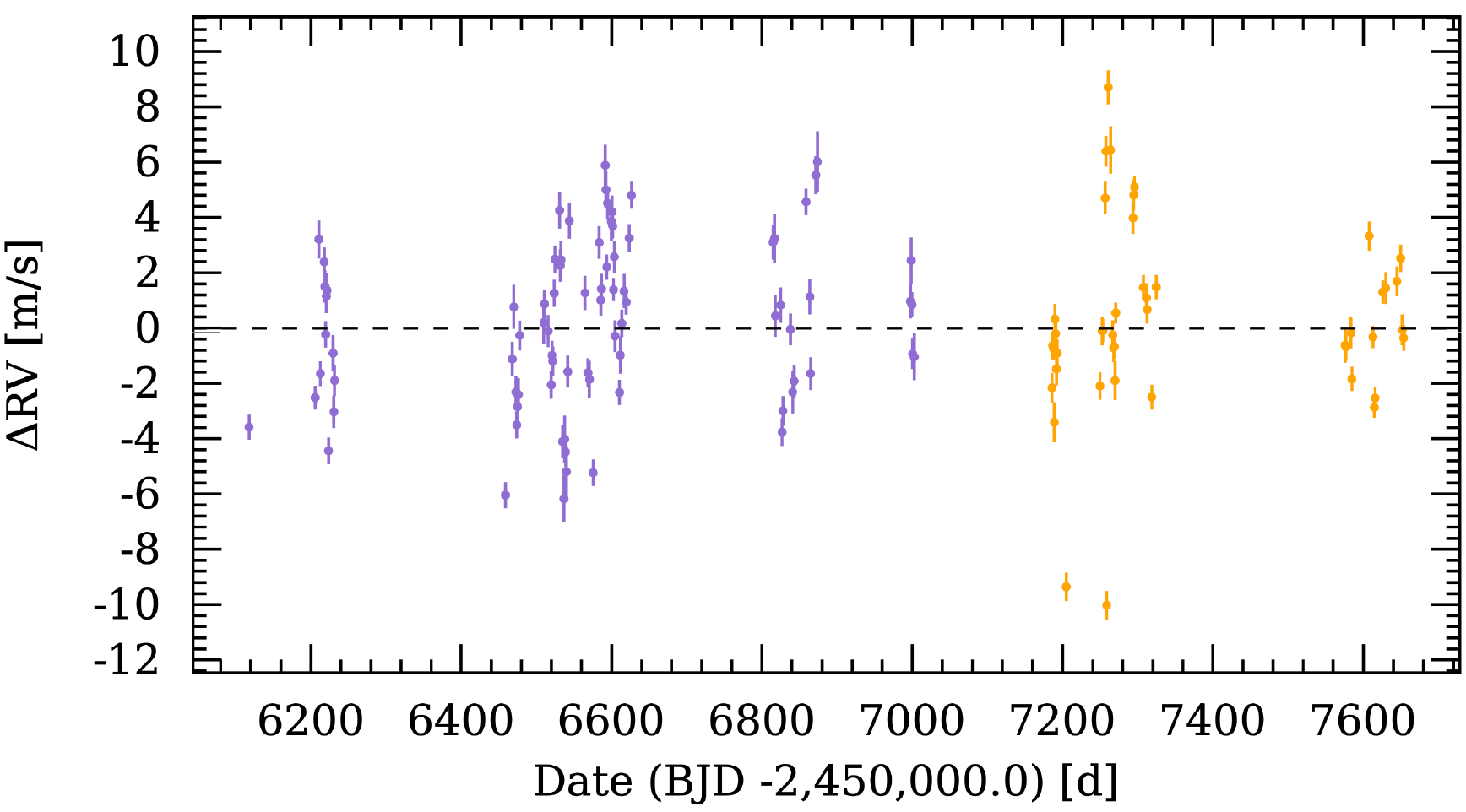}
  \end{subfigure}
  \hfill
  \begin{subfigure}{0.49\textwidth}
    \centering
    \includegraphics[width=\textwidth]{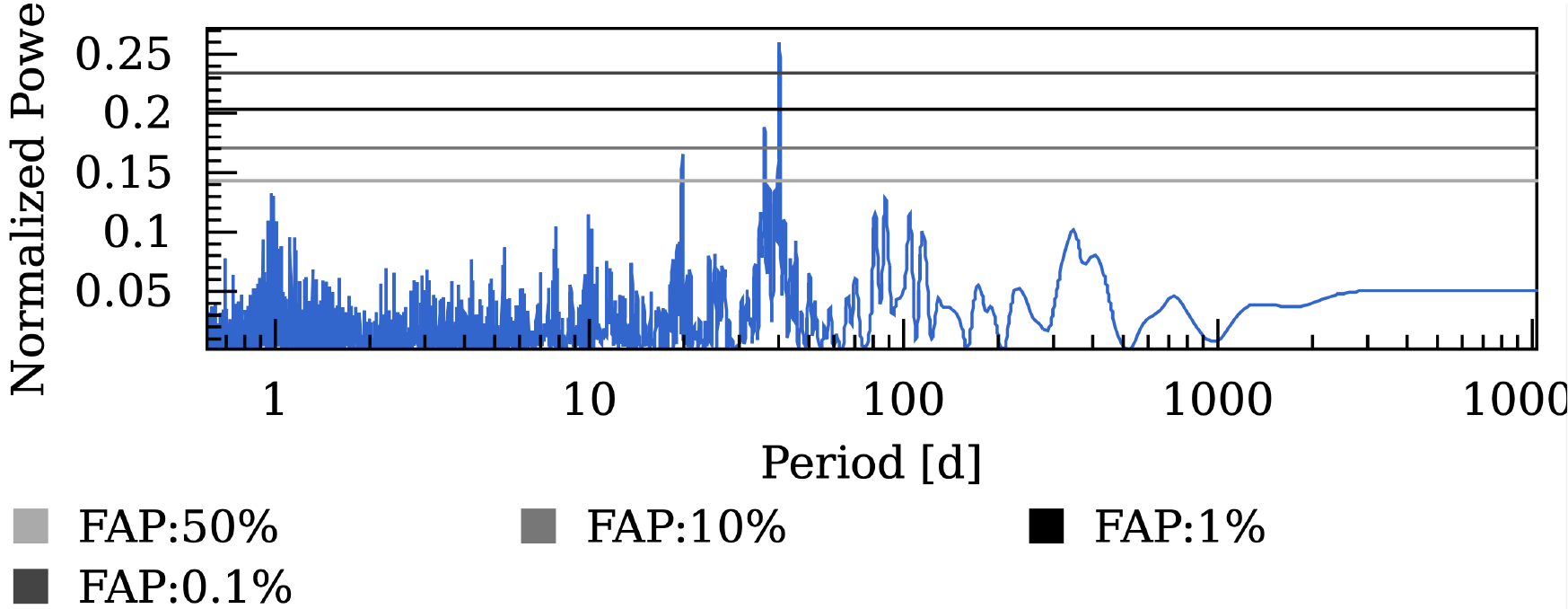}
  \end{subfigure}
  \hfill
  \begin{subfigure}{0.49\textwidth}
    \centering
    \includegraphics[width=\textwidth]{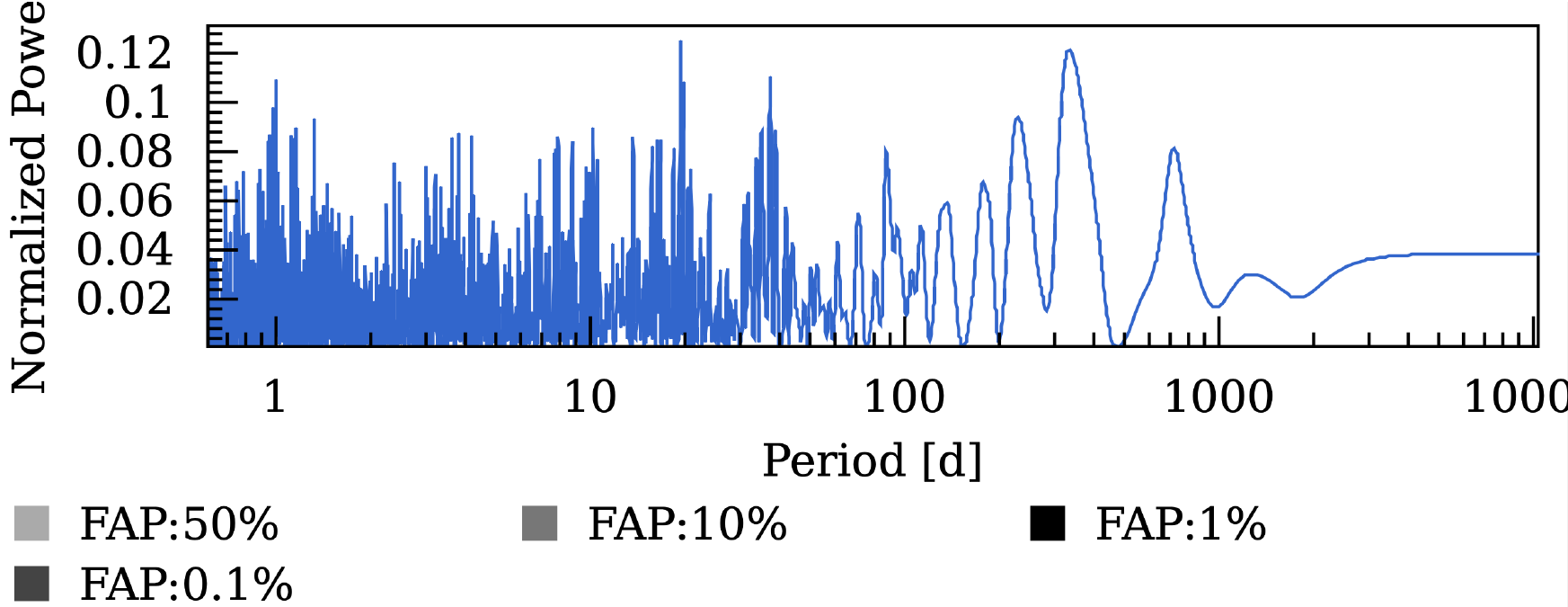}
  \end{subfigure}
  \hfill
  \caption{\label{gj880}RV time series and GLS periodogram of GJ~880. Left: original series, right: residuals.}
\end{figure}

\newpage

\section{Tables} \label{append1}

\begin{longtable}{r|c|r|c|r|c}
\caption{\label{progidtable}Programme ID and PIs} \\
PI & Prog. ID &PI & Prog. ID &PI & Prog. ID \\ \hline 
\hline
\endhead

\multirow{6}{*}{Bonfils}&198.C-0873 &\multirow{14}{*}{Lo Curto}&196.C-1006&\multirow{2}{*}{Udry} & 192.C-0852\\
& 191.C-0873 & & 099.C-0458 & & 183.C-0972 \\ \cline{5-6}
 & 183.C-0437 & & 098.C-0366 & \multirow{5}{*}{Lagrange} & 192.C-0224 \\
 & 180.C-0886 & & 096.C-0460 & & 099.C-0205 \\
 & 082.C-0718 & & 095.C-0551 & & 098.C-0739 \\
 & 1102.C-0339 & & 093.C-0409 & & 089.C-0739 \\ \cline{1-2}
 \multirow{2}{*}{ Anglada-Escude} & 191.C-0505 & & 091.C-0034 & &0104.C-0418 \\ \cline{5-6}
 & 096.C-0082 & & 090.C-0421& Lannier & 097.C-0864 \\ \cline{5-6} \cline{1-2}
 Astudillo Defru & 100.C-0884 & & 089.C-0732 & \multirow{3}{*}{Diaz} & 098.C-0518 \\ \cline{1-2}
 Poretti &185.D-0056 & & 087.C-0831& & 096.C-0499 \\ \cline{1-2}
 Lachaume & 089.C-0006 & & 085.C-0019 & & 0100.C-0487 \\ \cline{1-2} \cline{5-6}
 \multirow{2}{*}{Haswell} & 097.C-0390 & & 0102.C-0558 & \multirow{2}{*}{Ruiz} & 090.C-0395 \\
 & 096.C-0876 & & 0103.C-0432 & & 084.C-0228 \\ \cline{1-2} \cline{5-6}
 Kuerster & 097.C-0090 & & 0101.C-0379 & Santos & 086.C-0284 \\ \cline{1-2} \cline{5-6}
 \multirow{2}{*}{Mayor} & 077.C-0364 & & 0100.C-0097 & Pepe & 089.C-0050 \\ \cline{3-6}
 & 072.C-0488 & Dall & 078.D-0245 & Hebrard & 078.C-0044 \\ \hline
 Albrecht & 095.C-0718 & Sterzik & 079.C-0463 & Galland & 076.C-0279 \\ \hline
 \multirow{3}{*}{Guenther} & 076.C-0010 & Melo & 076.C-0155 & Jeffers & 0104.C-0863 \\ \cline{3-6}
 & 075.C-0202 & Robichon & 074.C-0364 & Trifonov & 0100.C-0414 \\ \cline{3-6}
 & 074.C-0037 & Berdinas & 0101.D-0494 & Debernardi & 075.D-0614 \\
\end{longtable}

\begin{longtable}{ l | c c c|c c c |c|c|c }
\caption{\label{tablepente} Linear and quadratic trend model} \\
Star & \multicolumn{3}{c}{Slope } & \multicolumn{3}{c}{Quadratic term } & Peak-to-Peak amplitude& $\rm \sigma_{res}$ & Span \\
 & \multicolumn{3}{c}{ (m/s/year)} & \multicolumn{3}{c}{ (m/s/$\rm year^{2}$)} & (m/s) & (m/s) & (days) \\
\hline
\hline
\endhead

GJ234 &     -468.35 & $\pm$ &    4.49 &  & - & &    2874.03 &     29.56 &     2206 \\
GJ569 &     -1.53& $\pm$ &  &    & - & &    31.78 &     8.69 &    3177\\
GJ618A &    3.418& $\pm$ & 0.26 &  & - &  &   25.07 &     1.75 &    2429\\
GJ667C &    1.63& $\pm$ &  0.18 &  & - &  &   33.64 &     3.88 &    5559\\
GJ682 &     0.47 & $\pm$&  0.16 &  & - &  &   9.58 & 1.96 & 2896\\
GJ707 &     1.51 & $\pm$&  0.46 &  & - &  &   37.11 &     7.23 &    3682\\
GJ864 &     -52.04 & $\pm$&     1.20 & & - &  &   436.06 &     11.28 &     2912\\
GJ1135 &    87.03 & $\pm$& 7.34 &  & - &  &   241.66 &    16.12 &    1082\\
GJ3090 &    15.14 & $\pm$& 3.67 &  & - &  &   37.93 &     7.85 &    326\\
GJ3141 &    678.61 & $\pm$&     150.72 & &   - &   &   19.72 &    3.39 & 9\\
GJ3279 &    -724.70 & $\pm$&    91.95 & &    - &   &   17.10 &    2.33 & 9\\
GJ3804 &    -1.16 & $\pm$& 0.36 &  & - &  &   10.10 &     2.24 &    1082\\
GJ3813 &    23.26 & $\pm$& 2.44 &  & - &  &   168.35 &    19.05 &    1807\\
GJ4056 &    -456.73 & $\pm$&    168.32 &     & - & &    16.38 &    3.96 & 9\\
GJ4206 &    6.67 & $\pm$&  1.02 &  & - &  &   50.89 &     6.40 &    1413\\
GJ9568 &    -49.96 & $\pm$&     2.03 &  & - &  &   201.81 &     5.19 & 1447\\
GJ105B &    209.08 & $\pm$&     38.08 &     -0.70 & $\pm$& 0.13 &    12.76 &     2.19 & 2418\\
GJ126 &     141.91 & $\pm$&     60.58 &     -0.46 & $\pm$& 0.20 &    21.71 &     2.75 & 4522\\
GJ208 &     20460.44 & $\pm$&    9476.48 &    -69.53 & $\pm$&     32.19 &   49.52 &     10.73 &     803\\
GJ229 &     -27.76 & $\pm$&     9.17 & 0.09 & $\pm$&  0.03 & 17.30 &    2.54 & 4674\\
GJ333 &     1799873.14 & $\pm$&   441529.61 &   -5992.28 & $\pm$&     1469.82 &    25.80 &     5.39 & 30\\
GJ334 &     307.64 & $\pm$&     128.46 &    -1.03 & $\pm$ &     0.43 &    94.10 &     18.45 &     3778\\
GJ369 &     -70.20 & $\pm$&     57.39 &     0.23 & $\pm$&  0.19 &    8.45 & 1.87 & 1459\\
GJ480 &     -1733.78 & $\pm$&    961.02 &    5.74 & $\pm$&  3.19 &    23.13 &     5.02 & 844\\
GJ634 &     -5162.42 & $\pm$&    2235.89 &    17.18 & $\pm$& 7.44 &    15.91 &     4.11 & 538\\
GJ660 &     -1991.50 & $\pm$&    1219.15 &    6.38 & $\pm$&  4.00 &    184.55 &    12.56 &     1231\\
GJ680 &     103.69 & $\pm$&     16.45 &     -0.34 & $\pm$& 0.05 &    25.32 &     1.78 & 3008\\
GJ1001A &    -152.42 & $\pm$&    60.83 &     0.51 & $\pm$&  0.20 &    23.35 &     4.99 & 2687\\
GJ1252 &    -315944.04 & $\pm$&   77842.50 &   980.94 & $\pm$&     241.69 &    12.53 &     2.48 & 60\\
GJ2121 &    3950.66 & $\pm$&    929.67 &    -13.10 & $\pm$&     3.05 &    173.08 &    11.43 &     1446\\
GJ3212 &    -50520.55 & $\pm$&   6885.03 &    166.96 & $\pm$&     22.75 &    9.05 & 1.15 & 786\\
GJ3501 &    -191420.67 & $\pm$&   33619.01 &   630.27 & $\pm$&     110.76 &    2940.24 &    744.23 &    1836\\
GJ3528 &    2705.84 & $\pm$&    1189.34 &    -8.94 & $\pm$& 3.93 &    10.79 &     2.50 & 1601\\
GJ3530 &    -19454.02 & $\pm$&   10359.77 &   64.77 & $\pm$& 34.56 &        110.07 &17.733 &    500\\
GJ3634 &    180.64 & $\pm$&     129.98 &    -0.56 & $\pm$&  0.42 & 41.03 &     4.10 & 1424\\
GJ3700 &    470803.47 & $\pm$&   221255.89 &   -1564.93 & $\pm$&     735.02 &    7.87 & 1.25 & 9\\
GJ3779 &    152.86 & $\pm$&     85.22 &     -0.55 & $\pm$& 0.28 &    78.80 &     2.83 & 1804\\
GJ3871 &    -496.66 & $\pm$&    322.19 &    1.63 & $\pm$&  1.06 &    25.51 &     5.37 & 1492\\
GJ3874 &    -386.73 & $\pm$&    142.09 &    1.27 & $\pm$&  0.47 &    11.20 &     2.41 & 1395\\
GJ4079 &    -1508.31 & $\pm$&    569.57 &    4.97 & $\pm$&  1.88 &    26.40 &     5.30 & 1493\\
GJ4088 &    -105409.22 & $\pm$&   24383.58 &   347.91 & $\pm$&     80.47 &    29.19 &     4.14 & 727\\
GJ4332 &    -310773.31 & $\pm$&   127431.77 &   970.85 & $\pm$&     398.06 &    15.05 &     3.13 & 63\\
GJ4364 &    4365.07 & $\pm$&    1375.04 &    -14.41 & $\pm$&     4.53 &    4.42 & 0.85 & 496\\
GJ9066 &    -465.04 & $\pm$&    181.48 &    1.60 & $\pm$&  0.61 &    53.75 &     6.03 & 1780\\
GJ9118B &    14770.21 & $\pm$&    300.79 &    -48.86 & $\pm$&     1.00 &    217.15 &    2.32 & 1103\\
\hline

\end{longtable}
\tablefoot{Parameters fitted and used for the long-term correction.}

\begin{longtable}{ l | c| c | c | c}
\caption{\label{off} Offsets computed for the model chosen for each star} \\

Star & Model & Offset & Spect. & Vsini \\
 & & (m/s) & Type & (km/s) \\
\hline
\endhead

GJ~54.1 &    constant    &    -5.90 $\pm$   0.46  & M4 & 3.4 \\
GJ~87  &    constant    &    -1.89 $\pm$   0.58  & M1 & 2.0\\
GJ~126 & quadratic   &    -8.52 $\pm$   0.75  & M1 & - \\
GJ~163 &    constant    &    -4.97 $\pm$  1.334  & M3.5 & - \\
GJ~176 &    constant    &    -7.55  $\pm$  1.075 & M2 & - \\
GJ~191 &    constant    &    1.94 $\pm$   1.40  & M0 & 3.01 \\
GJ~213 & constant   &    -5.22 $\pm$   1.34  & M5 & 3.41 \\
GJ~229 & quadratic &  0.14 $\pm$   0.68  & M1 & 2.61 \\
GJ~273 &    constant    &    2.88 $\pm$   0.94  & M4 & 2.68 \\
GJ~317 &    sinusoidal   &    \textit{-3.3} $\pm$   0.27  & M3.5 & - \\
GJ~367 &    constant    &    -7.34 $\pm$   0.81  & M2 & - \\
GJ~393 &    constant    &    3.87 $\pm$   1.1   & M2 & 2.28 \\
GJ~447 &    constant    &    -5.17 $\pm$   0.79  & M5 & 2.1 \\
GJ~514 &    constant    &    -1.56 $\pm$   0.72  & M1 & 2.55 \\
GJ~536 &    constant    &    38.56 $\pm$   0.99  & M0 & 4.7 \\
GJ~551 &    constant    &  -3.48 $\pm$     0.43  & M6 & 2.7 \\
GJ~588 &    constant    &    -7.64 $\pm$   1.08  & M3 & 2.35 \\
GJ~628 &    constant    &    -8.80 $\pm$   0.73  & M4 & 2.89 \\
GJ~654 &    constant    &    -1.64 $\pm$   0.98  & M2 & 5.7 \\
GJ~667C &    linear &    -4.51 $\pm$   1.92  & M2 & - \\
GJ~674 &    constant    &    -5.80 $\pm$   1.30  & M3 & 2.59 \\
GJ~693 &    constant    &    -3.44 $\pm$   0.41  & M4 & 3.3 \\
GJ~701 &    constant    &    -2.13 $\pm$   0.89  & M1 & 2.47 \\
GJ~752A &    constant    &    -12.76 $\pm$  1.46  & M3 & 2.34 \\
GJ~754 &    constant    &    -3.00 $\pm$   0.64  & M4.5 & - \\
GJ~803 &    -    &    \textit{-3.3} $\pm$   0.27  & M1 & 8.5 \\
GJ~849 &    sinusoidal   &    20.23 $\pm$   4.82  & M3 & - \\
GJ~876 &    -    &    \textit{-3.3} $\pm$   0.27  & M3.5 & 2.75 \\
GJ~880 &    constant    &    -2.06 $\pm$   1.47  & M1.5 & 2.62 \\
GJ~887 &    constant    &    -10.84 $\pm$  2.91  & M2 & 2.7 \\
GJ~1214 &    constant    &    1.50 $\pm$   2.42  & M4 & - \\
GJ~2066 &    constant    &    -2.53 $\pm$   0.43  & M2 & - \\
GJ~3135 &    constant    &    -3.30 $\pm$   0.57  & M3 & 3.7 \\
GJ~3323 &    constant    &    -2.94 $\pm$   0.60  & M4 & 3.8 \\
GJ~9482 &    sinusoidal   &    -27.10 $\pm$  15.70  & M0 & - \\
GJ~9592 &    constant    &    -1.19 $\pm$   0.59  & M2 & 2.51 \\
LP816-60    &    constant    &    -6.09 $\pm$   0.65  & M4 & 2.7 \\
\hline

\end{longtable}
\tablefoot{Offset values obtained according to the chosen long-term model.}




\begin{longtable}{ l | c| c| c|c |c |c }
\caption{\label{tableprotconnus} Identified signals as rotation signatures} \\
Star & FAP & P (d) & $P_{E}$ (d) & $P_{P}$ (d) & Method & Reference \\
\hline
\hline
\endhead

GJ~87 & 0.004 & 53.15 & 72.6 & 53.05 & H$\alpha$ & \cite{mignon23a} \\
\hline
GJ~205 & 0.0 & 27.2 & 21.1 & 20. & Ph & \cite{hebrard2016modelling} \\
    &    & &   & 33.4 & Ph  & \cite{suarez2016} \\
    &    & &&  35    & H$\alpha$, Ca & \cite{mignon23a} \\
\hline
GJ~229 & 0.0 & 27.60 & 25.3 & 27.3 & Ph & \cite{suarez2016} \\ 
    &    & &   &27.6 & H$\alpha$, NaD, Ca & \cite{mignon23a} \\ \cline{2-4}
    & 0.007 & 26.51 & & 26.3 & Ca & \cite{mignon23a} \\
\hline
GJ~358 & 0.0 & 24.96 & 21.2 & 25.4 & Ph & \cite{hebrard2016modelling} \\
\hline
GJ~388 & 0.0 & 2.23 & - & 2.23 & VR & \cite{morin2008large} \\
\hline
GJ~393 & 0.0 & 38.5 & 44.6 & 37.2 & Ca & \cite{mignon23a} \\
& 0.0 & 33.7 &  &  &  & \\
\hline
GJ~406 & 0.0 & 2.69 &  - & & Ph & \cite{alonso2019} \\
\hline
GJ~447 & 0.0 & 61.9 &  100 & 123 & VR & \cite{bonfils2018radial} \\
& & &  & 111 & H$\alpha$ & \cite{mignon23a} \\
\hline
GJ~479 & 0.0 & 23.10 & 28.7 & 24 & Ph & \cite{hebrard2016modelling} \\ 
&    & &  &  22.5 & Ch & \cite{suarez2015} \\\cline{2-4}
 & 0.01 & 11.27 & & & & \\
\hline
GJ~514 & 0.0 & 31.4 & 33.9 & & Ch & \cite{suarez2015} \\ 
 & 0.0 & 32.4 & & 34.0 & Ca, NaD & \cite{mignon23a} \\\cline{2-4}
 & 0.0 & 15.15 & & 15.1 & H$\alpha$, NaD, Ca & \cite{mignon23a} \\
\hline
GJ~536& 0.0 & 43.9 &  32.6 & 43.1 & Ph & \cite{suarez2016} \\
    &    & &   & 39. & H$\alpha$, Ca & \cite{mignon23a} \\
    &    & &   & 49. & NaD & \cite{mignon23a} \\
\hline
GJ~551 & 0.0 & 114 & 28.4 & & Ca, NaD, H$\alpha$ & \cite{mignon23a} \\ \cline{2-4}
 & 0.0 & 44.6 & & 83 & & \cite{kiraga2007age} \\
\hline
GJ~581 & 0.0 & 67 & 114 & 127 & Ca, NaD, H$\alpha$ & \cite{mignon23a} \\
\hline
GJ~628 & 0.0 & 48.3 & 92.6 & 93 & VR & \cite{astudillo2017harps} \\
& 0.0 & 186 & & & & \\
\hline
GJ~667C & 0.0 & 90.9 & 73 & 90 & VR & \cite{robertson2014} \\
& 0.001 & 38.9 & & & & \\
 & 0.003 & 82.4 & & & & \\

\hline
GJ~674& 0.0 & 36.68 & 33.3 & 33.9 & Ch & \cite{suarez2015} \\ \cline{2-4}
 & 0.007 & 35.62 & & & Ph & \cite{suarez2016} \\
 &   & &  &  35.2 & H$\alpha$, Ca & \cite{mignon23a} \\ \cline{2-4}
 & 0.009 & 16.9 & & & & \\
\hline
GJ~740& 0.002 & 17.58 & 23.4 & & - & \cite{toledo2021super} \\
&    & &  &  18.2 & Ca & \cite{mignon23a} \\
&    & &  & 37.2 & H$\alpha$ & \cite{mignon23a} \\
\hline
GJ~832 & 0.005 & 36.73 & 50 & 35.8 & Ca & \cite{mignon23a} \\
& & & & 35.4 & H$\alpha$ & \cite{mignon23a} \\
\hline
GJ~849 & 0.0 & 35.8 & 58 & 39.2 & Ch & \cite{suarez2015} \\
& & & & 39.2 & H$\alpha$, Ca & \cite{mignon23a} \\
\hline
GJ~880& 0.001 & 40.17 & 25.8 & 37.5 & Ch & \cite{suarez2015} \\
& & & & 37 & H$\alpha$, NaD, Ca & \cite{mignon23a} \\
\hline
GJ~1132 & 0.0 & 62.6 & 63 & 125 & VR & \cite{bonfils2018radial} \\
\hline
GJ~3148& 0.0 & 11.14 & 16.2 & & - & - \\
 & 0.0 & 5.67 & & & - & - \\
\hline
GJ~4303& 0.002 & 17.63 & 15.5 & 17.5 & H$\alpha$, NaD, Ca & \cite{mignon23a} \\
\hline
\end{longtable}
\tablefoot{The three periods are i) the period found in the frequency analysis of the RV time series (P), ii) the estimation of the rotation period obtained by the relation with the mean activity level of \cite{astudillo2017magnetic} (P$\rm _E$), and iii) the period found in the literature (P$\rm _P$). The 0.0 FAP value corresponds to a value under the 10$^{-3}$ threshold.}


\begin{longtable}{l|c|c|c|c c c c c}
\caption{\label{parameters_sample}Stellar parameters} \\
\hline
Star & Mass & Spectral & Distance & \multicolumn{5}{c}{ Magnitudes } \\
 & (M$\rm _{\odot}$) & Type & (pc) & V & K & J & G & H \\
  
\hline
\endhead

\hline \multicolumn{8}{r}{\textit{Continued on next page}} \\
\endfoot

\endlastfoot

BD-15869 & 0.6089 & M0 & 25.0996 & 10.881 & 7.198 & 8.053 & 10.1618 & 7.414   \\
CD-2417578 & 0.5642 & M0V & 23.1946 & 10.847 & 7.253 & 8.103 & 10.1758 & 7.448  \\
CD-246144 & 0.6686 & M0 & 17.7978 & 9.715 & 6.169 & 7.021 & 9.0853 & 6.372   \\
CD-281745 & 0.606 & M0V & 15.6683 & 10.07 & 6.19 & 7.04 & 9.2505 & 6.42 \\
CD-301270 & 0.5649 & M1V & 22.047 & 10.998 & 7.115 & 7.986 & 10.2204 & 7.356  \\
CD-4114656 & 0.6001 & M0V & 23.6596 & 10.544 & 7.139 & 7.936 & 9.9321 & 7.303  \\
G80-21 & 0.4841 & M3.0V & 16.8022 & 11.537 & 6.933 & 7.804 & 10.4962 & 7.174  \\
G~161-7 & 0.192 & M5.0V & 7.4129 & 13.8 & 7.733 & 8.605 & 11.9686 & 8.074    \\
G~161-71 & 0.2811 & M5.0V & 13.1277 & 13.646 & 7.601 & 8.496 & 11.9081 & 7.919  \\
G~272-10 & 0.2409 & - & 15.7114 & 13.009 & 8.35 & 9.181 & 11.8578 & 8.63    \\
GJ~1 & 0.3936 & M2V & 4.3453 & 8.562 & 4.535 & 5.328 & 7.6819 & 4.732  \\
GJ~7 & 0.4524 & M0V & 23.7425 & 11.67 & 7.856 & 8.655 & 10.9187 & 8.143 \\
GJ~12 & 0.2418 & M4V & 12.2143 & 12.6 & 7.807 & 8.619 & 11.3963 & 8.068 \\
GJ~16 & 0.5275 & M1V & 16.7799 & 10.864 & 6.741 & 7.564 & 10.0009 & 6.922    \\
GJ~46 & 0.3566 & M3V & 11.9735 & 11.801 & 6.892 & 7.763 & 10.5653 & 7.203    \\
\hline
\end{longtable}
\tablefoot{Properties of the 425 stars in the sample available online.}

\begin{small}
\begin{longtable}{l|c|c|c|c|c|c|c|c|c|c}
\caption{\label{tabtot} Parameters of the long-term analysis.} \\
\hline
Star & $N_{mes}$ & $\chi^2_{0*}$ & P($\chi^2_0$) & P($Fvalue_{0}$) & $\chi^2_{1*}$ & P($F_{0/1}$) & FAP (\%) & $\chi^2_{2*}$ & P($F_{0/2}$) & P($F_{1/2}$) \\
\hline
\hline
\endhead

\hline \multicolumn{11}{r}{\textit{Continued on next page}} \\
\endfoot

\endlastfoot

CD-246144 & 13 & \textbf{2.51} & \textbf{0.001} & \textbf{0.097} & \textbf{2.664} & 0.996 & 64.8 & \textbf{2.902} & 0.997 &\textbf{1.0} \\
CD-281745 & 15 & 21.90 &\textbf{0.0} &\textbf{0.0} & 16.405 &\textbf{0.0026} & 7.8 & 17.056 &\textbf{0.002} &\textbf{0.939} \\
CD-4114656 & 21 & 13.26 &\textbf{0.0} &\textbf{0.0} & 13.304 & 0.687 & 38.4 & 13.991 & 0.712 &\textbf{1.0} \\
G80-21 & 12 & 522.44 &\textbf{0.0} &\textbf{0.0} & 550.06 & 0.974 & 61.7 & 576.72 & 0.758 &\textbf{0.904} \\
GJ~1 & 46 & 4.79 &\textbf{0.0} &\textbf{0.0} & 4.617 &\textbf{0.0009} & 10.8 & 3.962 &\textbf{0.0} &\textbf{0.0} \\
GJ~7 & 10 &\textbf{2.14} &\textbf{0.011} &\textbf{0.147} &\textbf{2.073} & 0.52 & 32.2 &\textbf{2.323} & 0.581 &\textbf{1.0} \\
GJ~54.1 & 256 & 3.88 &\textbf{0.0} &\textbf{0.0} & 3.89 & 1.0 & 45.2 & 3.497 &\textbf{0.0} &\textbf{0.0} \\
GJ~85 & 11 & 9.67 &\textbf{0.0} &\textbf{0.002} & 9.32 & 0.474 & 29.0 & 6.025 &\textbf{0.003} &\textbf{0.004} \\
GJ~87 & 154 &\textbf{1.83} &\textbf{0.0} &\textbf{0.0} &\textbf{1.822} &\textbf{0.0} & 15.9 &\textbf{1.806} &\textbf{0.0} &\textbf{0.0} \\
GJ~91 & 26 & 5.29 &\textbf{0.0} &\textbf{0.0} & 5.163 & 0.17 & 24.3 & 5.317 & 0.102 &\textbf{0.999} \\
GJ~93 & 19 &\textbf{2.19} &\textbf{0.001} &\textbf{0.082} &\textbf{2.244} & 0.946 & 88.8 &\textbf{2.001} &\textbf{0.008} &\textbf{0.011} \\
GJ~105B & 22 & 4.63 &\textbf{0.0} &\textbf{0.001} & 4.692 & 0.859 & 91.9 &\textbf{1.794} &\textbf{0.0} &\textbf{0.0} \\
GJ~118 & 20 & 4.63 &\textbf{0.0} &\textbf{0.001} &\textbf{2.604} &\textbf{0.0} &\textbf{0.1} &\textbf{2.244} &\textbf{0.0} &\textbf{0.002} \\
GJ~126 & 32 & 3.61 &\textbf{0.0} &\textbf{0.002} &\textbf{2.253} &\textbf{0.0} &\textbf{0.0} &\textbf{2.328} &\textbf{0.0} &\textbf{1.0} \\
GJ~157B & 16 & 340.68 &\textbf{0.0} &\textbf{0.0} & 302.66 &\textbf{0.0438} & 81.7 & 323.21 & 0.054 &\textbf{1.0} \\
GJ~163 & 179 & 14.03 &\textbf{0.0} &\textbf{0.0} & 14.08 & 1.0 & 63.1 & 13.70 &\textbf{0.0} &\textbf{0.0} \\
GJ~173 & 16 &\textbf{2.67} &\textbf{0.0} &\textbf{0.066} &\textbf{2.477} & 0.123 & 17.4 &\textbf{1.449} &\textbf{0.0} &\textbf{0.0} \\
GJ~176 & 115 & 17.84 &\textbf{0.0} &\textbf{0.0} & 15.94 &\textbf{0.0} &\textbf{0.2} & 15.763 &\textbf{0.0} &\textbf{0.0} \\
GJ~179 & 22 & 40.10 &\textbf{0.0} &\textbf{0.0} & 31.973 &\textbf{0.0001} & 2.2 & 7.461 &\textbf{0.0} &\textbf{0.0} \\
GJ~180 & 49 & 6.21 &\textbf{0.0} &\textbf{0.0} & 6.279 & 0.999 & 63.7 & 6.402 & 0.996 &\textbf{1.0} \\
GJ~182 & 16 & 6687.65 &\textbf{0.0} &\textbf{0.0} & 7101.31 & 1.0 & 84.6 & 5700.64 &\textbf{0.006} &\textbf{0.003} \\
GJ~191 & 228 & 4.64 &\textbf{0.0} &\textbf{0.0} & 4.649 & 1.0 & 87.6 & 4.073 &\textbf{0.0} &\textbf{0.0} \\
GJ~203 & 10 &\textbf{2.168} &\textbf{0.01} & 0.28 &\textbf{1.987} & 0.321 & 26.1 &\textbf{1.419} &\textbf{0.011} &\textbf{0.021} \\
GJ~205 & 103 & 287.40 &\textbf{0.0} &\textbf{0.0} & 194.733 &\textbf{0.0} &\textbf{0.0} & 186.496 &\textbf{0.0} &\textbf{0.0} \\
GJ~208 & 18 & 52.128 &\textbf{0.0} &\textbf{0.0} & 29.734 &\textbf{0.0} &\textbf{0.4} & 23.326 &\textbf{0.0} &\textbf{0.001} \\
GJ~213 & 122 & 4.44 &\textbf{0.0} &\textbf{0.0} & 4.362 &\textbf{0.0} & 7.3 & 4.18 &\textbf{0.0} &\textbf{0.0} \\
GJ~221 & 110 & 24.57 &\textbf{0.0} &\textbf{0.0} & 23.796 &\textbf{0.0} & 6.6 & 23.172 &\textbf{0.0} &\textbf{0.0} \\
GJ~229 & 200 & 10.41 &\textbf{0.0} &\textbf{0.0} & 8.467 &\textbf{0.0} &\textbf{0.0} & 8.101 &\textbf{0.0} &\textbf{0.0} \\
GJ~234 & 20 & >10$\rm ^4$ &\textbf{0.0} &\textbf{0.0} & 290.983 &\textbf{0.0} &\textbf{0.0} & 158.446 &\textbf{0.0} &\textbf{0.0} \\
GJ~250B & 12 &\textbf{1.514} & 0.078 & 0.389 &\textbf{1.604} & 0.987 & 75.1 &\textbf{1.584} & 0.434 & 0.471 \\
GJ~273 & 315 & 7.26 &\textbf{0.0} &\textbf{0.0} & 7.019 &\textbf{0.0} &\textbf{0.2} & 7.011 &\textbf{0.0} &\textbf{0.003} \\
GJ~282C & 10 & 8288.96 &\textbf{0.0} &\textbf{0.0} & 145.849 &\textbf{0.0} &\textbf{0.0} & 24.053 &\textbf{0.0} &\textbf{0.0} \\
GJ~299 & 24 & 4.19 &\textbf{0.0} &\textbf{0.001} & 3.769 &\textbf{0.0032} & 10.3 & 3.535 &\textbf{0.0} &\textbf{0.019} \\
GJ~300 & 39 & 10.27 &\textbf{0.0} &\textbf{0.0} & 10.255 & 0.508 & 34.0 & 9.387 &\textbf{0.0} &\textbf{0.0} \\
GJ~317 & 137 & 210.70 &\textbf{0.0} &\textbf{0.0} & 153.206 &\textbf{0.0} &\textbf{0.0} & 145.64 &\textbf{0.0} &\textbf{0.0} \\
GJ~333 & 11 & 6.419 &\textbf{0.0} &\textbf{0.003} & 4.825 &\textbf{0.0305} & 9.9 &\textbf{2.946} &\textbf{0.0} &\textbf{0.003} \\
GJ~334 & 26 & 214.704 &\textbf{0.0} &\textbf{0.0} & 182.354 &\textbf{0.0001} & 4.8 & 151.038 &\textbf{0.0} &\textbf{0.0} \\
GJ~341 & 57 & 4.538 &\textbf{0.0} &\textbf{0.0} & 3.809 &\textbf{0.0} &\textbf{0.0} & 3.713 &\textbf{0.0} &\textbf{0.001} \\
GJ~357 & 48 & 7.575 &\textbf{0.0} &\textbf{0.0} & 7.729 & 1.0 & 91.3 & 7.897 & 1.0 &\textbf{1.0} \\
GJ~358 & 34 & 24.089 &\textbf{0.0} &\textbf{0.0} & 22.244 &\textbf{0.0003} & 8.0 & 22.655 &\textbf{0.0} &\textbf{0.996} \\
GJ~361 & 101 & 10.066 &\textbf{0.0} &\textbf{0.0} & 9.922 &\textbf{0.0} & 16.5 & 10.01 &\textbf{0.0} &\textbf{1.0} \\
GJ~367 & 24 & 6.625 &\textbf{0.0} &\textbf{0.0} & 6.902 & 1.0 & 98.1 & 6.525 &\textbf{0.042} &\textbf{0.029} \\
GJ~369 & 45 & 4.051 &\textbf{0.0} &\textbf{0.0} &\textbf{2.771} &\textbf{0.0} &\textbf{0.0} &\textbf{2.733} &\textbf{0.0} & 0.068 \\
GJ~377 & 11 & >10$\rm ^4$ &\textbf{0.0} &\textbf{0.0} &>10$\rm ^4$ &\textbf{0.0006} & 79.0 & >10$\rm ^4$ &\textbf{0.0} & 0.204 \\
GJ~382 & 33 & 25.708 &\textbf{0.0} &\textbf{0.0} & 22.708 &\textbf{0.0} & 6.8 & 23.124 &\textbf{0.0} &\textbf{0.993} \\
GJ~388 & 56 & 261.576 &\textbf{0.0} &\textbf{0.0} & 245.797 &\textbf{0.0} & 7.2 & 247.617 &\textbf{0.0} &\textbf{0.969} \\
GJ~390 & 46 & 6.233 &\textbf{0.0} &\textbf{0.0} & 6.329 & 1.0 & 60.3 & 6.386 & 0.715 &\textbf{0.962} \\
GJ~393 & 175 & 4.824 &\textbf{0.0} &\textbf{0.0} & 4.776 &\textbf{0.0} & 15.6 & 4.755 &\textbf{0.0} &\textbf{0.0} \\
GJ~406 & 34 & 17.355 &\textbf{0.0} &\textbf{0.0} & 17.427 & 0.758 & 61.7 & 17.507 & 0.104 &\textbf{0.71}\\
GJ~422 & 40 & 6.606 &\textbf{0.0} &\textbf{0.0} & 6.754 & 1.0 & 74.6 & 6.927 & 1.0 &\textbf{1.0} \\
GJ~433 & 86 & 8.065 &\textbf{0.0} &\textbf{0.0} & 8.158 & 1.0 & 90.9 & 8.074 &\textbf{0.003} &\textbf{0.002} \\
GJ~438 & 19 &\textbf{1.482} & 0.06 & 0.319 &\textbf{1.438} & 0.271 & 25.6 &\textbf{1.522} & 0.309 &\textbf{1.0} \\
GJ~443 & 18 & 31.888 &\textbf{0.0} &\textbf{0.0} & 30.783 & 0.255 & 24.3 & 25.461 &\textbf{0.0} &\textbf{0.002} \\
GJ~447 & 156 & 3.353 &\textbf{0.0} &\textbf{0.0} &\textbf{2.895} &\textbf{0.0} &\textbf{0.0} &\textbf{2.832} &\textbf{0.0} &\textbf{0.0} \\
GJ~465 & 23 &\textbf{2.3} &\textbf{0.0} &\textbf{0.058} &\textbf{2.111} &\textbf{0.0123} & 10.6 &\textbf{2.209} &\textbf{0.015} &\textbf{1.0} \\
GJ~476 & 12 &\textbf{1.734} &\textbf{0.035} &\textbf{0.134} &\textbf{1.621} & 0.291 & 16.1 &\textbf{1.728} & 0.24 &\textbf{0.977} \\
GJ~479 & 58 & 11.504 &\textbf{0.0} &\textbf{0.0} & 11.676 & 1.0 & 70.1 & 11.108 &\textbf{0.0} &\textbf{0.0} \\
 il GJ~480 & 37 & 12.896 &\textbf{0.0} &\textbf{0.0} & 10.854 &\textbf{0.0} & 1.6 & 9.904 &\textbf{0.0} &\textbf{0.0} \\
GJ~486 & 12 & 3.424 &\textbf{0.0} &\textbf{0.039} & 3.428 & 0.683 & 36.8 & 3.658 & 0.549 &\textbf{0.979} \\
GJ~488 & 10 & 10.315 &\textbf{0.0} &\textbf{0.002} & 11.168 & 0.995 & 71.4 & 9.791 & 0.196 & 0.145 \\
GJ~510 & 13 &\textbf{2.89} &\textbf{0.0} &\textbf{0.112} & 3.066 & 0.997 & 94.2 & 3.331 & 0.995 &\textbf{1.0} \\
GJ~514 & 160 & 7.484 &\textbf{0.0} &\textbf{0.0} & 7.327 &\textbf{0.0} & 5.3 & 7.2 &\textbf{0.0} &\textbf{0.0} \\
GJ~526 & 32 & 8.553 &\textbf{0.0} &\textbf{0.0} & 8.72 & 0.998 & 59.2 & 8.989 & 0.993 &\textbf{1.0} \\
GJ~536 & 195 & 15.579 &\textbf{0.0} &\textbf{0.0} & 40.643 & 1 & 100.0 & 14.397 &\textbf{0.0} &\textbf{0.0} \\
GJ~550.3 & 42 & 415.302 &\textbf{0.0} &\textbf{0.0} & 261.46 &\textbf{0.0} &\textbf{0.0} & 229.617 &\textbf{0.0} &\textbf{0.0} \\
GJ~551 & 246 & 3.841 &\textbf{0.0} &\textbf{0.0} & 3.723 &\textbf{0.0} & 8.3 & 3.505 &\textbf{0.0} &\textbf{0.0} \\
GJ~555 & 14 & 5.88 &\textbf{0.0} &\textbf{0.003} & 4.016 &\textbf{0.0013} & 3.1 & 3.929 &\textbf{0.001} & 0.372 \\
GJ~569A & 24 & 49.812 &\textbf{0.0} &\textbf{0.0} & 38.996 &\textbf{0.0} & 1.4 & 40.748 &\textbf{0.0} &\textbf{1.0} \\
GJ~570B & 340 & >10$\rm ^4$ &\textbf{0.0} &\textbf{0.0} & >10$\rm ^4$ &\textbf{0.0} &\textbf{0.0} & >10$\rm ^4$ &\textbf{0.0} &\textbf{0.0} \\
GJ~581 & 242 & 70.351 &\textbf{0.0} &\textbf{0.0} & 70.602 & 1.0 & 77.0 & 70.883 & 1.0 &\textbf{1.0} \\
GJ~588 & 75 & 5.186 &\textbf{0.0} &\textbf{0.0} & 4.007 &\textbf{0.0} &\textbf{0.0} & 3.709 &\textbf{0.0} &\textbf{0.0} \\
GJ~606 & 23 & 3.373 &\textbf{0.0} &\textbf{0.001} & 3.36 & 0.552 & 65.9 & 3.295 &\textbf{0.033} & 0.227\\
GJ~618A & 20 & 25.464 &\textbf{0.0} &\textbf{0.0} &\textbf{2.5} &\textbf{0.0} &\textbf{0.0} &\textbf{2.556} &\textbf{0.0} &\textbf{0.902} \\
GJ~620 & 22 & 4.54 &\textbf{0.0} &\textbf{0.003} & 4.4 & 0.189 & 36.9 & 3.832 &\textbf{0.0} &\textbf{0.001} \\
GJ~628 & 189 & 7.717 &\textbf{0.0} &\textbf{0.0} & 7.748 & 1.0 & 81.7 & 7.789 & 1.0 &\textbf{1.0} \\
GJ~634 & 23 & 4.649 &\textbf{0.0} &\textbf{0.004} & 3.687 &\textbf{0.0} & 5.3 & 3.045 &\textbf{0.0} &\textbf{0.0} \\
GJ~637 & 17 &\textbf{1.524} & 0.055 & 0.4 &\textbf{1.614} & 1.0 & 87.2 &\textbf{1.697} & 0.998 &\textbf{0.999} \\
GJ~645 & 10 &\textbf{0.528} & 0.809 & 0.789 &\textbf{0.372} &\textbf{0.0299} & 4.0 &\textbf{0.355} &\textbf{0.013} & 0.348 \\
GJ~654 & 193 & 4.836 &\textbf{0.0} &\textbf{0.0} & 4.785 &\textbf{0.0} & 14.1 & 4.78 &\textbf{0.0} & 0.078 \\
GJ~660 & 15 & 420.748 &\textbf{0.0} &\textbf{0.0} & 51.503 &\textbf{0.0} &\textbf{0.0} & 45.789 &\textbf{0.0} &\textbf{0.042} \\
GJ~667C & 248 & 13.551 &\textbf{0.0} &\textbf{0.0} & 9.669 &\textbf{0.0} &\textbf{0.0} & 9.707 &\textbf{0.0} &\textbf{1.0} \\
GJ~674 & 212 & 43.018 &\textbf{0.0} &\textbf{0.0} & 43.111 & 1.0 & 50.1 & 42.871 &\textbf{0.0} &\textbf{0.0} \\
GJ~676A & 125 & 4218.153 &\textbf{0.0} &\textbf{0.0} & 2228.731 &\textbf{0.0} &\textbf{0.0} & 2235.63 &\textbf{0.0} &\textbf{0.997} \\
GJ~680 & 39 & 25.65 &\textbf{0.0} &\textbf{0.0} & 4.326 &\textbf{0.0} &\textbf{0.0} &\textbf{2.15} &\textbf{0.0} &\textbf{0.0} \\
GJ~682 & 21 & 3.56 &\textbf{0.0} &\textbf{0.002} &\textbf{2.393} &\textbf{0.0} & 0.8 &\textbf{2.468} &\textbf{0.0} &\textbf{0.985} \\
GJ~686 & 20 & 5.61 &\textbf{0.0} &\textbf{0.0} & 5.486 & 0.319 & 24.8 & 5.422 &\textbf{0.038} & 0.37 \\
GJ~693 & 179 & 4.18 &\textbf{0.0} &\textbf{0.0} & 4.17 & 1.0 & 80.3 & 4.16 &\textbf{0.0} &\textbf{0.0} \\
GJ~696 & 42 & 5.26 &\textbf{0.0} &\textbf{0.0} & 5.15 &\textbf{0.032} & 29.3 & 4.54 &\textbf{0.0} &\textbf{0.0} \\
GJ~699 & 117 & 5.28 &\textbf{0.0} &\textbf{0.0} & 5.201 &\textbf{0.0} & 14.9 & 4.762 &\textbf{0.0} &\textbf{0.0} \\
GJ~701 & 153 & 4.79 &\textbf{0.0} &\textbf{0.0} & 4.77 &\textbf{0.0006} & 21.8 & 4.80 &\textbf{0.0} &\textbf{1.0} \\
GJ~707 & 17 & 51.62 &\textbf{0.0} &\textbf{0.0} & 31.74 &\textbf{0.0} & 1.4 & 33.75 &\textbf{0.0} &\textbf{1.0} \\
GJ~723 & 10 & 12.441 &\textbf{0.0} &\textbf{0.001} & 9.03 &\textbf{0.0384} & 7.4 & 5.479 &\textbf{0.001} &\textbf{0.006} \\
GJ~724 & 27 & 12.48 &\textbf{0.0} &\textbf{0.0} & 12.77 & 0.996 & 55.0 & 11.86 &\textbf{0.003} &\textbf{0.004} \\
GJ~729 & 29 & 93.37 &\textbf{0.0} &\textbf{0.0} & 83.06 &\textbf{0.0002} & 9.2 & 70.97 &\textbf{0.0} &\textbf{0.0} \\
GJ~739 & 19 & 5.42 &\textbf{0.0} &\textbf{0.001} & 4.83 &\textbf{0.0171} & 9.1 & 4.82 &\textbf{0.004} &\textbf{0.502} \\
GJ~740 & 54 & 19.21 &\textbf{0.0} &\textbf{0.0} & 18.73 &\textbf{0.0017} & 12.3 & 19.01 &\textbf{0.001} &\textbf{1.0} \\
GJ~752A & 147 & 11.30 &\textbf{0.0} &\textbf{0.0} & 10.90 &\textbf{0.0} & 0.9 & 10.65 &\textbf{0.0} &\textbf{0.0} \\
GJ~754 & 182 & 3.67 &\textbf{0.0} &\textbf{0.0} & 3.67 & 0.194 & 26.9 & 3.69 & 0.204 &\textbf{1.0} \\
GJ~784 & 39 & 15.41 &\textbf{0.0} &\textbf{0.0} & 15.32 & 0.359 & 58.4 & 14.95 &\textbf{0.001} &\textbf{0.025} \\
GJ~800A & 30 &\textbf{2.886} &\textbf{0.0} &\textbf{0.006} &\textbf{2.623} &\textbf{0.0004} & 5.8 &\textbf{2.546} &\textbf{0.0} & 0.056 \\
GJ~803 & 48 & 4993.01 &\textbf{0.0} &\textbf{0.0} & 5494.86 & 1 & 100.0 & 4804.53 &\textbf{0.0} &\textbf{0.0} \\
GJ~816 & 19 & 4.238 &\textbf{0.0} &\textbf{0.005} & 3.593 &\textbf{0.0036} & 7.6 & 3.659 &\textbf{0.002} &\textbf{0.83} \\
GJ~821 & 11 &\textbf{1.456} & 0.099 & 0.428 &\textbf{1.416} & 0.52 & 28.2 &\textbf{1.369} & 0.146 & 0.375 \\
GJ~825 & 20 & 8.322 &\textbf{0.0} &\textbf{0.0} & 6.605 &\textbf{0.0003} & 1.9 & 6.947 &\textbf{0.0} &\textbf{1.0} \\
GJ~832 & 60 & 116.00 &\textbf{0.0} &\textbf{0.0} & 21.90 &\textbf{0.0} &\textbf{0.0} & 21.68 &\textbf{0.0} &\textbf{0.045} \\
GJ~842 & 10 & 4.537 &\textbf{0.0} &\textbf{0.039} &\textbf{2.327} &\textbf{0.0028} & 2.1 &\textbf{2.583} &\textbf{0.004} &\textbf{0.999} \\
GJ~846 & 55 & 15.78 &\textbf{0.0} &\textbf{0.0} & 14.40 &\textbf{0.0} & 1.5 & 14.66 &\textbf{0.0} &\textbf{1.0} \\
GJ~849 & 71 & 185.51 &\textbf{0.0} &\textbf{0.0} & 181.67 &\textbf{0.0001} & 22.4 & 180.21 &\textbf{0.0} &\textbf{0.033} \\
GJ~855 & 35 & 10.04 &\textbf{0.0} &\textbf{0.0} & 10.28 & 1.0 & 98.5 & 10.37 & 0.731 &\textbf{0.859} \\
GJ~864 & 15 & 8193.9 &\textbf{0.0} &\textbf{0.0} & 58.573 &\textbf{0.0} &\textbf{0.0} & 61.134 &\textbf{0.0} &\textbf{0.96} \\
GJ~871.1A & 28 & 5062.77 &\textbf{0.0} &\textbf{0.0} & 4954.25 & 0.171 & 33.7 & 4666.37 &\textbf{0.0} &\textbf{0.008} \\
GJ~876 & 256 & >10$\rm ^4$ &\textbf{0.0} &\textbf{0.0} & >10$\rm ^4$ &\textbf{0.0} &\textbf{0.0} & >10$\rm ^4$ &\textbf{0.0} &\textbf{1.0} \\
GJ~877 & 42 & 13.18 &\textbf{0.0} &\textbf{0.0} & 13.04 & 0.179 & 25.3 & 13.31 & 0.111 &\textbf{1.0} \\
GJ~880 & 140 & 12.81 &\textbf{0.0} &\textbf{0.0} & 12.83 & 0.932 & 41.0 & 12.377 &\textbf{0.0} &\textbf{0.0} \\
GJ~887 & 154 & 21.143 &\textbf{0.0} &\textbf{0.0} & 20.291 &\textbf{0.0} & 1.0 & 19.605 &\textbf{0.0} &\textbf{0.0} \\
GJ~891 & 28 & 4.541 &\textbf{0.0} &\textbf{0.0} & 4.664 & 1.0 & 62.7 & 4.843 & 1.0 &\textbf{1.0} \\
GJ~900 & 13 & 425.316 &\textbf{0.0} &\textbf{0.0} & 401.119 & 0.286 & 23.8 & 386.728 & 0.057 & 0.307 \\
GJ~908 & 87 & 4.41 &\textbf{0.0} &\textbf{0.0} & 4.135 &\textbf{0.0} & 1.6 & 4.138 &\textbf{0.0} &\textbf{0.643} \\
GJ~1001A & 25 & 7.792 &\textbf{0.0} &\textbf{0.0} & 7.112 &\textbf{0.0045} & 8.4 & 5.931 &\textbf{0.0} &\textbf{0.0} \\
GJ~1009 & 12 &\textbf{2.165} &\textbf{0.007} & 0.247 &\textbf{1.567} &\textbf{0.0113} & 6.3 &\textbf{1.7} &\textbf{0.014} &\textbf{0.999} \\
GJ~1018 & 15 & 6.902 &\textbf{0.0} &\textbf{0.001} & 4.536 &\textbf{0.0003} & 2.5 & 4.634 &\textbf{0.0} &\textbf{0.784} \\
GJ~1032 & 10 & 4.645 &\textbf{0.0} &\textbf{0.055} & 4.092 & 0.221 & 21.5 & 4.48 & 0.224 &\textbf{0.99} \\
GJ~1046 & 20 & >10$\rm ^4$ &\textbf{0.0} &\textbf{0.0} & >10$\rm ^4$ & 0.714 & 41.9 & >10$\rm ^4$ &\textbf{0.0} &\textbf{0.0} \\
GJ~1050 & 10 &\textbf{1.484} & 0.096 & 0.239 &\textbf{1.454} & 0.59 & 32.3 &\textbf{1.31} & 0.106 & 0.192 \\
GJ~1061 & 111 & 397.06 &\textbf{0.0} &\textbf{0.0} & 168.883 &\textbf{0.0} &\textbf{0.0} & 147.826 &\textbf{0.0} &\textbf{0.0} \\
GJ~1075 & 35 & 11.051 &\textbf{0.0} &\textbf{0.0} & 11.33 & 1.0 & 88.5 & 11.19 & 0.153 & 0.178 \\
GJ~1084 & 10 & 79.49 &\textbf{0.0} &\textbf{0.0} & 83.49 & 0.929 & 71.0 & 88.52 & 0.712 &\textbf{0.888} \\
GJ~1132 & 117 & 4.20 &\textbf{0.0} &\textbf{0.0} & 4.185 &\textbf{0.0333} & 42.1 & 3.403 &\textbf{0.0} &\textbf{0.0} \\
GJ~1135 & 18 & 1787.75 &\textbf{0.0} &\textbf{0.0} & 179.97 &\textbf{0.0} &\textbf{0.0} & 177.16 &\textbf{0.0} & 0.354 \\
GJ~1214 & 112 & 6.35 &\textbf{0.0} &\textbf{0.0} & 6.361 & 0.941 & 46.1 & 6.406 & 0.594 &\textbf{1.0} \\
GJ~1236 & 11 &\textbf{2.58} &\textbf{0.002} &\textbf{0.109} &\textbf{2.819} & 1.0 & 91.4 &\textbf{1.8} &\textbf{0.009} &\textbf{0.004} \\
GJ~1248 & 10 &\textbf{0.57} & 0.77 & 0.879 &\textbf{0.585} & 0.831 & 69.5 &\textbf{0.639} & 0.742 &\textbf{0.986} \\
GJ~1284 & 22 & >10$\rm ^4$ &\textbf{0.0} &\textbf{0.0} & >10$\rm ^4$ & 1.0 & 91.2 & >10$\rm ^4$ & 0.997 &\textbf{1.0} \\
GJ~2003 & 41 &\textbf{2.772} &\textbf{0.0} &\textbf{0.006} &\textbf{2.792} & 0.914 & 49.0 &\textbf{2.465} &\textbf{0.0} &\textbf{0.0} \\
GJ~2060 & 11 & >10$\rm ^4$ &\textbf{0.0} &\textbf{0.0} & 7084.75 &\textbf{0.0019} & 73.2 & 7859.99 &\textbf{0.003} &\textbf{1.0} \\
GJ~2066 & 111 &\textbf{2.516} &\textbf{0.0} &\textbf{0.0} &\textbf{2.421} &\textbf{0.0} & 1.9 &\textbf{2.419} &\textbf{0.0} & 0.342-\\
GJ~2121 & 28 & 121.59 &\textbf{0.0} &\textbf{0.0} & 37.70 &\textbf{0.0} &\textbf{0.0} & 22.55 &\textbf{0.0} &\textbf{0.0} \\
GJ~3009 & 11 & 13.84 &\textbf{0.0} &\textbf{0.002} & 15.20 & 1.0 & 96.7 & 15.47 & 0.781 &\textbf{0.673} \\
GJ~3018 & 25 & 4.151 &\textbf{0.0} &\textbf{0.001} & 3.938 &\textbf{0.0387} & 16.2 & 3.744 &\textbf{0.0} &\textbf{0.032} \\
GJ~3053 & 224 &\textbf{2.517} &\textbf{0.0} &\textbf{0.0} &\textbf{2.524} & 1.0 & 93.5 &\textbf{2.535} & 1.0 &\textbf{1.0} \\
GJ~3082 & 42 & 4.173 &\textbf{0.0} &\textbf{0.0} & 4.256 & 1.0 & 86.1 & 4.295 & 0.823 &\textbf{0.945} \\
GJ~3090 & 25 & 7.013 &\textbf{0.0} &\textbf{0.0} & 6.914 & 0.329 & 34.5 & 6.744 &\textbf{0.01} & 0.146 \\
GJ~3102 & 26 & 14.23 &\textbf{0.0} &\textbf{0.0} & 8.03 &\textbf{0.0} &\textbf{0.0} & 4.54 &\textbf{0.0} &\textbf{0.0} \\
GJ~3135 & 168 &\textbf{1.61} &\textbf{0.0} &\textbf{0.0} &\textbf{1.578} &\textbf{0.0} & 7.2 &\textbf{1.568} &\textbf{0.0} &\textbf{0.0} \\
GJ~3141 & 10 & 12.86 &\textbf{0.0} &\textbf{0.001} & 4.029 &\textbf{0.0001} &\textbf{0.3} & 4.95 &\textbf{0.0} &\textbf{1} \\
GJ~3148 & 62 & 26.445 &\textbf{0.0} &\textbf{0.0} & 26.126 &\textbf{0.0229} & 31.1 & 25.764 &\textbf{0.0} &\textbf{0.01} \\
GJ~3189 & 10 &\textbf{1.80} &\textbf{0.035} & 0.384 &\textbf{1.928} & 0.977 & 67.1 &\textbf{2.144} & 0.965 &\textbf{1.0} \\
GJ~3205 & 20 &\textbf{2.58} &\textbf{0.0} &\textbf{0.063} &\textbf{2.398} & 0.05 & 17.2 &\textbf{2.53} & 0.063 &\textbf{1.0} \\
GJ~3212 & 12 &\textbf{2.56} &\textbf{0.001} &\textbf{0.144} &\textbf{2.462} & 0.418 & 28.3 &\textbf{2.675} & 0.421 &\textbf{1.0} \\
GJ~3218 & 45 & 10.82 &\textbf{0.0} &\textbf{0.0} & 11.016 & 1.0 & 67.5 & 10.792 &\textbf{0.013} &\textbf{0.019} \\
GJ~3221 & 28 & 5.42 &\textbf{0.0} &\textbf{0.0} & 5.48 & 0.896 & 59.6 & 5.66 & 0.8 &\textbf{1.0} \\
GJ~3256 & 29 & 7.88 &\textbf{0.0} &\textbf{0.0} & 6.61 &\textbf{0.0} & 2.1 & 6.815 &\textbf{0.0} &\textbf{1.0} \\
GJ~3279 & 11 & 14.72 &\textbf{0.0} &\textbf{0.001} & 9.139 &\textbf{0.005} & 0.8 &\textbf{2.669} &\textbf{0.0} &\textbf{0.0} \\
GJ~3293 & 201 & 14.04 &\textbf{0.0} &\textbf{0.0} & 14.05 & 0.811 & 90.3 & 14.12 & 0.774 &\textbf{1.0} \\
GJ~3307 & 20 & >10$\rm ^4$ &\textbf{0.0} &\textbf{0.0} & 1960.572 &\textbf{0.0} &\textbf{0.0} & 29.07 &\textbf{0.0} &\textbf{0.0} \\
GJ~3323 & 148 &\textbf{2.581} &\textbf{0.0} &\textbf{0.0} &\textbf{2.591} & 1.0 & 78.4 &\textbf{2.609} & 1.0 &\textbf{1.0} \\
GJ~3331 & 19 & 5468.52 &\textbf{0.0} &\textbf{0.0} & 5099.43 & 0.072 & 69.0 & 5248.26 &\textbf{0.037} &\textbf{0.947} \\
GJ~3341 & 134 &\textbf{2.57} &\textbf{0.0} &\textbf{0.0} &\textbf{2.603} & 1.0 & 75.0 &\textbf{2.602} & 0.216 & 0.391 \\
GJ~3356 & 10 &\textbf{0.54} & 0.797 & 0.859 &\textbf{0.502} & 0.364 & 25.7 &\textbf{0.541} & 0.306 &\textbf{0.952} \\
GJ~3379 & 14 & 2434.22 &\textbf{0.0} &\textbf{0.0} & 2294.17 & 0.245 & 24.1 & 2485.33 & 0.296 &\textbf{1.0} \\
GJ~3403 & 14 & 4.67 &\textbf{0.0} &\textbf{0.012} & 5.023 & 1.0 & 96.4 & 5.186 & 0.912 &\textbf{0.858} \\
GJ~3404 & 13 &\textbf{0.87} & 0.502 & 0.45 &\textbf{0.883} & 0.787 & 84.5 &\textbf{0.826} & 0.107 & 0.185 \\
GJ~3455 & 12 &\textbf{2.67} &\textbf{0.001} &\textbf{0.164} &\textbf{2.757} & 0.887 & 92.3 &\textbf{2.279} &\textbf{0.036} &\textbf{0.037} \\
GJ~3501 & 11 & >10$\rm ^4$ &\textbf{0.0} &\textbf{0.0} & >10$\rm ^4$ & 0.344 & 53.2 & >10$\rm ^4$ &\textbf{0.0} &\textbf{0.0} \\
GJ~3502 & 22 &\textbf{1.78} &\textbf{0.009} & 0.222 &\textbf{1.327} &\textbf{0.0} & 1.3 &\textbf{0.918} &\textbf{0.0} &\textbf{0.0} \\
GJ~3528 & 15 & 3.28 &\textbf{0.0} &\textbf{0.039} & 3.402 & 0.965 & 63.0 &\textbf{2.55} &\textbf{0.002} &\textbf{0.002} \\
GJ~3530 & 24 & 166.35 &\textbf{0.0} &\textbf{0.0} & 79.45 &\textbf{0.0} &\textbf{0.0} & 70.89 &\textbf{0.0} &\textbf{0.001} \\
GJ~3543 & 94 & 4.811 &\textbf{0.0} &\textbf{0.0} & 4.805 & 0.354 & 37.5 & 4.758 &\textbf{0.0} &\textbf{0.001} \\
GJ~3563 & 12 &\textbf{1.52} & 0.075 & 0.245 &\textbf{1.457} & 0.393 & 76.2 &\textbf{1.491} & 0.192 &\textbf{0.734} \\
GJ~3634 & 73 & 20.30 &\textbf{0.0} &\textbf{0.0} & 4.212 &\textbf{0.0} &\textbf{0.0} & 4.222 &\textbf{0.0} &\textbf{0.804} \\
GJ~3643 & 21 &\textbf{1.391} & 0.084 & 0.207 &\textbf{1.24} &\textbf{0.0076} & 6.0 &\textbf{1.174} &\textbf{0.0} & 0.06 \\
GJ~3700 & 11 &\textbf{1.91} &\textbf{0.021} & 0.259 &\textbf{0.642} &\textbf{0.0001} &\textbf{0.0} &\textbf{0.437} &\textbf{0.0} &\textbf{0.007} \\
GJ~3708 & 24 &\textbf{2.32} &\textbf{0.0} &\textbf{0.089} &\textbf{2.401} & 1.0 & 77.1 &\textbf{2.376} & 0.294 & 0.343 \\
GJ~3709 & 25 &\textbf{2.34} &\textbf{0.0} &\textbf{0.076} &\textbf{2.424} & 1.0 & 79.5 &\textbf{2.378} & 0.202 & 0.204 \\
GJ~3728 & 17 & 15.51 &\textbf{0.0} &\textbf{0.0} & 15.029 & 0.311 & 26.2 & 16.031 & 0.361 &\textbf{1.0} \\
GJ~3759 & 12 &\textbf{1.61} & 0.055 & 0.387 &\textbf{1.748} & 1.0 & 94.9 &\textbf{1.79} & 0.796 &\textbf{0.738} \\
GJ~3779 & 20 & 91.49 &\textbf{0.0} &\textbf{0.0} &\textbf{2.114} &\textbf{0.0} &\textbf{0.0} &\textbf{1.728} &\textbf{0.0} &\textbf{0.0} \\
GJ~3804 & 18 &\textbf{2.85} &\textbf{0.0} &\textbf{0.032} &\textbf{1.803} &\textbf{0.0} &\textbf{0.3} &\textbf{1.878} &\textbf{0.0} &\textbf{0.992} \\
GJ~3813 & 21 & 193.97 &\textbf{0.0} &\textbf{0.0} & 34.26 &\textbf{0.0} &\textbf{0.0} & 35.52 &\textbf{0.0} &\textbf{0.997} \\
GJ~3822 & 67 & 17.376 &\textbf{0.0} &\textbf{0.0} & 16.919 &\textbf{0.0} & 9.9 & 15.319 &\textbf{0.0} &\textbf{0.0} \\
GJ~3871 & 36 & 9.653 &\textbf{0.0} &\textbf{0.0} & 6.649 &\textbf{0.0} &\textbf{0.0} & 6.39 &\textbf{0.0} &\textbf{0.006} \\
GJ~3874 & 22 &\textbf{2.553} &\textbf{0.0} &\textbf{0.05} &\textbf{2.474} & 0.187 & 26.0 &\textbf{1.887} &\textbf{0.0} &\textbf{0.0} \\
GJ~3885 & 12 &\textbf{0.668} & 0.712 & 0.846 &\textbf{0.682} & 0.829 & 44.1 &\textbf{0.723} & 0.631 &\textbf{0.954} \\
GJ~3918 & 29 &\textbf{2.34} &\textbf{0.0} &\textbf{0.03} &\textbf{2.394} & 0.999 & 72.1 &\textbf{2.443} & 0.839 &\textbf{0.988} \\
GJ~4001 & 21 & >10$\rm ^4$ &\textbf{0.0} &\textbf{0.0} & 1516.222 &\textbf{0.0} &\textbf{0.0} & 583.754 &\textbf{0.0} &\textbf{0.0} \\
GJ~4056 & 11 & 8.51 &\textbf{0.0} &\textbf{0.003} & 5.719 &\textbf{0.0102} & 2.1 & 5.035 &\textbf{0.002} & 0.117 \\
GJ~4079 & 21 & 10.64 &\textbf{0.0} &\textbf{0.0} & 11.127 & 1.0 & 80.5 & 8.42 &\textbf{0.0} &\textbf{0.0} \\
GJ~4088 & 14 & 23.69 &\textbf{0.0} &\textbf{0.0} & 14.96 &\textbf{0.0005} &\textbf{0.4} & 13.113 &\textbf{0.0} &\textbf{0.044} \\
GJ~4160 & 18 &\textbf{1.004} & 0.384 & 0.714 &\textbf{1.036} & 0.979 & 60.2 &\textbf{0.954} &\textbf{0.035} &\textbf{0.043} \\
GJ~4206 & 28 & 38.53 &\textbf{0.0} &\textbf{0.0} & 13.47 &\textbf{0.0} &\textbf{0.0} & 13.986 &\textbf{0.0} &\textbf{1.0} \\
GJ~4254 & 31 & >10$\rm ^4$ &\textbf{0.0} &\textbf{0.0} & >10$\rm ^4$ & 0.984 & 77.5 & >10$\rm ^4$ &\textbf{0.0} &\textbf{0.0} \\
GJ~4303 & 72 & 84.56 &\textbf{0.0} &\textbf{0.0} & 32.141 &\textbf{0.0} &\textbf{0.0} & 15.646 &\textbf{0.0} &\textbf{0.0} \\
GJ~4310 & 10 & 10.32 &\textbf{0.0} &\textbf{0.002} & 11.099 & 0.987 & 83.9 & 11.712 & 0.798 &\textbf{0.865} \\
GJ~4332 & 33 & 5.59 &\textbf{0.0} &\textbf{0.001} & 5.502 & 0.189 & 62.4 & 5.43 &\textbf{0.003} & 0.181 \\
GJ~4353 & 14 & 6.17 &\textbf{0.0} &\textbf{0.01} & 5.896 & 0.311 & 23.0 & 5.846 & 0.082 & 0.477 \\
GJ~4364 & 11 &\textbf{0.494} & 0.86 & 0.906 &\textbf{0.521} & 0.961 & 94.5 &\textbf{0.574} & 0.95 &\textbf{1.0} \\
GJ~9010 & 10 & 3.60 &\textbf{0.0} &\textbf{0.158} & 3.685 & 0.811 & 73.4 &\textbf{2.8} &\textbf{0.038} &\textbf{0.036} \\
GJ~9018 & 50 & 5.57 &\textbf{0.0} &\textbf{0.0} & 5.562 & 0.511 & 33.6 & 5.471 &\textbf{0.0} &\textbf{0.023} \\
GJ~9050 & 15 &\textbf{2.94} &\textbf{0.0} &\textbf{0.083} & 3.025 & 0.918 & 53.2 &\textbf{2.719} &\textbf{0.041} & 0.053 \\
GJ~9066 & 25 & 36.24 &\textbf{0.0} &\textbf{0.0} & 7.768 &\textbf{0.0} &\textbf{0.0} & 6.197 &\textbf{0.0} &\textbf{0.0} \\
GJ~9103A & 44 & 12.38 &\textbf{0.0} &\textbf{0.0} & 12.098 &\textbf{0.0207} & 21.6 & 11.507 &\textbf{0.0} &\textbf{0.0} \\
GJ~9118B & 11 & 1153.76 &\textbf{0.0} &\textbf{0.0} & 657.914 &\textbf{0.0024} & 9.7 &\textbf{2.812} &\textbf{0.0} &\textbf{0.0} \\
GJ~9133 & 14 &\textbf{1.116} & 0.27 & 0.567 &\textbf{0.659} &\textbf{0.0002} & 1.7 &\textbf{0.606} &\textbf{0.0} & 0.111 \\
GJ~9137 & 19 & 10.19 &\textbf{0.0} &\textbf{0.0} & 9.083 &\textbf{0.0164} & 26.0 & 8.396 &\textbf{0.001} &\textbf{0.038} \\
GJ~9349 & 18 &\textbf{2.73} &\textbf{0.0} &\textbf{0.033} &\textbf{2.579} & 0.132 & 23.6 &\textbf{2.736} & 0.155 &\textbf{1.0} \\
GJ~9375 & 13 & 10.71 &\textbf{0.0} &\textbf{0.0} & 10.866 & 0.786 & 47.4 & 10.335 & 0.135 & 0.244 \\
GJ~9425 & 86 & 156.12 &\textbf{0.0} &\textbf{0.0} & 155.236 &\textbf{0.048} & 27.7 & 147.663 &\textbf{0.0} &\textbf{0.0} \\
GJ~9568 & 15 & 165.99 &\textbf{0.0} &\textbf{0.0} & 3.739 &\textbf{0.0} &\textbf{0.0} & 3.967 &\textbf{0.0} &\textbf{0.999} \\
GJ~9588 & 33 & 9011.62 &\textbf{0.0} &\textbf{0.0} & 9262.826 & 1.0 & 94.3 & 8866.913 &\textbf{0.01} &\textbf{0.009} \\
GJ~9590 & 14 & >10$\rm ^4$ &\textbf{0.0} &\textbf{0.0} & >10$\rm ^4$ &\textbf{0.0002} & 13.7 & >10$\rm ^4$ &\textbf{0.0} &\textbf{0.998} \\
GJ~9592 & 148 & 6.52 &\textbf{0.0} &\textbf{0.0} & 6.471 &\textbf{0.0} & 27.5 & 6.427 &\textbf{0.0} &\textbf{0.0} \\
GJ~9647 & 11 & 3.32 &\textbf{0.0} &\textbf{0.117} & 3.07 & 0.3 & 28.1 & 3.255 & 0.227 &\textbf{0.928} \\
HD223889 & 11 & 6.53 &\textbf{0.0} &\textbf{0.006} & 6.927 & 0.978 & 68.5 & 7.665 & 0.978 &\textbf{1.0} \\
L32-8 & 19 & >10$\rm ^4$ &\textbf{0.0} &\textbf{0.0} & >10$\rm ^4$ &\textbf{0.0237} & 20.7 & >10$\rm ^4$ &\textbf{0.023} &\textbf{1.0} \\
LP776-25 & 12 & 120.20 &\textbf{0.0} &\textbf{0.0} & 126.752 & 0.978 & 81.9 & 57.257 &\textbf{0.0} &\textbf{0.0} \\
LP816-60 & 153 & 4.63 &\textbf{0.0} &\textbf{0.0} & 4.371 &\textbf{0.0} &\textbf{0.0} & 4.322 &\textbf{0.0} &\textbf{0.0} \\

\hline
\end{longtable}
\tablefoot{Parameters of the LT analysis for the 200 stars with at least ten measurements.}
\end{small}


\begin{longtable}{ l | c| c|c || l | c| c|c || l | c| c|c}
\caption{\label{tableperiode}Average activity level and estimated rotation period.} \\

\hline
Star & N$\rm _{mes}$ & $\rm log_{r'HK}$ & P(d) & Star & N$\rm _{mes}$ & $\rm log_{r'HK}$ & P (d) & Star & N$\rm _{mes}$ & $\rm log_{r'HK}$ & P (d) \\

\hline
\endhead

\hline
\endfoot

\hline
\endlastfoot

CD-246144    &    13   &    -4.619 &    23.6 & CD-281745    &    15   &    -4.383 &    16.5 & CD-4114656  &    21   &    -4.669 &    25.5 \\
G80-21 &    12   &    -3.972 &    <10   & GJ~1 &    46   &    -5.505 &    90.8 & GJ~54.1    &    256   &    -4.597 &    22.9  \\
GJ~79  &    12   &    -    &    - & GJ~85    &    11   &    -6.197 &    260.2  & GJ~87  &    154   &    -5.358 &    72.6 \\
GJ~91  &    26   &    -4.889 &    35.6  & GJ~93 &    19   &    -5.103 &    49.3 & GJ~105B    &    22   &    -5.543 &    96.2  \\
GJ~118 &    20   &    -5.278 &    64.4 & GJ~126  &    32   &    -4.997 &    42.0  & GJ~157B   &    16   &    -4.436 &    17.9 \label{GJ~157} \\
GJ~163 &    179   &    -5.398 &    77.2  & GJ~173    &    16   &    -5.120 &    50.6 & GJ~176    &    115   &    -4.813 &    31.8  \\
GJ~179 &    22   &    -5.210 &    58.1 & GJ~180  &    49   &    -5.151 &    53.0  & GJ~182 &    16   &    -3.896 &    <10 \\
GJ~191 &    228   &    -5.826 &    148.0  & GJ~205    &    103   &    -4.543 &    21.1  & GJ~208 &    18   &    -4.086 &    10.5  \\
GJ~213 &    122   &    -5.891 &    163.4 & GJ~221 &    110   &    -4.719 &    27.5  & GJ~229 &    200   &    -4.664 &    25.3  \\
GJ~234 &    20   &    -4.346 &    15.6  & GJ~250B    &    12   &    -4.924 &    37.6  & GJ~273 &    315   &    -5.475 &    86.8  \\
GJ~299 &    24   &    -5.482 &    87.8  & GJ~300    &    39   &    -5.308 &    67.3  & GJ~317 &    137   &    -5.262 &    62.8  \\
GJ~333 &    11   &    -5.443 &    82.8  &  GJ~334   &    26   &    -4.373 &    16.3  & GJ~341 &    57   &    -4.674 &    25.7  \\
GJ~357 &    48   &    -5.557 &    98.4  & GJ~358    &    34   &    -4.546 &    21.2  & GJ~361 &    101   &    -4.849 &    33.5  \\
GJ~367 &    24   &    -4.964 &    39.9  &  GJ~369   &    45   &    -4.964 &    39.9  & GJ~377 &    11   &    -5.074 &    47.2  \\
GJ~382 &    33   &    -4.639 &    24.4  & GJ~388    &    56   &    -4.185 &    12.2  & GJ~390 &    46   &    -4.651 &    24.8  \\ 
GJ~393 &    175   &    -5.037 &    44.6  & GJ~406    &    34   &    -4.574 &    22.1  & GJ~422 &    40   &    -5.669 &    116.6  \\
GJ~433 &    86   &    -5.079 &    47.6  & GJ~438    &    19   &    -5.202 &    57.3  & GJ~443 &    18   &    -4.986 &    41.3  \\
GJ~447 &    156   &    -5.365 &    73.4  & GJ~465    &    23   &    -5.660 &    115.1  & GJ~476    &    12   &    -5.198 &    57.0  \\
GJ~479 &    58   &    -4.746 &    28.7   & GJ~480    &    37   &    -5.177 &    55.2  &  GJ~486    &    12   &    -5.686 &    119.6  \\
GJ~510 &    13   &    -4.537 &    20.9  &  GJ~514   &    160   &    -4.857 &    33.9   & GJ~526    &    32   &    -5.088 &    48.2  \\
GJ~536 &    195   &    -4.832 &    32.6   & GJ~550.3   &    42   &    -4.452 &    18.3  & GJ~551 &    246   &    -4.741 &    28.4  \\
GJ~555 &    14   &    -5.502 &    90.4  &  GJ~569A   &    24   &    -4.263 &    13.8   & GJ~570B    &    340   &    -4.808 &    31.5  \\
GJ~581 &    242   &    -5.657 &    114.4  & GJ~588    &    75   &    -5.128 &    51.2  & GJ~606 &    23   &    -4.644 &    24.5  \\
GJ~618A &    20   &    -5.283 &    64.8  & GJ~620    &    22   &    -4.568 &    21.9   & GJ~628    &    189   &    -5.518 &    92.7  \\
GJ~634 &    23   &    -5.764 &    134.7  & GJ~637    &    17   &    -5.173 &    54.8  &  GJ~654    &    193   &    -5.266 &    63.1   \\
GJ~660 &    15   &    -5.092 &    48.5  &  GJ~667C   &    248   &    -5.367 &    73.6   & GJ~674    &    212   &    -4.844 &    33.3  \\
GJ~676A &    125   &    -4.583 &    22.4   & GJ~680    &    39   &    -5.048 &    45.3  & GJ~682 &    21   &    -5.371 &    74.2   \\
GJ~686 &    20   &    -5.120 &    50.6  &  GJ~693   &    179   &    -5.529 &    94.2   & GJ~696    &    42   &    -4.490 &    19.4  \\
GJ~699 &    117   &    -5.536 &    95.3   & GJ~701    &    153   &    -4.928 &    37.8  & GJ~707 &    17   &    -4.445 &    18.1   \\
GJ~724 &    27   &    -4.694 &    26.5  &  GJ~729   &    29   &    -4.354 &    15.8   & GJ~739    &    19   &    -5.106 &    49.5  \\
GJ~740 &    54   &    -4.612 &    23.4   & GJ~752A   &    147   &    -5.067 &    46.7  &  GJ~754  &    182   &    -5.405 &    78.1   \\
GJ~784 &    39   &    -4.534 &    20.8  &  GJ~800A   &    30   &    -4.884 &    35.3   & GJ~803    &    48   &    -3.870 &    <10   \\
GJ~816 &    19   &    -5.131 &    51.5   & GJ~821    &    11   &    -5.614 &    107.2  &  GJ~825    &    20   &    -4.693 &    26.4   \\
GJ~832 &    60   &    -5.118 &    50.4  &  GJ~846   &    55   &    -4.522 &    20.4   & GJ~849    &    71   &    -5.212 &    58.2  \\
GJ~855 &    35   &    -4.601 &    23.0   & GJ~864    &    15   &    -4.779 &    30.1  &  GJ~871.1A  &    28   &    -4.174 &    12.0   \\
GJ~876 &    256   &    -5.401 &    77.6  & GJ~877    &    42   &    -5.290 &    65.6   & GJ~880    &    140   &    -4.676 &    25.8  \\
GJ~887 &    154   &    -4.739 &    28.4   & GJ~891    &    28   &    -4.988 &    41.4  &  GJ~900    &    13   &    -4.237 &    13.2  \\
GJ~908 &    87   &    -5.276 &    64.1  &  GJ~1001A  &    25   &    -5.391 &    76.4   & GJ~1009    &    12   &    -4.757 &    29.2  \\
GJ~1018 &    15   &    -5.284 &    65.0   & GJ~1046   &    20   &    -5.280 &    64.5  & GJ~1061    &    111   &    -5.697 &    121.7 \\
GJ~1075 &    35   &    -4.343 &    15.5  & GJ~1132   &    117   &    -5.266 &    63.2   & GJ~1135    &    18   &    -4.721 &    27.6  \\
GJ~1214 &    112   &    -5.563 &    99.2   & GJ~1236   &    11   &    -5.392 &    76.6  & GJ~1284 &    22   &    -4.243 &    13.4 \\
GJ~2003 &    41   &    -5.683 &    119.1  & GJ~2060    &    11   &    -3.970 &    <10   & GJ~2066    &    111   &    -5.098 &    49.0  \\
GJ~2121 &    28   &    -5.400 &    77.5   & GJ~3009   &    11   &    -5.482 &    87.8  & GJ~3018    &    25   &    -4.708 &    27.0   \\
GJ~3053 &    224   &    -5.488 &    88.6  & GJ~3082    &    42   &    -4.704 &    26.9   & GJ~3090    &    25   &    -4.38  &    16.4  \\
GJ~3102 &    26   &    -4.240 &    13.3   & GJ~3135   &    168   &    -5.414 &    79.1  & GJ~3148 &    62   &    -4.369 &    16.2   \\
GJ~3192 &    18   &    0.0   &    0.0  &  GJ~3205   &    20   &    -5.193 &    56.6   & GJ~3212    &    12   &    -5.072 &    47.1  \\
GJ~3218 &    45   &    -4.581 &    22.3   & GJ~3221   &    28   &    -4.484 &    19.2  & GJ~3256    &    29   &    -4.607 &    23.2   \\
GJ~3279 &    11   &    -    &    -    & GJ~3293   &    201   &    -5.022 &    43.6 & GJ~3307    &    20   &    -4.704 &    26.9  \\
GJ~3323 &    148   &    -4.671 &    25.6   & GJ~3331   &    19   &    -3.880 &    <10   & GJ~3341    &    134   &    -5.259 &    62.5   \\
GJ~3379 &    14   &    -4.907 &    36.6  & GJ~3403   &    14   &    -4.634 &    24.2   & GJ~3404    &    13   &    -5.494 &    89.4  \\
GJ~3455 &    12   &    -5.629 &    109.8 & GJ~3501 &    11   &    -4.927 &    37.8  & GJ~3502    &    22   &    -5.304 &    66.9   \\
GJ~3528 &    15   &    -5.068 &    46.7  & GJ~3530   &    24   &    -5.064 &    46.5   & GJ~3543    &    94   &    -4.594 &    22.7  \\
GJ~3563 &    12   &    -5.386 &    75.8   & GJ~3634   &    73   &    -5.099 &    49.0  & GJ~3643    &    21   &    -5.479 &    87.4   \\
GJ~3700 &    11   &    -5.2755 &    64.0  & GJ~3708    &    24   &    -5.376 &    74.7   & GJ~3709    &    25   &    -5.472 &    86.4  \\
GJ~3728 &    17   &    -4.380 &    16.4   & GJ~3759   &    12   &    -4.891 &    35.7  & GJ~3779    &    20   &    -    &    -    \\
GJ~3804 &    18   &    -5.552 &    97.6  & GJ~3813   &    21   &    -4.883 &    35.3   & GJ~3822    &    67   &    -4.475 &    19.0  \\
GJ~3871 &    36   &    -4.745 &    28.6   & GJ~3874   &    22   &    -4.947 &    38.9  & GJ~3885    &    12   &    -5.267 &    63.2   \\
GJ~3918 &    29   &    -5.173 &    54.9  & GJ~4001   &    21   &    -4.568 &    21.9   & GJ~4056    &    11   &    -5.254 &    62.0  \\
GJ~4079 &    21   &    -4.312 &    14.8   & GJ~4088   &    14   &    -4.956 &    39.5  & GJ~4160    &    18   &    -    &    -    \\
GJ~4206 &    28   &    -4.6598 &    25.1  & GJ~4254    &    31   &    -4.545 &    21.1   & GJ~4303    &    72   &    -4.340 &    15.5  \\
GJ~4332 &    33   &    -4.539 &    20.9   & GJ~4353   &    14   &    -4.396 &    16.8  & GJ~4364    &    11   &    -4.810 &    31.6   \\
GJ~9018 &    50   &    -4.681 &    26.0  & GJ~9050   &    15   &    -4.850 &    33.6   & GJ~9066    &    25   &    -4.610 &    23.3  \\
GJ~9103A    &    44   &    -5.144 &    52.5   & GJ~9118B   &    11   &    -4.574 &    22.1  &  GJ~9133   &    14   &    -5.160 &    53.8   \\
GJ~9137 &    19   &    -5.230 &    59.8  & GJ~9349   &    18   &    -5.104 &    49.4   & GJ~9375    &    13   &    -4.515 &    20.2  \\
GJ~9425 &    86   &    -4.590 &    22.6   & GJ~9568   &    15   &    -5.575 &    101.0  & GJ~9588 &    33   &    -5.305 &    67.0   \\
GJ~9592 &    148   &    -4.747 &    28.7  & GJ~9647    &    11   &    -4.768 &    29.6   & HD223889   &    11   &    -4.829 &    32.5  \\
L32-8  &    19   &    -5.271 &    63.7   & LP776-25   &    12   &    -3.996 &    <10 &  LP816-60  &    153   &    -5.236 &    60.3  \\
\end{longtable}
\tablefoot{Estimation of the rotation period (\textit{P}) derived from the empiric law determined in \cite{astudillo2017magnetic} with the average level of activity (\textit{$log_{r'HK}$}).}

\end{appendix}


\end{document}